\documentclass[a4paper]{aa}  
\usepackage{graphicx}
\usepackage{natbib}
\usepackage{txfonts}
\begin{document}
   \title{Low Surface Brightness Galaxies around the HDF-S}

   \subtitle{I. Object extraction and photometric results}

   \author{L. Haberzettl
          \inst{1}
          \and
          D.J. Bomans\inst{1}
          \and
          R.-J. Dettmar\inst{1}
          \and
          M. Pohlen\inst{2,1}
          }

   \offprints{L. Haberzettl}

   \institute{Astronomical Institute, Ruhr-University Bochum,
              Universit\"atsstrasse 150, 44780 Bochum, Germany\\
              \email{lutz.haberzettl@astro.rub.de}
              \and
             Kapteyn Astronomical Institute, University of Groningen, P.O. Box
              800, NL-9700 AV Groningen, The Netherlands\\
}
   \date{}

% \abstract{}{}{}{}{} 
% 5 {} token are mandatory
 
  \abstract 
  % context heading (optional) % 
  {} %leave it empty if necessary {} 
  %aims heading (mandatory) 
  {{\rm The aim of this study is to
  extend the parameter space for Low Surface Brightness (LSB) galaxies to
  reach lower central surface brightnesses, smaller sizes and higher number
  densities.}}  
  % in order to provide better estimates %on the volume
  %densities of these objects.  %if comparing to earlier surveys. 
  {{\rm This study reports on photometric results of a
  search for LSB galaxies in a 0.76\,deg$^2$ field centered on the Hubble Deep
  Field-South (HDF-S). We present results from photometric analysis of the
  derived sample galaxies and compare number densities to results of former
  surveys. We used public data from the NOAO Deep Wide-Field survey and the 
  multi-wavelength Goddard Space Flight Center survey. The former 
  reaches a limiting surface brightness of 
  $\mu_{B_{W}}$\,$\sim$\,29\,mag\,arcsec$^{-2}$ and is therefore
  one of the most sensitive ground based data sets systematically
  analyzed for LSB galaxies. 
  The search was performed with two methods. For faint objects, with a
  blue central surface brightness of
  $\mu_{B_{W}}$\,$\geq$\,24\,mag\,arcsec$^{-2}$, we applied a spatial
  filtering method in combination with an object search by eye. For
  brighter objects ($\mu_{B_{W}}$\,$\geq$\,22\,mag\,arcsec$^{-2}$) we used an
  automatic search routine. To reduce the contamination by High Surface 
  Brightness (HSB) galaxies at higher redshift, mimicking LSBs due to the 
  ``Tolman Dimming'' effect, we placed a lower dimater limit of 10\farcs8 
  and compared the colors of our cadidate galaxies with the redshift tracks 
  of 5 ``standard'' HSB galaxy types.}}  
  % % results heading (mandatory) 
  {{\rm We report the
  detection of 37 galaxies with low apparent central surface brightness
  ($\mu_{B_{W}}$\,$\leq$\,22\,mag\,arcsec$^{-2}$). Using color-color diagrams
  we were able to derive a subsample of 9 LSB galaxy candidates with
  intrinsic central surface brightnesses below
  $\mu_{0,B_{W}}$\,=\,22.5\,mag\,arcsec$^{-2}$ and diameters larger than the
  preselected size limit of 10\farcs8. We selected three additional LSB
  candidates due to there extreme low blue central surface birghntess
  ($\mu_{B_{W}}$\,$\leq$\,25\,mag\,arcsec$^{-2}$). These galaxies were only
  found in the larger and more sensitive NOAO data. So finally we derived a
  sample of 12 LSB galaxy candidates and therfore this survey results in a
  four times higher surface density than other CCD based surveys for field
  galaxies before.  %We also extend the parameter %space of LSB galaxies down
  to smaller scale-length and fainter total %magnitudes.  
  }} 
  {}

   \keywords{Surveys -- galaxies: photometry -- galaxies: fundamental
   parameters(colors, radii)}

   \maketitle
%
%________________________________________________________________

\section{Introduction}

  In the early 70's, the results of galaxy surveys were strongly biased by
  selection effects, which led for example to the so called Freeman
  Law. From a study of 32 disk galaxies \citet{1970ApJ...160..811F} found
  that all galaxies have nearly the same  disk central surface brightness
  of $\mu_\mathrm{0,B}$\,=\,21.65$\pm$0.3\,mag\,arcsec$^{-2}$. The
  distribution 
  in surface brightness appeared to follow a Gaussian distribution, and
  much fainter and much brighter objects did not appear to
  exist. However, more sensitive observations in the following years were able
  to show that this distribution was biased and that surface brightness
  selection effects suppressed the detection of the population of LSB galaxies
  \citep{1983MNRAS.205.1253D,1990MNRAS.244....8D}.
  During the last years it became increasingly clear that LSB galaxies
  represent an important part of the local galaxy population.  
  %In the early 70th the LSB galaxies were not recognized. 
  %Selection effects, 
  %biasing the results of galaxy surveys, led for example to the so called
  %Freeman Law \citep{1970ApJ...160..811F}, where he found a constant
  %central surface brightness of about $\rm \mu_{0,B}\sim
  %21.65\pm0.3\;mag\;arcsec^{-2}$ studying 32 disk galaxies. The distribution
  %in surface brightness appeared to follow a Gaussian distribution. Much
  %fainter and much brighter 
  %objects did not exist. However, more sensitive observations in the
  %following years, were able to show that this distribution does not
  %exist and that the surface brightness selection effects suppressed the
  %detection of the population of LSB galaxies
  %\citep{1983MNRAS.205.1253D,1990MNRAS.244....8D}.  

\indent
  Early searches in photographic catalogs, like the UGC catalog
  \citep{ucg1973}, already showed that galaxies with a central
  disk surface brightness fainter than
  $\mu_\mathrm{0,B}$\,$\ge$\,23\,mag\,arcsec$^{-2}$ (this is 
  more than 3\,$\sigma$ fainter than the Freeman value) do exist in
  significant numbers. The amount of Low Surface Brightness (LSB)
  galaxies in the diameter limited UGC 
  catalog is much higher, compared to magnitude limited catalogs
  \citep{1997ARAA..35..267I}.
  
  \indent
  After recognizing these selection effects, the use of new amplification
  techniques and new emulsions led to more sensitive photographic
  surveys. Searches for galaxies using these surveys (e.g., visual inspections
  on POSS-II plates) could be done to a much deeper surface brightness limit
  of $\mu_\mathrm{lim,B}$\,$\approx$\,27.5\,mag\,arcsec$^{-2}$
  \citep{1988AJ.....95.1389S,1992AJ....103.1107S,1988ApJ...330..634I} 
  %and therefore, 
  resulting in higher surface densities for 
  %the 
  LSB galaxies. Due to these advancements the surface density of cluster
  LSB galaxies was increased to 
  $\sim$\,11 per square degree \citep{1988ApJ...330..634I}. For field LSB
  galaxies the surface density increased to $\sim$\,0.2 LSBs per square degree
  \citep{1990AA...228...42B}.  
  Galaxy clusters are more dominated by the dwarf elliptical LSBs
  \citep[][]{2003MNRAS.341..981S,2005MNRAS.357..819S}, whereas
  the general field is dominated by gas rich galaxies.

\indent
  With the advent of large CCDs, much more sensitive surveys became  
  possible. These surveys resulted in even higher values for the surface
  densities of LSB galaxies. The 'Texas survey'
  \citep{1997AJ....113.1212O,1997AJ....114.2448O} for example reports a
  surface density of 4 LSB 
  galaxies per square degree in the general field, which is 20 times the
  old value from \citet{1990AA...228...42B}. 

\indent
  Studies of the amount of LSB galaxies in catalogs like the UGC first showed
  that LSB galaxies are not necessarily HI poor dwarfs
  \citep{1982ApJ...263...94R}.
  A population of gas rich LSB disk galaxies exists, which even contains the
  largest and HI richest objects known today \citep[e.g., Malin
  I;][]{1987AJ.....94...23B}. In the last couple of years it was demonstrated 
  %could be shown
  that 
  %the 
  LSB galaxies represent an important part of the local galaxy
  population \citep[][]{2004A&A...428..823O,2004MNRAS.355.1303M}.

\indent
  However, until now the formation and evolution processes of the
  population of LSB galaxies are not well understood.
  %known.
  One evolutionary scenario that described the existence
  of LSB galaxies as the result of faded High Surface Brightness
  (HSB)
  galaxies could be ruled out since they do not have extremely red
  colors suggested by this
  scenario. LSBs are found to exist over the whole color range of
  HSBs \citep{1994AJ....107..530M}, although they mainly have blue
  colors.  
  However, one possible explanation for the evolution of LSBs is based
  on the star formation activity in these galaxies.
  From current sets of data it appears that star formation in LSBs propagate
  with a much lower rate than in HSB galaxies  
  \citep{2000AA...357..397V}. A possible explanation of this reduced star 
  formation could be the low HI surface density found in LSB
  galaxies. In most cases the HI density does not reach the empirical
  threshold of \citet{1989ApJ...344..685K} above which star formation should
  occur \citep{1993AJ....106..548V,1997AJ....114.1858P}.

\indent
  In this paper, we describe a search for LSB galaxies using deep ground
  based CCD mosaic imaging data of a region including the HDF-S, as well as
  all flanking fields. The main goal of this search is to enlarge the
  parameter space known for LSB galaxies. With our data we reach smaller
  disk scale-lengths, fainter total  magnitudes, and a larger sample
  volume than previous surveys, 
  however, it covers a relatively small survey area of
  0.76 square degree. The resulting sample is expected to have a significant 
  contamination of redshifted, cosmologically dimmed HSB galaxies, which
  we try to eliminate by comparing the colors of the
  selected LSB galaxy candidates to those of five standard,
  redshifted HSB galaxies (see Sect.~\ref{photred}).
  The structure of the paper is as follows: in Section 2 we present the
  photometric data used for this search, Section 3 describes the analysis we
  have done including search methods and phtometry, in Section 4 we present
  the results of our analyssis, while we end in Section 5 with summary and
  conclusions.
  
\section{Data}
  For our study we used two different public data
  sets of deep CCD mosaic imaging programs centered on  the Hubble Deep Field
  South. The first data set was obtained in 1998 (19.-24.09.) at the CTIO as
  a pilot field for the NOAO Deep Wide-Field survey \citep{noao:99}. 
  The observations were done using the Blanco-4m-Telescope, equipped with
  the Big Throughput Camera (BTC). The BTC is a mosaic camera build out of 4
  CCD detectors. Each CCD has 2048\,$\times$\,2048 pixels$^2$ with a pixel size
  of  24\,$\mathrm{\mu m}$ corresponding to a scale of 0\farcs43 per 
  pixel. Due to dithering and a large cross-shaped gap between the
  individual CCDs, the final effective field of view is about
  0.76\,$\mathrm{\sq\degr}$. 
 % \footnote{For more technical information see BTC Web page \\
 %http://www.astro.lsa.umich.edu/btc/btc.html}. 
\begin{table}
\begin{center}
\begin{tabular}{l c r }
\hline
\hline
Filter  &Origin       & Total Exp. Time\\
\hline
$B_\mathrm{W}$ &NOAO& 31920s          \\
$R$ & NOAO &  3400s          \\
$U$ & GSFC & 15600s          \\
$B$ & GSFC &  7200s          \\
$V$ & GSFC &  4500s          \\
$R$ & GSFC &  4800s          \\
$I$ & GSFC &  5100s          \\
\hline
\hline
\end{tabular}
\end{center}
\caption[ ]{Exposures times of the observations in the different filters.}
\label{exposure}
\end{table}  
  The NOAO data consists of observations in two filter bands ($R$ and
  $B_\mathrm{W}$). For the search we used the $B_\mathrm{W}$ filter, which is
  broader than 
  the typical Johnson $B$ filter and shows an extension into the wavelength
  region of the $U$ filter. The observations were conducted partly to get a
  better understanding of the broader, non-standard $B_\mathrm{W}$ filter,
  which was later also used for the NOAO Deep Wide-Field Survey. 
  %The data was made
  %public and is available from the web pages of NOAO
  %(http://www.noao.edu/noao/noaodeep/hdfsinfo.html).

  \indent
  The final $B_\mathrm{W}$ image was generated by NOAO, using dithering
  and 
  combining 38 exposures. The resulting image covers a continuous field of
  56\arcmin\,$\times$\,56\arcmin ($\sim$\,7800\,$\times$\,7800 pixels). The
  exposure-time for 
  each individual exposure was 840s, with a mean seeing of 1.7\arcsec.
  The object detection limit in surface brightness of the final combined
  image is about 27\,mag\,arcsec$^{-2}$ and the limiting surface brightness of
  the data is about 29\,mag\,arcsec$^{-2}$.
  The dithering results in a non-uniform sensitivity of the image leading 
  to a lower sensitivity at the edges of the field and around the holes in
  the image. However, our sample has just one candidate located in a region
  with lower sensitivity (\object{LSB J22311-60160}). Which is reasonable
  because LSB objects would not appear in regions of low signal to noise. All
  other selected 
  objects are located in regions with uniform sensitivity. Therefore, we do
  not account for this effect, which, however, provides us 
  with lower limits for our results.
  The R band data which were also observed by NOAO are less sensitive and
  cover a significantly smaller field compared to the $B_\mathrm{W}$ band data
  %why they
  and were therefore, not used for our studies.

%  \indent
%  The R image was produced by combining 2-4 exposures and covers only
%  $35\arcmin \times 54\arcmin$ ($4852 \times 7570$ pixels). The effective
%  exposure time varies from 1700s to 3400s over the field. The mean seeing was
%  about 1.7\arcsec. Due to the low sensitivity and the smaller size of the
%  R--field we did not use these data.
  \indent
  In order to get multi-color information of the objects detected in the
  $B_\mathrm{W}$ field, we used a second dataset, made available by
  the STIS instrument team  at Goddard Space Flight Center (GSFC). This dataset
  includes observations in five filters ($U$, $B$, $V$, $R$, and $I$). The
  observations were also obtained in September 1998 using the Blanco 4m
  Telescope equipped with the BTC. Compared to the NOAO $B_\mathrm{W}$
  data the GSFC data are less sensitive in surface brightness
  ($\sim$\,0.5\,mag\arcsec$^{-2}$) and detection limit ($\sim$\,0.5\,mag) 
  and cover a smaller field of view, but they are well 
  calibrated and span a larger color range \citep{1998AAS...193.7507T}. 
  The GSFC field is 47$\farcm$4$\times$46$\farcm$0 (6592\,$\times$\,6400
  pixels) in size with a scale of 0\farcs43 per pixel. 
  Throughout this paper we
  use the term ``multi-color'' for those objects which have measurements in
  the NOAO $B_\mathrm{W}$--band, and the GSFC $U$,$B$,$V$,$R$,$I$--bands.

  \indent
  Exposure times and filters of all CCD mosaic fields are 
  listed in Table~\ref{exposure}. The central positions of the NOAO
  $B_\mathrm{W}$ 
  field is $RA$\,=\,22$^\mathrm{h}$32$^\mathrm{m}$59$^\mathrm{s}$.5,
  $DEC$\,=\,-60$\mathrm{\degr}$35$\mathrm{\arcmin}$33$\mathrm{\arcsec}$,
  whereas the GSFC observations are centered at
  $RA$\,=\,22$^\mathrm{h}$33$^\mathrm{m}$35$^\mathrm{s}$.1,\\
  $DEC$\,=\,-60$\mathrm{\degr}$33$\mathrm{\arcmin}$45$\mathrm{\arcsec}$. All
  coordinates are given for epoch J2000.

\section{Analysis}

\subsection{Search Method and Selection Criteria}
\label{ser}
  As mentioned before, the search for LSB galaxy candidates was done using the
  $B_\mathrm{W}$ data of the NOAO, being the most sensitive of the data at our
  disposal. To identify the LSB galaxies we used different search
  methods for objects with a $\mu_\mathrm{0,B_{W}}$ brightner or fainter
  than 24.5\,mag\,arcsec$^{-2}$. 
  %We identified the LSB candidates with a combination of two methods.

  \indent
  The faint LSB galaxy candidates, with
  $\mu_\mathrm{0,B_{W}}$\,$\ge$\,24.5\,mag\,arcsec$^{-2}$  were extracted
  with a digital filter method adapted from 
  the one described by \citet{1998AJ....116.2287A}. We cut the full image
  into 9
  regions in order to speed up the filtering process and to make
  the handling of the image easier. To subtract bright stars and
  galaxies,  
  %from the images 
  which influence the filtering every region was searched with
  the source extraction program SExtractor \citep{1996AAS..117..393B},
  optimized for detecting small and bright objects. In addition to the object
  list, SExtractor can return a background image and an object image
  including only the extracted objects.

 % \indent
  To create a background--image SExtractor removes all sources using 
  a $\kappa\sigma$-clipping algorithm and estimates the background
  by derving the mode in each mesh of a grid overlaid on the original
  image. The mesh size is variable, but 
  should not be to small otherwise the background could be affected by the
  presence of residuals of objects. If the meshes are to large the small
  scale variations in the background could not be reproduced. 
  Typical values of the mesh size for most images are in the range of
  32 to 128 pixels.  
  For our search we chose a mesh size of 64 pixels, as it is recommended by
  \citet{1996AAS..117..393B}. 
  %in sExtractor language (erledigt)
  For the object image (needed to remove the detected objects from the science
  data) we used a small  detection filter (default.conv), in combination with a
  detection-threshold of 3\,$\sigma$ and a minimum detection area of 5 pixels.

  \indent
  We then generated a background subtracted and source free image by
  subtracting the two SExtractor produced images from the original. As a next
  step we filtered these images with a median filter using a fixed kernel size
  of 25 pixels corresponding to our chosen diameter limit of
  10.8\,$\mathrm{\arcsec}$. One has to keep in mind that this 25
  pixel kernel sets a strong selection criterion against much smaller and
  larger objects. We searched the resulting images by eye for local brightness
  maxima, which represent extreme LSB galaxy candidates.

  \indent
  For brighter LSB galaxy candidates with
  $\mu_\mathrm{{0,B_{W}}}$\,$\le$\,24.5\,mag\,arcsec$^\mathrm{{-2}}$ a
  small SExtractor parameter study showed 
  that we could use a simpler approach, using the output of SExtractor
  directly, since such candidate objects were noticed to be present in
  the SExtractor produced object-images and catalogs and therefore
  were subtracted from the original images during the search for
  faint LSB galaxies $\mu_\mathrm{{0,B_{W}}}$\,
  $\ge$\,24.5\,mag\,arcsec$^\mathrm{{-2}}$ (see above).
  To improve the search for LSB galaxies we choose a filter for
  SExtractor which is optimized to find faint and large objects. For our
  final search we used a modified
  tophat-filter with a kernel width of 21 pixels (comparable to the diameter
  selection criterion used, see below).  
  In the resulting new SExtractor-tables we searched for galaxies with low
  central surface brightnesses.

  \indent
  In order to derive a candidate sample of LSB galaxies we applied several
  selection criteria. As a first step we selected only galaxies with
  a central surface brightness below
  $\mu_\mathrm{{0,B_{W}}}$\,=\,22\,mag\,arcsec$^\mathrm{{-2}}$, which
  is just 1\,$\sigma$ below the so--called Freeman value of
  $\mu_\mathrm{{0,B}}$\,$\sim$\,21.65\,$\pm$\,0.3\,mag\,arcsec$^\mathrm{{-2}}$
  and therefore covers an overlap in $\mu_\mathrm{{0,B}}$
  with HSB galaxies. In addition to the surface brightness criterion,
  we also selected for galaxy diameter (D$_{29}$). We only included 
  galaxies which have diameters larger than 10.8\,arcsec. We used this 
  relative large diameter limit in order to avoid a substantial 
  contamination by high redshift galaxies, which are cosmologically 
  dimmed into the LSB surface brightness
  range (see Sect.~\ref{photred}). While we may thus introduce a
  bias against dwarf or relatively distant LSB galaxies, we expect it to 
  result in a cleaner sample of LSB galaxies. 

  \indent
  Using the two described search methods and applying our two selection
  criteria ($\mu_\mathrm{{0,B_{W}}}\,\ge\,$22.0\,mag\,arcsec$^{-2}$,
  D$_{29}$\,$\ge$\,10.8\,arcsec), we were able to derive a total sample of 37
  galaxies in the 
  larger and more sensitive $B_\mathrm{W}$ image of the NOAO.
  After extracting these LSB candidates from this image we 
  tried to locate them in the smaller and less sensitive
  multi-color data of the GSFC. This left us with an overlapping,
  multi-color sample of
  19 candidates. The remaining 18 candidates were at locations in the NOAO
  $B_\mathrm{W}$ image which were not covered by or did not have
  the required sensitivity in the GSFC data. 

\subsection{Photometry and Profile Fitting}
\label{photpro}
  After the selection of the candidate galaxies, we extracted their
  photometric parameters. For this step we first subtracted the background
  light from the 
  NOAO $B_\mathrm{W}$-image using the SExtractor background image and then
  fitted isophotes with ellipses using the 
  {\it ellipse} task included in the {\it IRAF/STSDAS} package. We also fitted
  the $B$-band data of the GSFC in the same way. For the photometric
  calibration of the $U$,$B$,$V$,$R$,$I$ images we used the calibration
  from the Goddard Space Flight Center/STIS team (parameter see
  Table~\ref{calibpar}): 
\begin{eqnarray}
 m\,=\,-2.5\cdot \log\,CPS\,+\,C\,+\,X \cdot AIRMASS\,+\,K \cdot COLOR
\label{calib}
\end{eqnarray}
  In Eqn.~\ref{calib} we use counts per second (CPS) for the flux.

  In order to fit ellipses with the {\it ellipse} task we allowed to vary the
  center position, the ellipticity as well as the position angle for the fitted
  ellipses. For the ellipses we choose a logarithimic spacing in radial
  direction with a step size of 0.1, meaning the next ellipse is fitted
  going inward from the position of the first one, at 1/(1\,+\,step
  size)$\cdot$SMA in pixel (SMA = Semi Major Axis) (see {\it IRAF} help for
  {\it geompar}).    
\begin{table}
\begin{center}
\begin{tabular}{c c c c c c}
\hline
\hline
Filter&C&X&AIRMASS&K&COLOR\\
\hline
$U$&23.032&-0.392&1.172&0.041&$U$-$B$\\
$B$&25.385&-0.204&1.168&0.133&$B$-$V$\\
$V$&25.561&-0.108&1.248&0.022&$B$-$V$\\
$R$&25.748&-0.049&1.180& -- & -- \\
$I$&24.899&-0.033&1.187&0.051&$V$-$I$\\
\hline
\end{tabular}
\end{center}
\caption{Parameters for the photometric calibration (see
  Eq.~\ref{calib}) of the GSFC data. Where C represents the photometric
  calibration constant, X the extinction coefficient and K the color
  coefficient.}
\label{calibpar}
\end{table}
  
 \indent
  From the ellipse fitting routine we derived the azimuthally averaged
  radial surface brightness distributions (see Fig.~\ref{appA}) and
  fitted them with a simple exponential law, since none show the
  presence of a significant de Vaucouleur bulge component (for more details
  see Sect.~\ref{radprof}).
  \begin{eqnarray}
  \Sigma\mathrm{\left(r\right)}=\Sigma_\mathrm{0}e^{-\frac{r}{\alpha}}
  \label{expon}
  \end{eqnarray}
  For this fit we excluded the innermost 0\farcs9 of the profile, 
  which are influenced by the seeing of $\sim$\,1.7 arcsec (diameter). 

\indent
  For several candidate galaxies we see a break in the outer parts of the
  profiles 
  with either an up-bending or down-bending shaped profile 
  (see figures in App.~\ref{appA}). 
  This break is visible in both independent datasets 
  (NOAO $B_\mathrm{W}$- and GSFC $B$-band) 
  indicating that this is not an effect of an incorrect skysubtraction.
  We only fitted the inner exponential part inside the break.
\begin{figure}
\centering
\includegraphics[width=8.5cm]{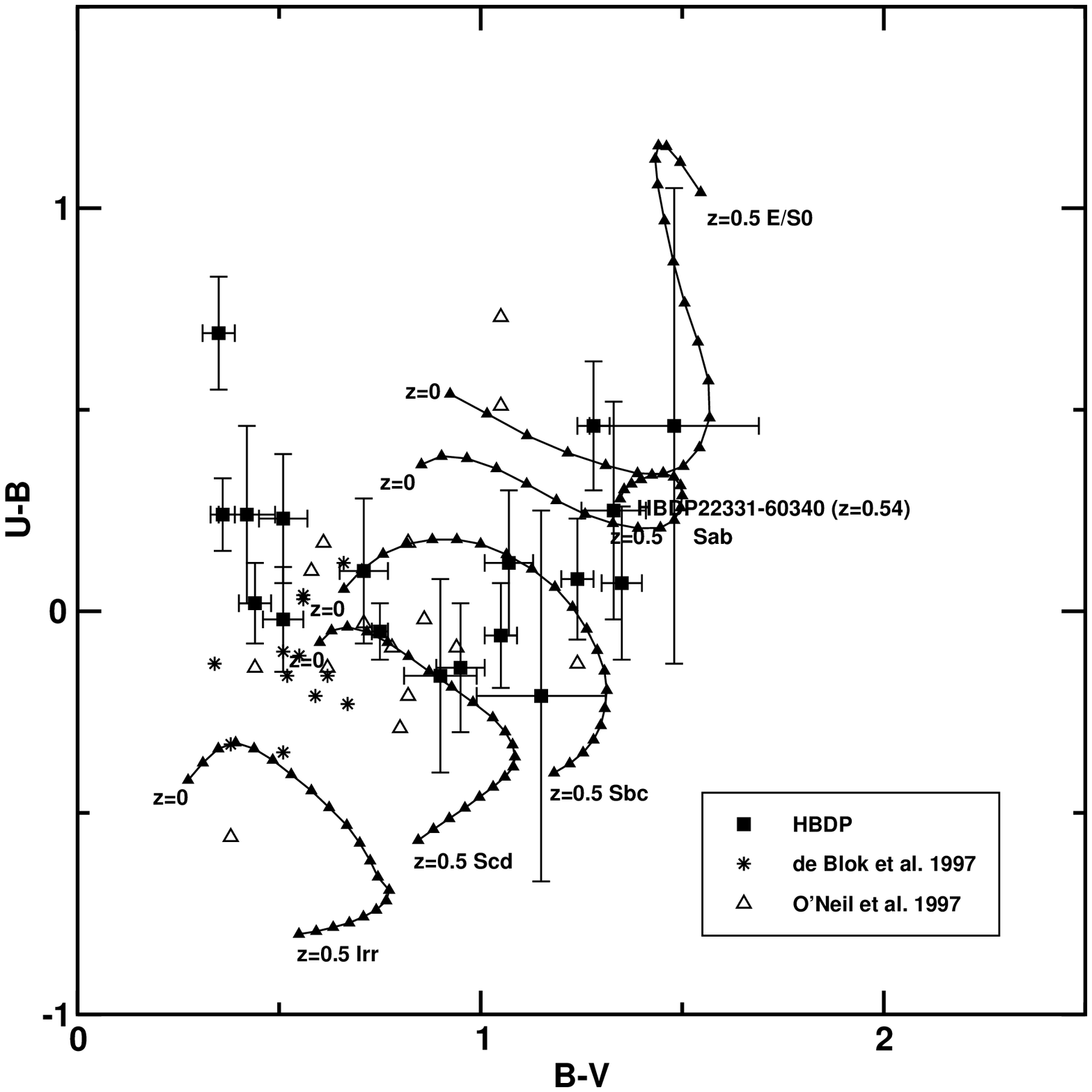}
\includegraphics[width=8.5cm]{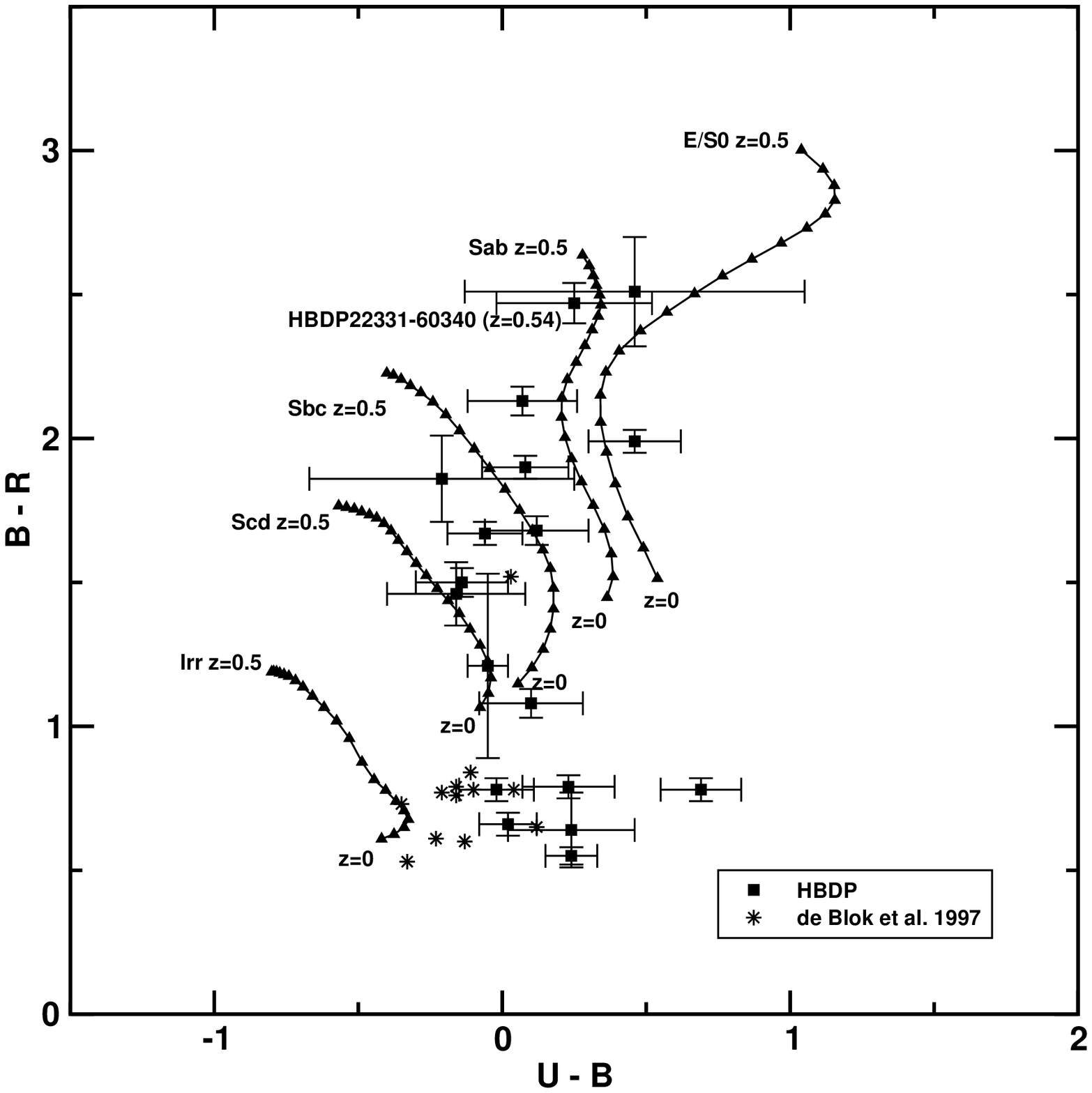}
\caption{Color-Color diagrams (left panel: $B-V$ vs. $U-B$, right panel:  $U-B$
  vs. $B-R$). In these diagrams we compare the location of our sample
  (LSB J)
  with the redshift tracks of five standard galaxy types (Irr, Scd, Sbc, Sab,
  E/S0). This tracks were adopted from a work of Liu \& Green (1998). We
  also plotted two sample of known LSB galaxies from O'Neil et al. (1997)
  (triangles) and de Blok et al. (1997) (stars). The diagrams show a clear
  separation between some of our sample galaxies and the redshift
  tracks. These galaxies are selected as LSB candidates.}
\label{UBBV}
\end{figure}
  From the exponential fits to the radial surface brightness we obtained the
  scale-length in the $B_\mathrm{W}$- and $B$-band filter for all galaxies in
  our sample. 

  \indent
  We also derived a simple estimate for the inclination angle, assuming a
  flat disk, and using the $r_\mathrm{minor}$ and $r_\mathrm{major}$ 
  radii of the outermost successful ellipse fits in the $B_\mathrm{W}$
  filter.  
  \begin{eqnarray}
  i\,=\,\cos^\mathrm{-1}\frac{r_\mathrm{minor}}{r_\mathrm{major}}
  \label{inc}
  \end{eqnarray}
  With this estimation we obtained the
  inclination corrected blue central surface brightness
  $\mu_\mathrm{0,B_{W_{corr}}}$. We used the same inclination angle 
  for the surface brightness correction in all photometric
  bands of the GSFC data. 
  \begin{eqnarray}
  \rm \mu_{0,X_{corr}}(0)=\mu_X(0)-2.5\log \left(\cos i \right)
  \label{incli}
  \end{eqnarray}

  \indent
  In a next step we extracted the total $B_\mathrm{W}$ and $B$ magnitudes for
  all selected objects. In this context we use total magnitudes for a circular
  aperture with a radius including all the flux, since we have choosen the
  aperture radius to be significant larger than the optical size of the
  galaxies.
  To compare our results with already existing surveys, we need to convert 
  the $B_\mathrm{W}$ magnitudes into Johnson $B$ magnitudes.  
  For all objects with counterparts in the GSFC $B$ data,  we used the
  magnitudes directly measured on this data set.
  Since we have no $B$-band information for several of the deep
  NOAO $B_\mathrm{W}$ image detections (they are not located in the area 
  covered by the GSFC data), we calculated a mean offset of
  $<$\,$B_\mathrm{W}$\,-\,$B$\,$>$\,=\,0.02\,$\pm$\,0.07\,mag, between the
  $B_\mathrm{W}$- and the $B$-filter and applied this conversion to the
  measurements of the galaxies only detected in the $B_\mathrm{W}$ data. Those
  estimated values are indicated by square 
  brackets (e.g.~Table~\ref{tabelle}).
  
  \indent
  For all sample galaxies, present in the GSFC data we measured the total
  $UVRI$ magnitudes in the same way, applying the aperture radius derived in
  the $B_\mathrm{W}$ filter. Results of the $B_\mathrm{W}$ and $B$ magnitudes
  together with coordinates, scale-length, and central surface brightnesses
  are listed in Table~\ref{tabelle}. This table is organized as followed.  

  \indent
  {\bf Column 1:} Galaxy names as used in this paper. 

  {\bf Column 2-3:} Right ascension and declination of the galaxies. We
   measured the coordinates, using peak intensities of the objects in the
   NOAO field. The astrometry was performed by the NOAO as part of the data
   reduction. 

  {\bf Column 4-7:} Measured total magnitudes in $B_\mathrm{W}$ and $B$. For
   the galaxies 
   without counterparts in the Goddard field we calculated total $B$ magnitudes
   using the measured mean offset between $B_\mathrm{W}$ and $B$ (values
   in square brackets). The errors in
   Cols. 5 and 7, are standard errors resulting from the photometric
   measurements within IRAF. 

  {\bf Column 8-11:} Central surface brightnesses and standard errors in mag
   arcsec$^\mathrm{-2}$ measured in the $B_\mathrm{W}$ (column 6, 7) and the
   $B$ filter
   (column 8,9). The errors are standard errors, resulting from the linear
   regression. Linear regressions were progressed using the analyzing 
   software {\it xmgrace}. For those galaxies were no detections are 
   available in the
   $B$-band data, we estimated $\mu_\mathrm{0,B}$ by applying a
   mean offset of 0.41$\pm$0.28, derived from galaxies detected in the
   B-dand of the GSFC data (values in square brackets). 

  {\bf Column 12:} Inclination angle  in degree, calculated from the 
   ratio of the major and the minor axis, obtained from the ellipse fit 
   (see Eq.~\ref{inc}) in the $B_\mathrm{W}$ filter.

  {\bf Column 13-14:} Inclination corrected $B_\mathrm{W}$ and $B$ band
   central surface brightness. For the correction we used the simple approach
   of Eq.~\ref{incli}. For those galaxies without $B$-band 
   information we estimated the inclination corrected $B$ central
   surface brightness 
   $\mu_\mathrm{0,B_{corr}}$ using the estimated $B$ central surface brightness
   $\mu_\mathrm{0,B}$ and the inclination angle derived from the
   $B_\mathrm{W}$ profiles (values in square brackets). 

  {\bf Column 15-18:} Disk scale-length in arcsec, obtained from
   the exponential fit. In the $B$ band only galaxies detected in the
   deep NOAO field with counterparts in the Goddard field are fitted.
   The errors (column 14,16) are calculated using standard
   errors resulting from the linear regression and the Gaussian error
   propagation. 

  {\bf Column 19:} The letter in this Col., indicates whether the object is
   located in both the Goddard and the NOAO field (b), or only 
   identified in the NOAO field (o).

\vspace{0.5cm}
\indent
  The results for the $UBVRI$ magnitudes and the colors
  including errors of the objects detected also in the GSFC data are
  listed in Table~\ref{color}.

  {\bf Column 1:} Name as used in Table~\ref{tabelle}.

  {\bf Column 2-6:} Measured total magnitudes in the $UBVRI$ filters for the
  galaxies also detected in the GSFC. The magnitudes were derived performing
  aperture photometry using the same aperture radius estimated from 
  the $B_\mathrm{W}$ data. 

  {\bf Column 7-14:} Colors are derived from the total $UBVRI$ 
  magnitudes. The
  errors in Cols. 8, 10, 12, 14 result from the Gaussian error
  propagation using the standard errors of the total magnitudes from the
  photometric measurements with IRAF. 

\setlength{\tabcolsep}{0.5mm}
\renewcommand{\arraystretch}{1.5}
\begin{table*}
\begin{center}
\begin{tabular}{c c c c c c c c c c c c c c c c c c c}
\hline
\hline
Name&RA(J2000)&DEC(J2000)&$B_\mathrm{W}$&$\delta
B_\mathrm{W}$&$B$&$\delta B$&$\mu_\mathrm{0,B_{W}}$&$\delta\mu_\mathrm{0,B_{W}}$&$\mu_\mathrm{0,B}$&$\delta\mu_\mathrm{0,B}$&$i$&$\mu_\mathrm{0,B_{W_{corr}}}$&$\mu_\mathrm{0,B_{corr}}$&$\alpha_{B_\mathrm{W}}$&$\delta\alpha_{B_\mathrm{W}}$&$\alpha_B$&$\delta\alpha_B$&\\
(1)&(2)&(3)&(4)&(5)&(6)&(7)&(8)&(9)&(10)&(11)&(12)&(13)&(14)&(15)&(16)&(17)&(18)&(19)\\
\hline
\object{LSB J22291-60303}&22:29:18.73 & -60:30:39.0 &20.31&0.16&[20.29]&[0.17]&22.95&0.08&[23.36]&[0.29]&52&23.48&[23.89]&1.55&0.04&--&--&o\\
\object{LSB J22291-60522}&22:29:17.15 & -60:52:25.8 &18.82&0.08&[18.80]&[0.11]&22.59&0.05&[23.00]&[0.28]&14&22.62&[23.03]&2.94&0.08&--&--&o\\
\object{LSB J22292-60540}&22:29:28.53 & -60:54:09.0 &19.08&0.09&[19.06]&[0.11]&22.36&0.02&[22.77]&[0.28]&52&22.89&[23.30]&1.87&0.03&--&--&o\\
\object{LSB J22293-60523}&22:29:35.92 & -60:52:32.0 &19.04&0.09&[19.02]&[0.11]&23.38&0.09&[23.79]&[0.29]&27&23.50&[23.91]&3.19&0.09&--&--&o\\
\object{LSB J22295-61001}&22:29:53.87 & -61:00:15.3 &18.99&0.08&[18.97]&[0.11]&22.82&0.02&[23.23]&[0.28]&69&23.93&[24.43]&6.03&0.34&--&--&o\\
\object{LSB J22300-60300}&22:30:09.75 & -60:30:06.6 &19.09&0.09&[19.07]&[0.11]&22.94&0.05&[23.35]&[0.28]&48&23.38&[23.79]&3.74&0.12&--&--&o\\
\object{LSB J22300-60380}&22:30:03.37 & -60:38:07.3 &18.94&0.09&[18.92]&[0.11]&22.92&0.01&[23.33]&[0.28]&18&22.97&[23.38]&3.53&0.03&--&--&o\\
\object{LSB J22301-60415}&22:30:17.59 & -60:41:50.9 &19.76&0.13&[19.74]&[0.15]&22.47&0.07&[22.88]&[0.29]&53&23.02&[23.43]&1.81&0.03&--&--&o\\
\object{LSB J22302-60352}&22:30:22.97 & -60:35:29.4 &18.24&0.06&[18.22]&[0.09]&22.67&0.08&[23.08]&[0.29]&25&22.78&[23.19]&4.34&0.35&--&--&o\\
\object{LSB J22302-60474}&22:30:23.08 & -60:47:48.9 &19.94&0.14&[19.92]&[0.16]&23.23&0.03&[23.64]&[0.28]&60&23.98&[24.39]&3.19&0.09&--&--&o\\
\object{LSB J22303-60514}&22:30:30.40 & -60:51:48.9 &20.08&0.15&[20.06]&[0.17]&22.47&0.05&[22.88]&[0.28]&55&23.07&[23.48]&2.01&0.11&--&--&o\\
\object{LSB J22304-61004}&22:30:42.63 & -61:00:41.8 &18.67&0.08&[18.65]&[0.11]&22.57&0.06&[22.98]&[0.29]&52&23.10&[23.51]&3.62&0.12&--&--&o\\
\object{LSB J22311-60160}&22:31:14.40 & -60:16:08.0 &23.80&0.81&[23.78]&[0.81]&26.52&0.29&[26.93]&[0.40]&39&26.79&[27.20]&2.72&0.54&--&--&o\\
\object{LSB J22311-60503}&22:31:13.04 & -60:50:34.4 &18.51&0.07&18.50&0.02&22.39&0.03&22.55&0.04&25&22.50&22.66&2.53&0.06&2.53&0.06&b\\
\object{LSB J22315-60481}&22:31:59.35 & -60:48:17.3 &20.60&0.17&20.61&0.06&23.50&0.04&23.51&0.06&64&24.40&24.41&2.09&0.04&2.01&0.07&b\\
\object{LSB J22320-60381}&22:32:02.65 & -60:38:14.0 &22.24&0.39&22.10&0.05&26.86&0.05&26.91&0.03&36&27.09&27.14&4.18&0.16&6.46&0.15&b\\
\object{LSB J22321-61015}&22:32:17.78 & -61:01:56.0 &20.13&0.15&[20.11]&[0.17]&23.30&0.03&[23.71]&[0.28]&65&24.24&[24.65]&4.34&0.35&--&--&o\\
\object{LSB J22322-60142}&22:32:27.40 & -60:14:25.6 &20.24&0.14&20.31&0.04&22.82&0.05&23.09&0.02&49&23.28&23.55&0.94&0.03&1.10&0.02&b\\
\object{LSB J22324-60520}&22:32:41.81 & -60:52:07.1 &19.62&0.12&19.60&0.04&22.25&0.05&22.30&0.10&50&22.73&22.78&1.57&0.05&1.60&0.07&b\\
\object{LSB J22325-60155}&22:32:52.19 & -60:15:58.4 &18.98&0.09&19.00&0.03&23.26&0.16&23.82&0.04&46&23.66&24.25&4.18&0.48&9.05&1.51&b\\
\object{LSB J22325-60211}&22:32:55.39 & -60:21:17.1 &20.14&0.15&20.10&0.05&23.18&0.01&23.25&0.05&50&23.66&23.73&3.19&0.09&2.31&0.20&b\\
\object{LSB J22330-60543}&22:33:03.19 & -60:54:38.4 &18.71&0.08&18.70&0.02&22.24&0.05&22.40&0.05&25&22.46&22.62&2.13&0.04&2.05&0.08&b\\
\object{LSB J22331-60340}&22:33:13.73 & -60:34:04.7 &21.01&0.22&21.13&0.07&24.41&0.08&24.49&0.06&26&24.64&24.72&2.13&0.08&2.05&0.04&b\\
\object{LSB J22332-60561}&22:33:26.60 & -60:56:16.6 &19.49&0.11&[19.51]&[0.13]&22.71&0.02&[23.12]&[0.28]&65&23.65&[24.06]&2.72&0.07&--&--&o\\
\object{LSB J22341-60475}&22:34:14.64 & -60:47:53.3 &19.59&0.12&19.57&0.04&22.51&0.04&22.58&0.01&58&23.20&23.27&2.86&0.15&2.86&0.08&b\\
\object{LSB J22342-60505}&22:34:20.23 & -60:50:51.2 &19.70&0.10&19.73&0.03&23.36&0.01&23.39&0.02&65&24.30&24.33&3.03&0.03&2.72&0.07&b\\
\object{LSB J22343-60222}&22:34:32.28 & -60:22:20.5 &19.18&0.09&19.10&0.03&23.30&0.03&23.44&0.02&59&24.02&24.16&4.53&0.19&4.34&0.17&b\\
\object{LSB J22345-60210}&22:34:57.78 & -60:21:07.8 &20.22&0.15&20.15&0.05&22.36&0.05&22.81&0.09&38&22.62&23.07&1.02&0.03&1.28&0.08&b\\
\object{LSB J22352-60420}&22:35:22.50 & -60:42:09.0 &20.19&0.15&20.22&0.05&23.16&0.03&23.47&0.03&59&23.88&24.19&3.74&0.13&3.74&0.13&b\\
\object{LSB J22353-60311}&22:35:34.37 & -60:31:11.6 &19.36&0.10&19.18&0.03&23.39&0.07&23.56&0.02&22&23.47&23.64&2.72&0.34&2.78&0.14&b\\
\object{LSB J22354-60122}&22:35:46.14 & -60:12:20.4 &19.82&0.13&19.77&0.04&22.51&0.04&22.95&0.01&50&22.99&23.43&2.31&0.10&2.72&0.07&b\\
\object{LSB J22355-60183}&22:35:58.34 & -60:18:39.9 &19.83&0.13&19.75&0.04&22.52&0.04&22.78&0.04&50&23.00&23.26&2.26&0.14&2.26&0.14&b\\
\object{LSB J22355-60390}&22:35:54.99 & -60:39:01.6 &20.39&0.16&20.28&0.05&23.57&0.06&23.66&0.08&68&24.64&24.73&2.72&0.14&3.10&0.09&b\\
\object{LSB J22360-60561}&22:36:00.81 & -60:56:17.4 &19.34&0.10&[19.32]&[0.12]&22.72&0.04&[23.13]&[0.28]&67&23.74&[24.15]&4.72&0.21&--&--&o\\
\object{LSB J22361-60223}&22:36:17.20 & -60:22:31.7 &19.82&0.13&19.86&0.04&22.85&0.08&23.13&0.01&49&23.31&23.59&2.13&0.08&2.47&0.06&b\\
\object{LSB J22361-60562}&22:36:17.87 & -60:56:27.2 &19.86&0.13&[19.84]&[0.15]&22.85&0.03&[23.26]&[0.28]&61&23.64&[24.05]&3.74&0.26&--&--&o\\
\object{LSB J22364-60405}&22:36:45.09 & -60:40:50.6 &22.74&--&[22.72]&--&25.41&0.28&[25.82]&[0.40]&--&--&--&2.22&0.59&--&--&o\\
%\hline         
\hline         
\end{tabular}
\caption{Surface photometry parameters for all the LSB galaxy candidates
found in the deep $B_\mathrm{W}$ image. The letter in Col. 19 indicates if
there was a counterpart in the GSFC multi-color data (b) or if the
object was only located in the larger deep $B_\mathrm{W}$ image (o). All
  values in square brackets are derived using a mean offset between $B_W$- and
  $B$-band data (see text).}
\label{tabelle}
\end{center}
\end{table*}  
\renewcommand{\arraystretch}{1}

\subsection{How many LSB candidates are HSB disks at higher redshift?}
\label{photred}

  Since the surface brightness $\mu$ of an object is only independent of
  distance in the nearby Universe, it is not trivial to decide whether
  our candidates are genuine LSB galaxies or dimmed ``normal'' HSB 
  galaxies at high redshift. At larger distances (z $>$ 0.1) 
  the surface brightness
  $\mu$ increases significantly with the redshift z \citep[e.g.][]{peacock99} 
  \begin{eqnarray}
  \mu_{\nu}\sim\frac{\mu_\mathrm{tot}}{(1+z)^4}
  \label{flhel}
  \end{eqnarray}
  where $\mu_\mathrm{tot}$ represents the measured and $\mu_{\nu}$
  the surface brightness corrected for cosmological dimming.
  For example the surface brightness for a galaxy will be reduced by
  1.8\,mag\,arcsec$^{-2}$ at a distance of z\,=\,0.5.
  So, without any distance information it is not obvious, if a galaxy with 
  $\mu_0$\,=\,23\,mag\,arcsec$^{-2}$ is a genuine low surface brightness
  galaxy  or a cosmological dimmed normal galaxy. Since LSB galaxies are 
  present for all Hubble types  \citep{1992AJ....103.1107S}, a pure
  morphological selection can be misleading. 
  
  \indent
  Subtracting the population of higher redshifted
  galaxies (z\,$\ge$0.15, see below) should provide us a reasonably clean
  sample of  genuine LSB galaxy candidates.
  Since spectroscopic redshifts for our LSB galaxy candidates are not
  available yet, we need a method to select the LSB galaxy candidates against
  the background of higher redshifted galaxies. The relatively large
  diameter limit, used for the selection of the galaxies, was a first step 
  to keep this contamination low.  
  A natural choice for further selection is the use of photometric redshifts 
  \citep{1985AJ.....90..418K,1986ApJ...303..154L}. This method is based on the
  change of galaxy colors due to the shift of spectral features (e.g. Balmer
  break, Lyman break) into redder filter bands with increasing redshift. 
  A typical spectral energy distribution of a galaxy (e.g., Sb), therefore,
  moves along a specific track in a color-color diagram
  \citep{1998AJ....116.1074L}. 
  The location and shape of these tracks depends on the knowledge of the
  spectral energy distribution (and, therefore, the stellar population mix) of
  the galaxy. Since we have only very limited knowledge about the star
  formation history of LSB galaxies and, therefore, their spectral energy
  distributions we cannot use standard photometric redshift methods to
  select our LSB galaxies directly. However, we know from deep pencil beam
  redshift surveys \citep[e.g.][]{1995ApJ...455...50L}, that the dominant
  contamination to our sample is caused by ``normal'' HSB disk galaxies at
  intermediate redshifts (z=0.1 to 0.5). Galaxies with distances
  z\,$\ge$\,0.5 have much smaller angular sizes than our diameter
  limit \citep{1996AJ....112..369G}. For normal galaxies with redshifts in the 
  range between z=0.15 to 1, the photometric redshift method works well. For
  example \citet{2000ApJ...538..493Y} obtained photometric redshifts with an
  accuracy around $\Delta$z\,$\sim$\,0.09 for redshifts z\,$<$\,1.0.
  \citet{2002MNRAS.330..889F} derived an error component for the photometric
  redshifts following $\sigma_\mathrm{z}$\,=\,0.065\,(1\,+\,$z$) and resulting
  in an accuracy of $\Delta$z\,$\le$\,0.10 for redshifts z\,$\le$\,0.5. 
  Due to this relatively large errors it is not possible to derive photometric
  redshifts for lower redshifted (z\,$<$\,0.15) galaxies.
  %{\bf S}ubtracting the population of higher
  %redshifted galaxies (z\,$\ge$0.15) should provide us a reasonably clean
  %sample of {\bf genuine} LSB galaxy candidates. 
  At this point it is important to mention that we do not aim at
  obtaining photometric redshifts for local galaxies (z\,$\le$\,0.15) in the
  following paragraphs. Due to the large uncertainties of the photometric
  redshifts this is not feasible. 
  However, as we are inevitably influenced by dimmed high redshift HSB 
  galaxies, the application of the photometric redshift method allows us 
  to extract a reduced sample of most probable intrinsically LSB 
  galaxies for which it is useful to derive spectroscopic redshifts. 
  
\renewcommand{\arraystretch}{1.5}
\begin{table*}
\begin{center}
\begin{tabular}{c c c c c c r c c c c c c c}
\hline
\hline
Name&$U$&$B$&$V$&$R$&$I$&$U-B$&$\Delta(U-B)$&$B-V$&$\Delta(B-V)$&$V-I$&$\Delta(V-I)$&$B-R$&$\Delta(B-R)$\\
(1)&(2)&(3)&(4)&(5)&(6)&(7)&(8)&(9)&(10)&(11)&(12)&(13)&(14)\\
\hline
\object{LSB J22311-60503}&18.45&18.50&17.75&17.29&16.76&-0.05&$\pm$0.07&0.75&$\pm$0.02&0.99&$\pm$0.02&1.21&$\pm$0.32\\
\object{LSB J22315-60481}&20.40&20.61&19.46&18.75&18.16&-0.21&$\pm$0.46&1.15&$\pm$0.16&1.30&$\pm$0.13&1.86&$\pm$0.15\\
\object{LSB J22322-60142}&20.76&20.31&18.83&17.80&17.26& 0.46&$\pm$0.59&1.48&$\pm$0.21&1.57&$\pm$0.12&2.51&$\pm$0.19\\
\object{LSB J22324-60520}&20.06&19.60&18.32&17.61&17.05& 0.46&$\pm$0.16&1.28&$\pm$0.04&1.27&$\pm$0.02&1.99&$\pm$0.04\\
\object{LSB J22325-60155}&19.02&19.00&18.56&18.34&17.92& 0.02&$\pm$0.10&0.44&$\pm$0.04&0.64&$\pm$0.03&0.66&$\pm$0.04\\
\object{LSB J22325-60211}&19.96&20.10&19.14&18.60&18.09&-0.14&$\pm$0.16&0.95&$\pm$0.06&1.05&$\pm$0.04&1.50&$\pm$0.05\\
\object{LSB J22330-60543}&18.94&18.70&18.34&18.15&17.75& 0.24&$\pm$0.09&0.36&$\pm$0.03&0.59&$\pm$0.04&0.55&$\pm$0.03\\
\object{LSB J22331-60340}&21.38&21.13&19.80&18.66&18.12& 0.25&$\pm$0.27&1.33&$\pm$0.08&1.68&$\pm$0.05&2.47&$\pm$0.07\\
\object{LSB J22341-60475}&19.51&19.57&18.51&17.90&17.31&-0.06&$\pm$0.13&1.05&$\pm$0.04&1.20&$\pm$0.02&1.67&$\pm$0.04\\
\object{LSB J22342-60505}&19.57&19.73&18.83&18.27&17.33&-0.16&$\pm$0.24&0.90&$\pm$0.09&1.50&$\pm$0.10&1.46&$\pm$0.11\\
\object{LSB J22343-60222}&19.34&19.10&18.68&18.46&17.93& 0.24&$\pm$0.22&0.42&$\pm$0.07&0.75&$\pm$0.06&0.64&$\pm$0.13\\
\object{LSB J22345-60210}&20.27&20.15&19.08&18.47&17.92& 0.12&$\pm$0.18&1.07&$\pm$0.06&1.16&$\pm$0.04&1.68&$\pm$0.05\\
\object{LSB J22352-60420}&20.32&20.22&19.51&19.14&18.68& 0.10&$\pm$0.18&0.71&$\pm$0.06&0.82&$\pm$0.04&1.08&$\pm$0.05\\
\object{LSB J22353-60311}&19.87&19.18&18.83&18.15&18.05& 0.69&$\pm$0.14&0.35&$\pm$0.04&0.77&$\pm$0.03&0.78&$\pm$0.04\\
\object{LSB J22354-60122}&19.75&19.77&19.26&18.99&18.53&-0.02&$\pm$0.13&0.51&$\pm$0.05&0.73&$\pm$0.04&0.78&$\pm$0.04\\
\object{LSB J22355-60183}&19.98&19.75&19.24&18.96&18.54& 0.23&$\pm$0.16&0.51&$\pm$0.06&0.70&$\pm$0.04&0.79&$\pm$0.04\\
\object{LSB J22355-60390}&20.35&20.28&18.93&18.15&17.57& 0.07&$\pm$0.19&1.35&$\pm$0.05&1.36&$\pm$0.03&2.13&$\pm$0.05\\
\object{LSB J22361-60223}&19.94&19.86&18.62&17.96&17.39& 0.08&$\pm$0.15&1.24&$\pm$0.04&1.23&$\pm$0.02&1.90&$\pm$0.04\\
\hline
%\hline
\end{tabular}
\caption[]{Total magnitudes and color indices for the objects found in the
Goddard Space Flight Center/STIS Field.}
\label{color}
\end{center}
\end{table*}
\renewcommand{\arraystretch}{1}
  In order to derive photometric redshifts for the higher redshifted galaxies
  within our sample, we used a multi-color system of
  \citet{1998AJ....116.1074L}. From this work we got the colors of five
  representative template galaxy Spectral Energy Distributions (SEDs) and a
  system of six optical and near IR broadband filters, including standard
  $U$, $B$, $V$, $R$, as used with the GSFC multi-color data. Four of
  the template SED's cover the basic range of galaxy types E/S0, Sbc, Scd and
  Irr (starburst) and resulted from the catalogs of integrated spectrometry of
  \citet{1992ApJS...79..255K} and \citet{1980ApJS...43..393C}. A fifth SED is
  a composite spectrum of a Sa and Sab galaxy from
  \citet{1996ApJ...467...38K}. For these five galaxy types colors were
  calculated using k-correction and covering a redshift range from z\,=\,0 to
  z\,=\,1.0 with a resolution of $\Delta z$\,=\,0.025. The calculation of the
  colors were done, assuming no intrinsic evolution of the galaxies.

  \indent 
  We used the colors of \citet{1998AJ....116.1074L} in our two color-color
  diagrams, $U-B$ vs. $B-V$ and $B-R$ vs. $U-B$ for further comparisons
  (Fig.~\ref{UBBV}). The derived tracks for each galaxy template were
  limited to redshifts between $z$\,=\,0 and $z$\,=\,0.5.   
 
  \indent
  Due to the smaller field size of the Goddard data, we only have multi-color
  information available for 18 of the 37 detected sample galaxies. Therefore,
  the further analysis is restricted to this much smaller subsample.

  \indent 
  The comparison in the color-color diagrams result in a subsample of 9
  galaxies with colors equivalent to photometric redshifts
  z\,$\ge$\,0.15, which we therefore excluded from our final sample. 
  For these
  galaxies the positions in the color-color
  diagram can be used as distance indicators since the accuracy of the
  photometric redshift determination is small enough to allow a rough
  estimation  redshift (discussion above). 
  For one of the higher redshifted galaxies, LSB J22331-60340, a
  spectroscopic redshift of 0.543 is available \citep{aat98}. 
  
  \indent
  The remaining sample, used for further analysis, consists of 9 genuine,
  local LSB galaxy candidates. The color-color comparison shows that 7 
  (marked in Table~\ref{lsbc}) of them have colors that are significantly 
  different from colors of HSB galaxies with similar Hubble types 
  (see Fig.~\ref{UBBV}). Hence, they cannot be redshifted galaxies.
  To test our assumption we added LSB galaxies with good CCD
  multi-color photometry from \citet{deblok.e...1997} and
  \citet{1997AJ....113.1212O} to the diagrams of Fig.~\ref{UBBV}. These
  spectroscopically confirmed LSB galaxies are also located in color
  space outside of the area defined by the redshifted HSB galaxies,
  showing consistency in color with our LSB galaxies candidates.

  \indent
  Two galaxies of our final selected LSB candidate sample
  (\object{LSB J22311-60503}, 
  \object{LSB J22324-60520}) have a location in the color-color
  diagrams consistent with those of HSBs with redshifts
  z\,$\le$\,0.15. Although their photometric redshift estimation has a large
  distance uncertainty ($\delta z\sim$0.08), their central surface
  brightnesses assuming z\,=\,0.15 is still below our
  LSB threshold of
  $\mu_\mathrm{0,B_{W}}\ge\,$22.0\,mag\,arcsec$^{-2}$.

\begin{figure*}
\centering
\resizebox{8cm}{!}{\includegraphics{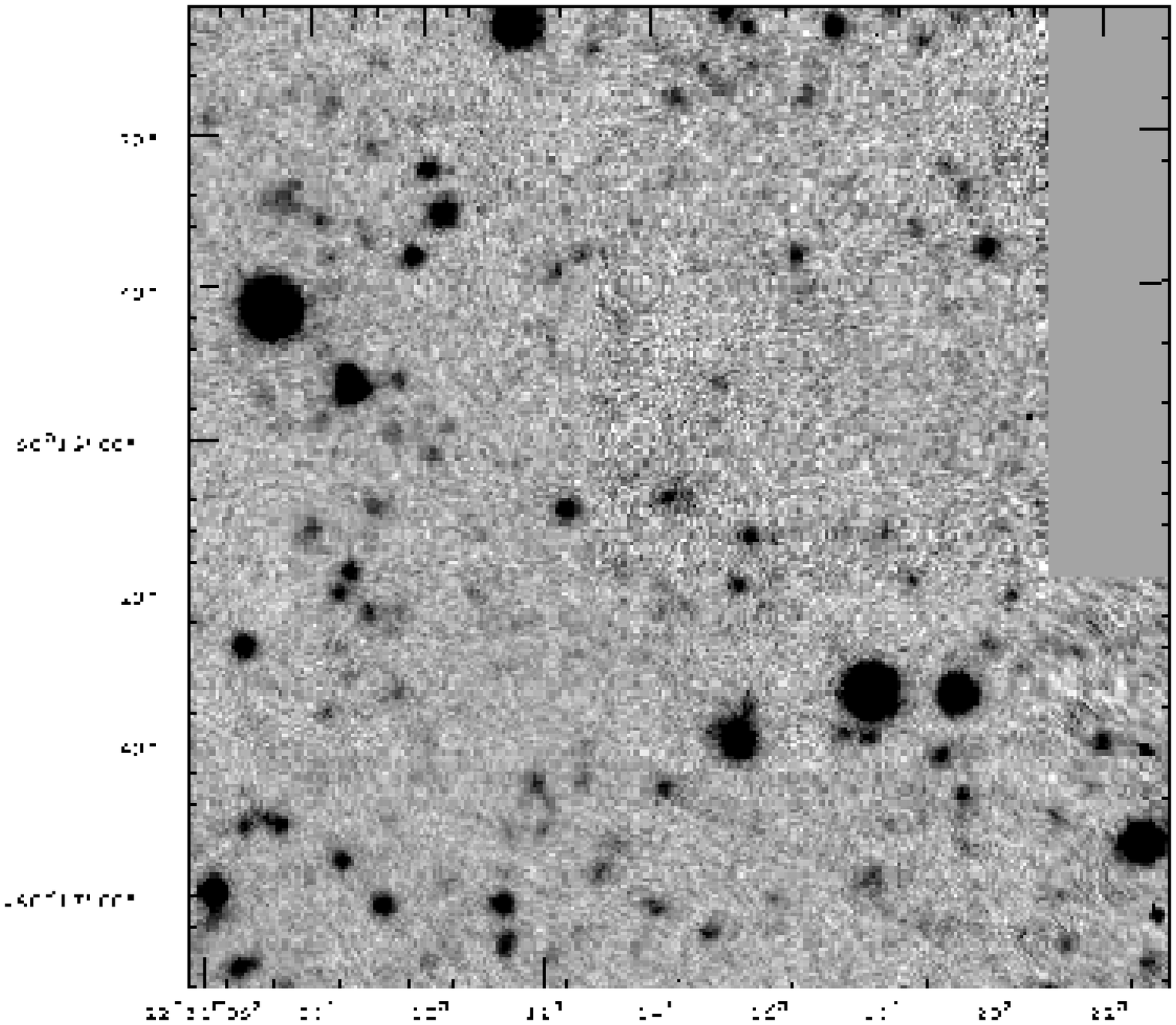}}
\resizebox{8cm}{!}{\includegraphics{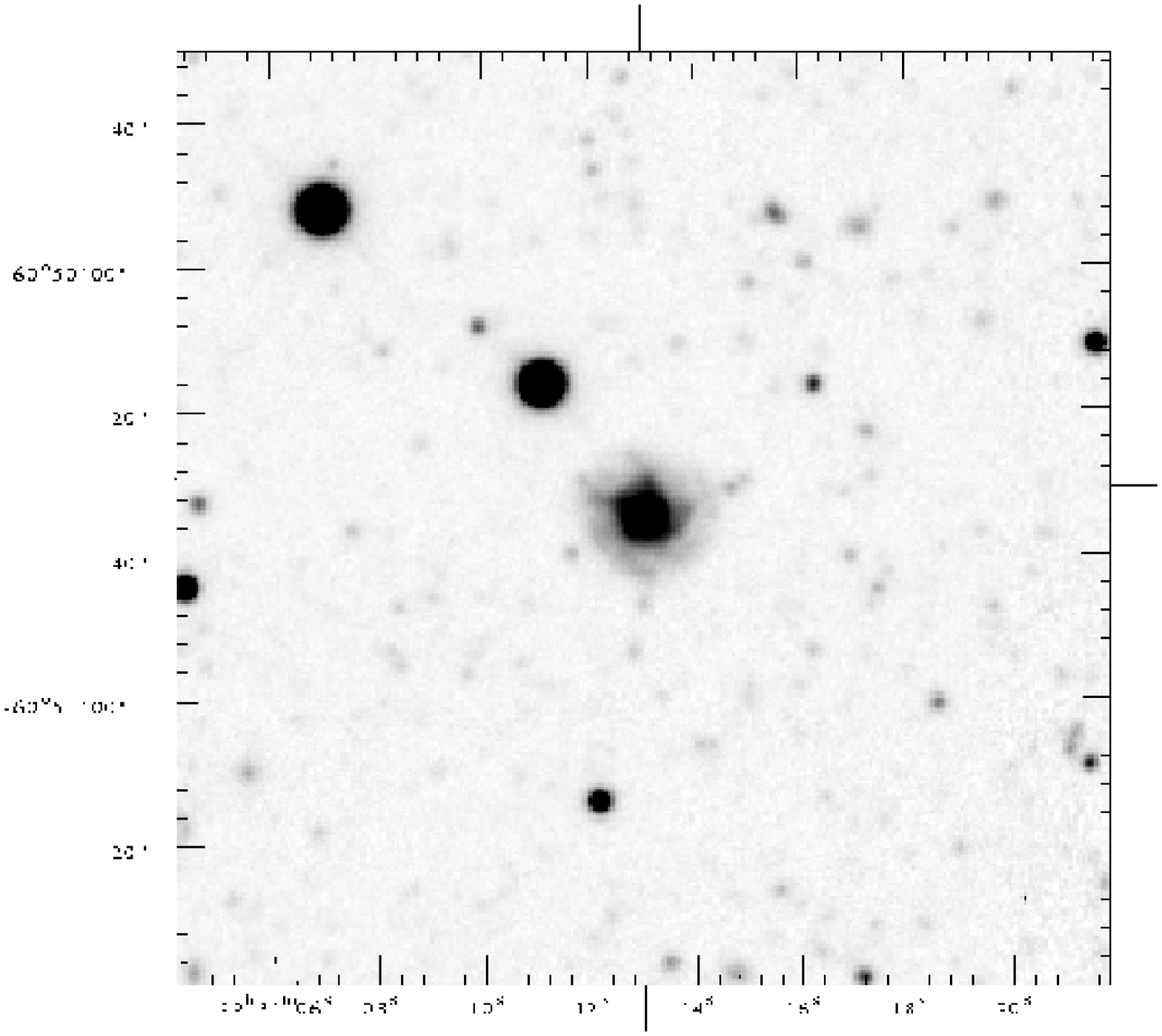}}
\resizebox{8cm}{!}{\includegraphics{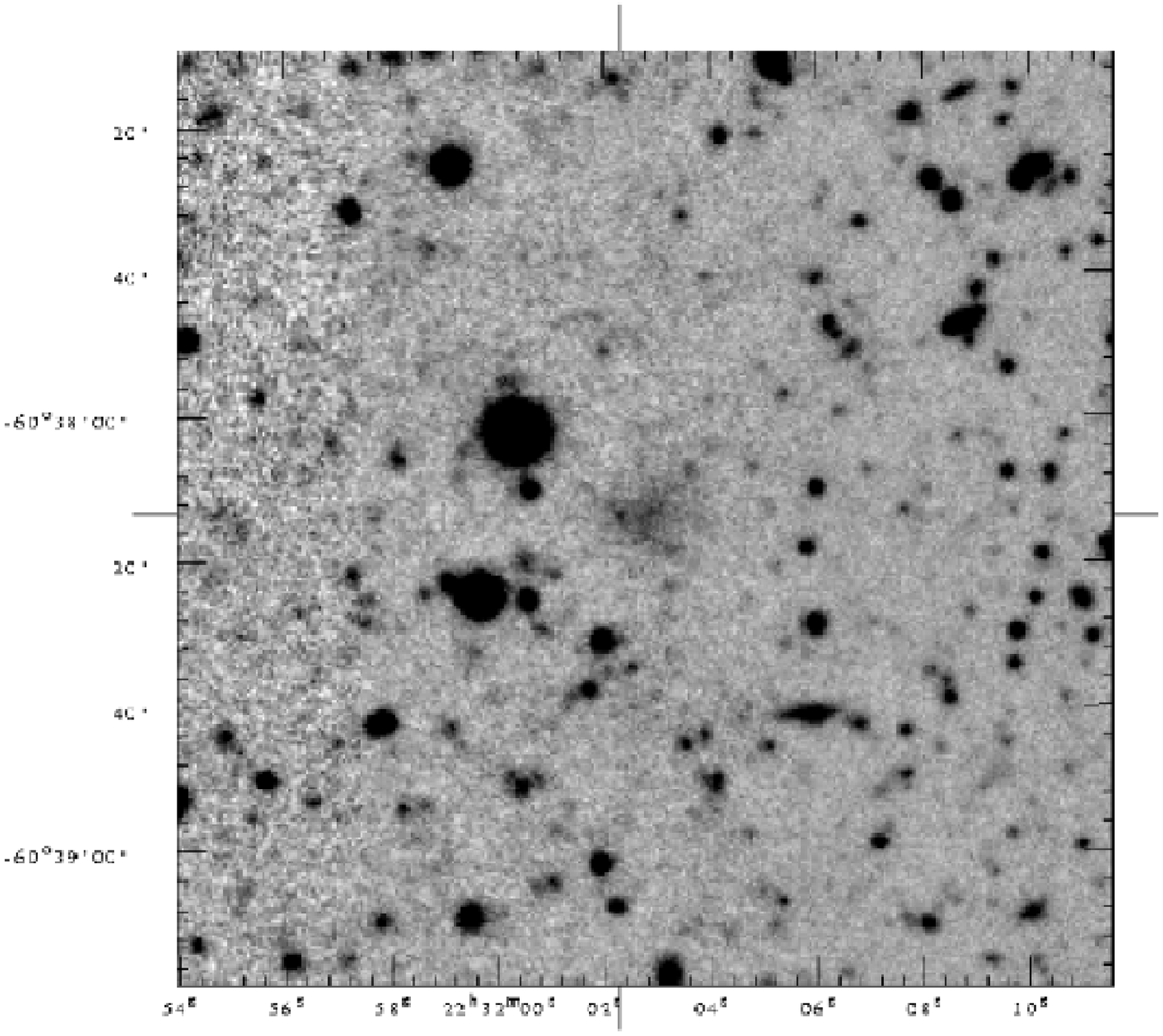}}
\resizebox{8cm}{!}{\includegraphics{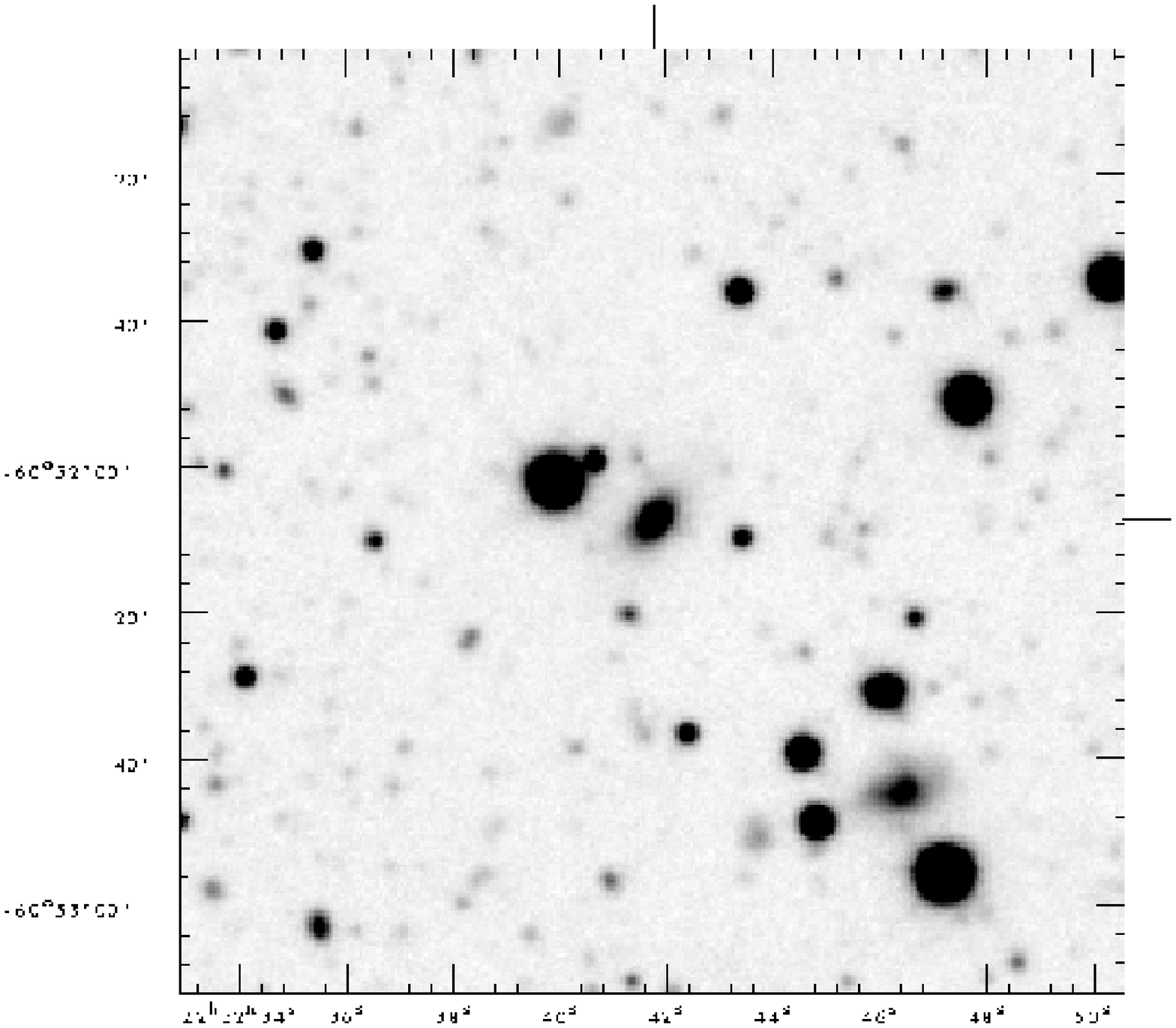}}
\resizebox{8cm}{!}{\includegraphics{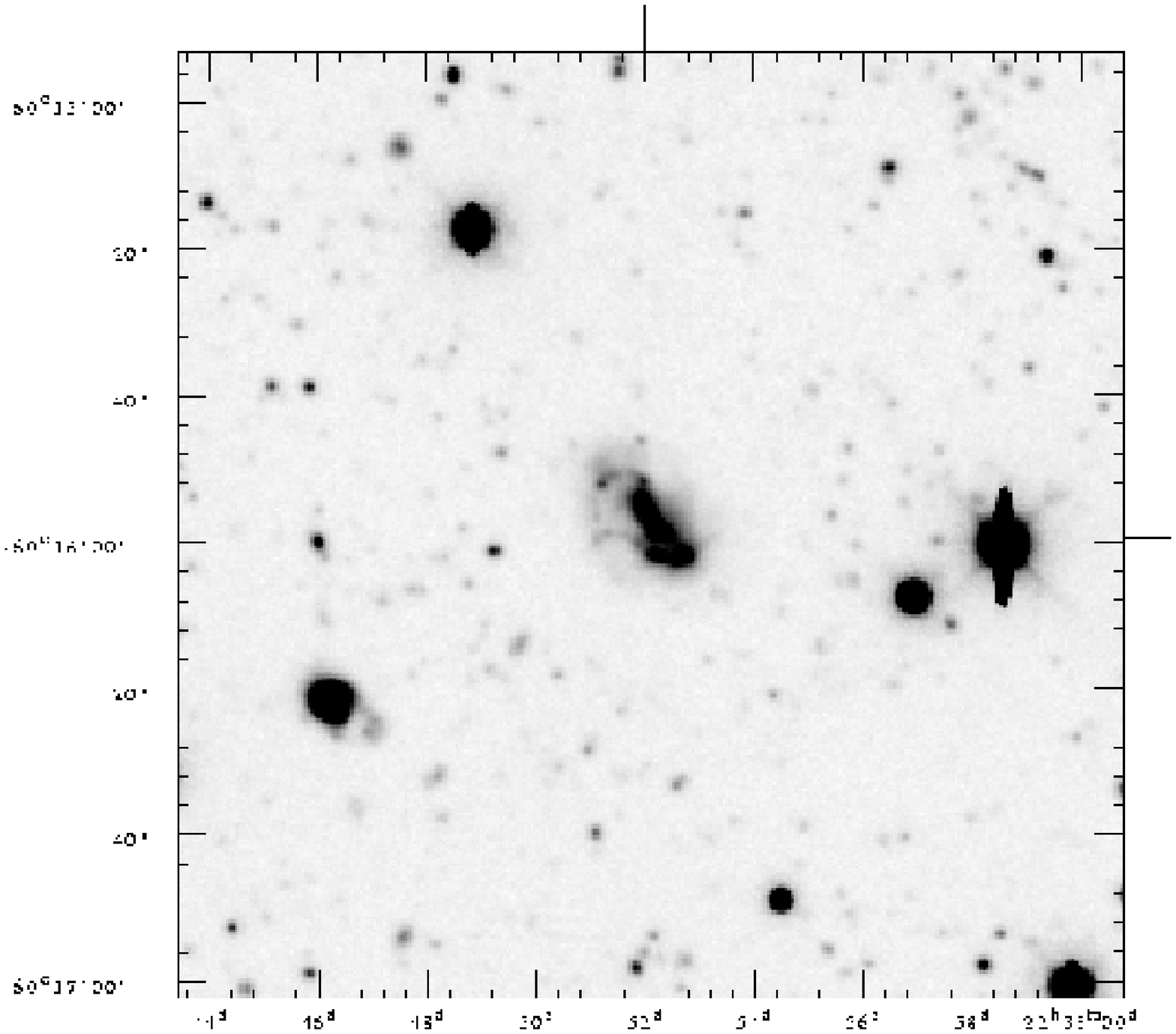}}
\resizebox{8cm}{!}{\includegraphics{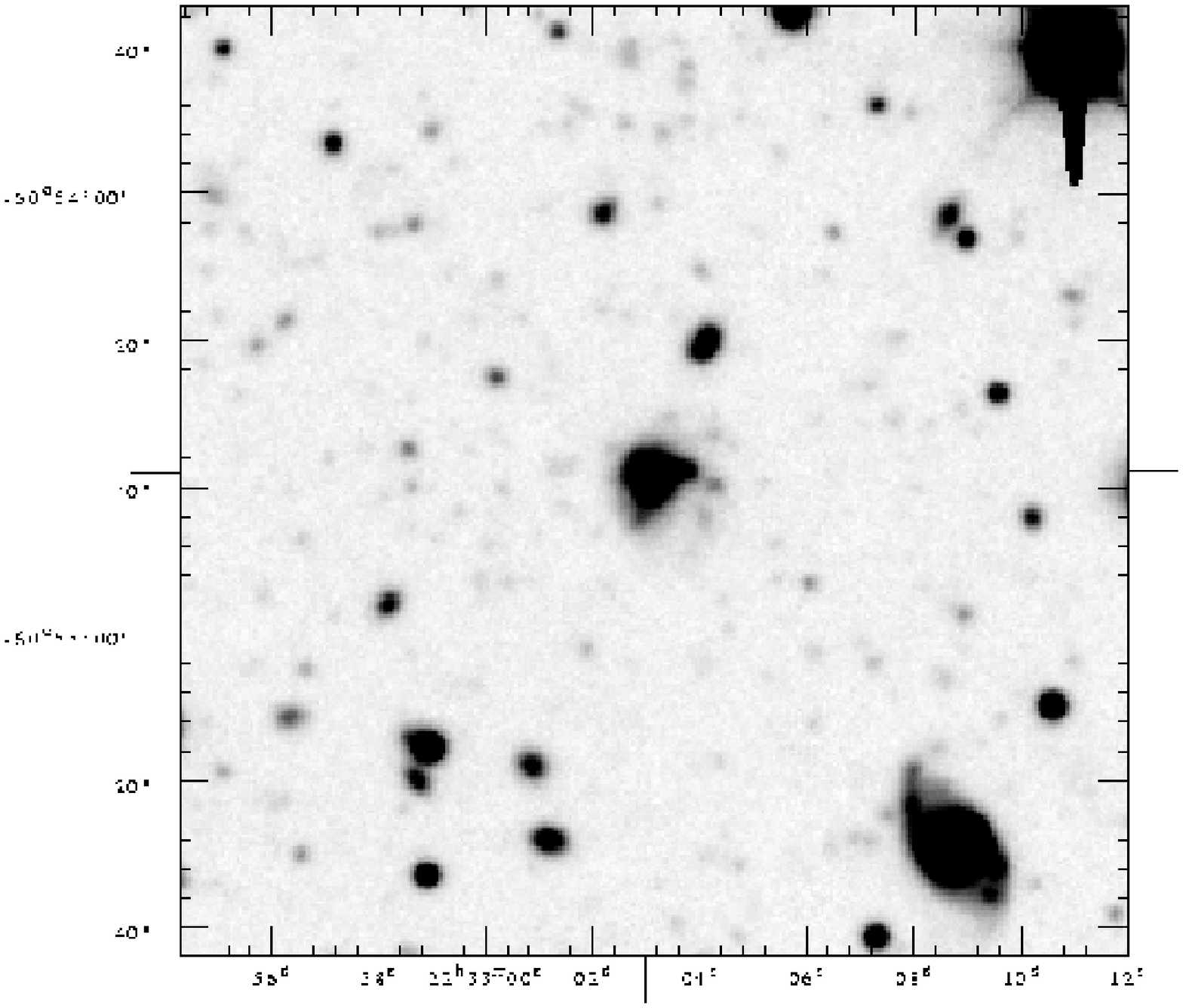}}
\caption{Images of 6, of our 12, LSB galaxy candidates ordered by RA 
  (\object{LSB J22311-60160}, \object{LSB J22311-60503}, 
  \object{LSB J22320-60381}, \object{LSB J22324-60520}, 
  \object{LSB J22325-60155}, and
  \object{LSB J22330-60543}; from left to right and top to bottom). The images
  are cutouts from the NOAO $B_\mathrm{W}$ data with a size of
  2\,$\times$\,2\,arcmin$^2$. The position of the galaxies is marked with four
  large ticks at the axes.}
\label{images1}
\end{figure*}

  \indent
  To summarize, with our selection criteria we are able to select 
  a sample of 9 highly probable LSB galaxy candidates out of a 
  sample of 18 candidate galaxies (50 \%) having multi-colour information 
  available. Apparently, our
  original size selection criterion of 10.8 arcsec (25 pixels) works
  reasonably well in suppressing the redshifted galaxy
  population. Additionally, we find three extreme low surface brightness
  candidates ($\mu_\mathrm{0,B_{W}}$\,$\ge$\,25.0\,mag\,arcsec$^{-2}$) for
  which  %no or only less 
  the GSFC multi-color data set is not available
  (\object{LSB J22311-60160}, 
  \object{LSB J22320-60381}, \object{LSB J22364-60405}). Due to the large
  diameter selection limit used for our sample, the distances of these galaxies
  are not likely to exceed the 0.5 redshift limit. Correcting
  their surface brightness assuming they are at z\,=\,0.5 leads to
  a 1.8\,mag\,arcsec$^{-2}$ higher surface brightness. Therefore, these
  galaxies can still be classified as LSB galaxies (e.g.,
  \object{LSB J22320-60381} with
  $\mu_\mathrm{0,B}$\,=\,26.91\,mag\,arcsec$^{-2}$ could be even at a redshift
  of z\,$\approx$\,1.8) and we decided to include these
  galaxies in our final LSB candidate sample. 
  This leaves us with a final sample of 12 possible LSB galaxy 
  candidates (see Table~\ref{lsbc}), of which we present the images 
  in Fig.~\ref{images1}. 
  Table~\ref{lsbc} is organized in the following way:

{\bf Column 1:} Galaxy names as used in Table~\ref{tabelle}.

{\bf Column 2\&4:} Central surface brightnesses as shown in
Tabel~\ref{tabelle}.

{\bf Column 3\&5:} Inclination corrected central surface 
brightness $\mu_\mathrm{0,B_{W}corr}$ and $\mu_\mathrm{0,Bcorr}$ derived
following Eqn.~\ref{incli}.

{\bf Column 6:} The letters indicates if the galaxy has an offset in one (co)
or both (cob) color-color diagrams or not (n). Galaxies without entry in
this column were not  detected in all filter bands of the GSFC data.

\setlength{\tabcolsep}{1.5mm}
\renewcommand{\arraystretch}{1.5}
\begin{table}
\begin{center}
\begin{tabular}{c c c c c c}
\hline
\hline
Name&$\mu_\mathrm{0,B_{W}}$&$\mu_\mathrm{0,B_{{W_{corr}}}}$&$\mu_\mathrm{0,B}$&$\mu_\mathrm{0,B_{corr}}$&\\
\hline
\object{LSB J22311-60160}&26.52&26.79&[26.93]&[27.20]&--\\
\object{LSB J22311-60503}&22.39&22.50&22.55&22.66&n\\
\object{LSB J22320-60381}&26.86&27.09&26.91&27.14&--\\
\object{LSB J22324-60520}&22.25&22.73&22.30&22.78&n\\
\object{LSB J22325-60155}&23.26&23.66&23.82&24.25&cob\\
\object{LSB J22330-60543}&22.24&22.46&22.40&22.62&cob\\
\object{LSB J22343-60222}&23.30&24.03&23.44&24.16&cob\\
\object{LSB J22352-60420}&23.16&23.88&23.47&24.19&co\\
\object{LSB J22353-60311}&23.39&23.47&23.56&23.64&cob\\
\object{LSB J22354-60122}&22.51&22.99&22.95&23.43&cob\\
\object{LSB J22355-60183}&22.52&23.00&22.78&23.26&cob\\
\object{LSB J22364-60405}&25.41& --  &[25.82]& --  &--\\
\hline
\hline
\end{tabular}
\end{center}
\caption{This table lists the color selected, high probability LSB galaxy 
candidates and there central surface brightnesses in
mag\,arcsec$^{-2}$. Values in square brackets are derived by applying the
  mean offset between $B_W$- and $B$-band data (see Sect.~\ref{photpro}).}
\label{lsbc}
\end{table}
\renewcommand{\arraystretch}{1}

\begin{figure*}
\centering
\resizebox{8cm}{!}{\includegraphics{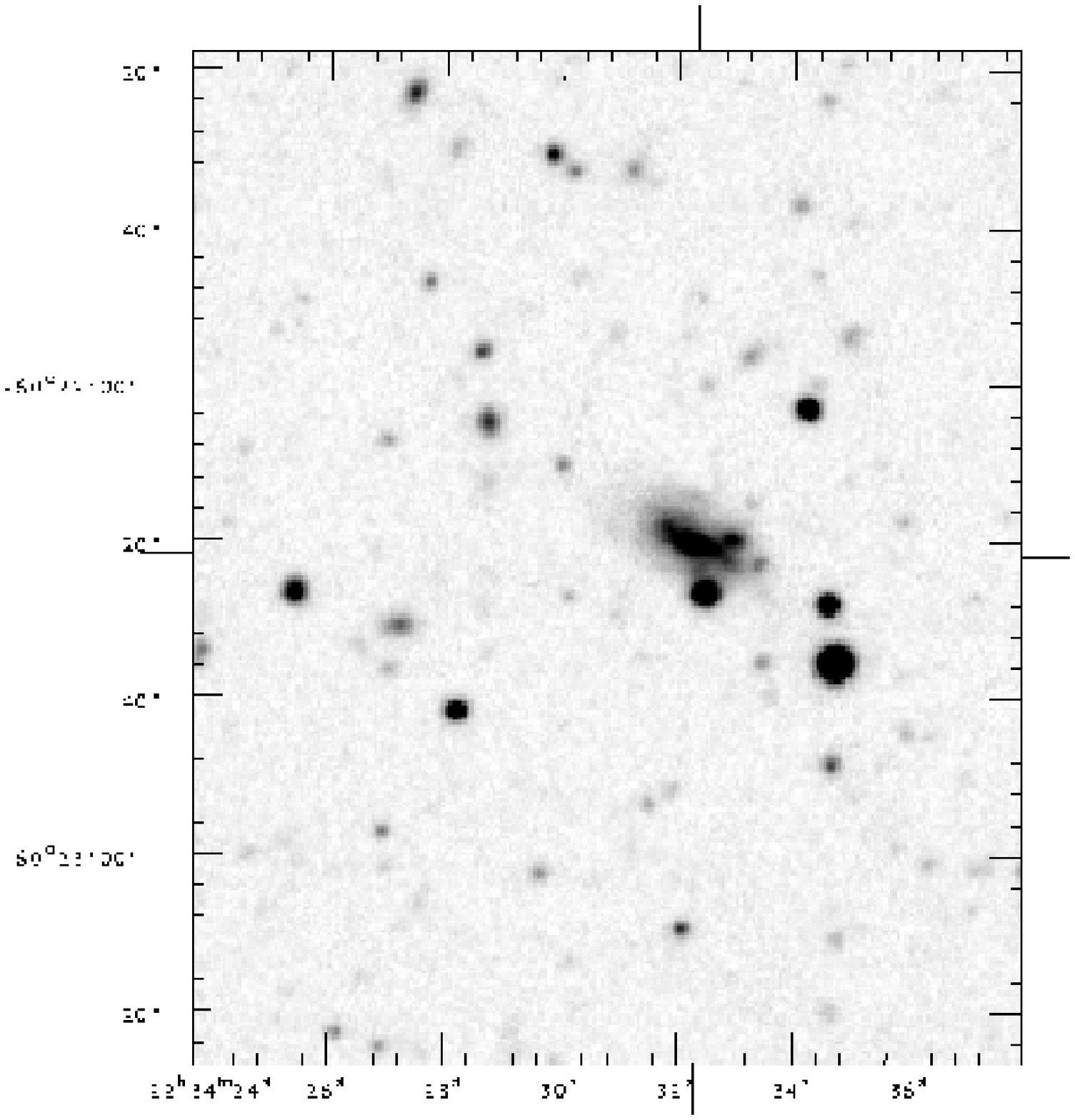}}
\resizebox{8cm}{!}{\includegraphics{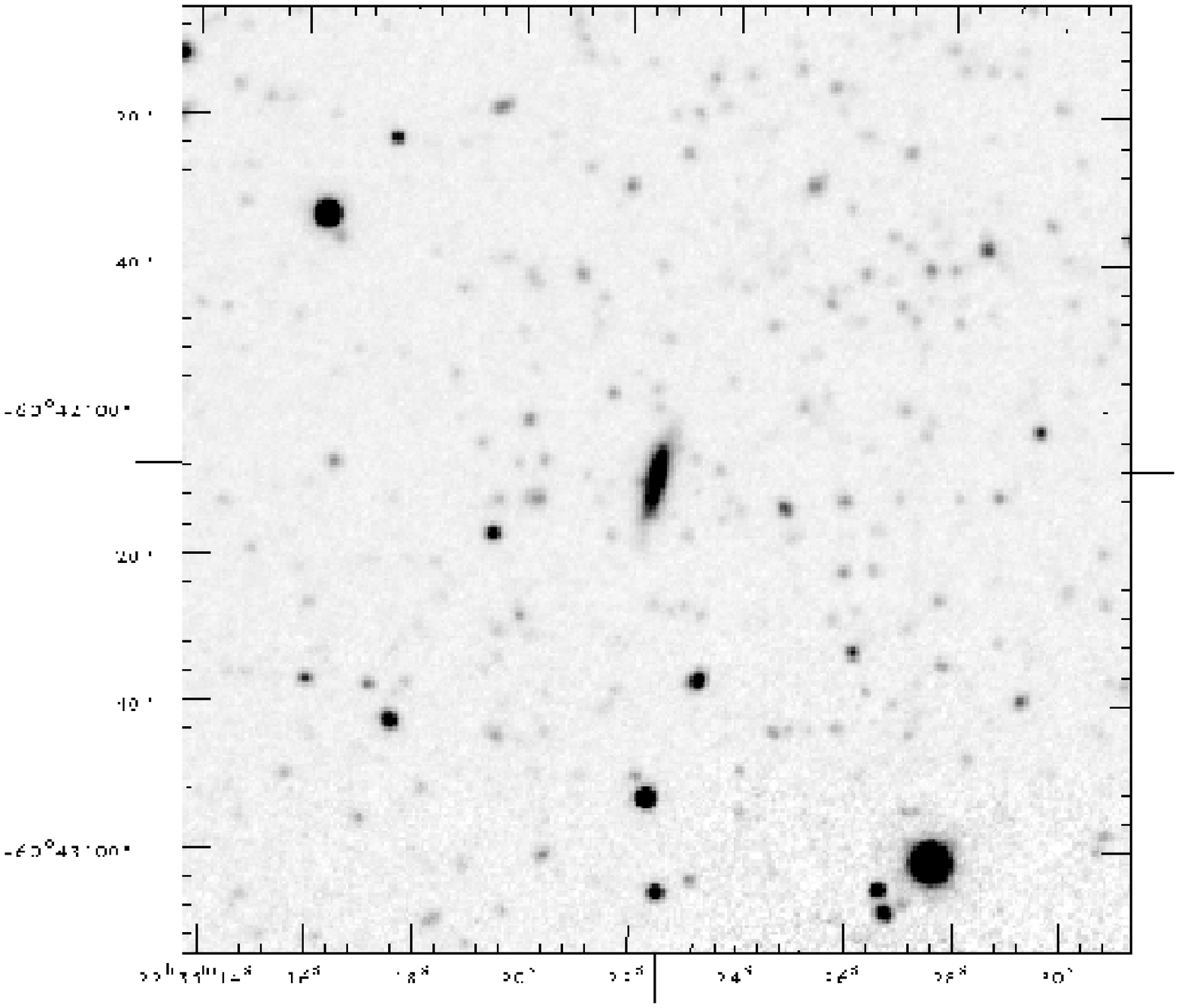}}
\resizebox{8cm}{!}{\includegraphics{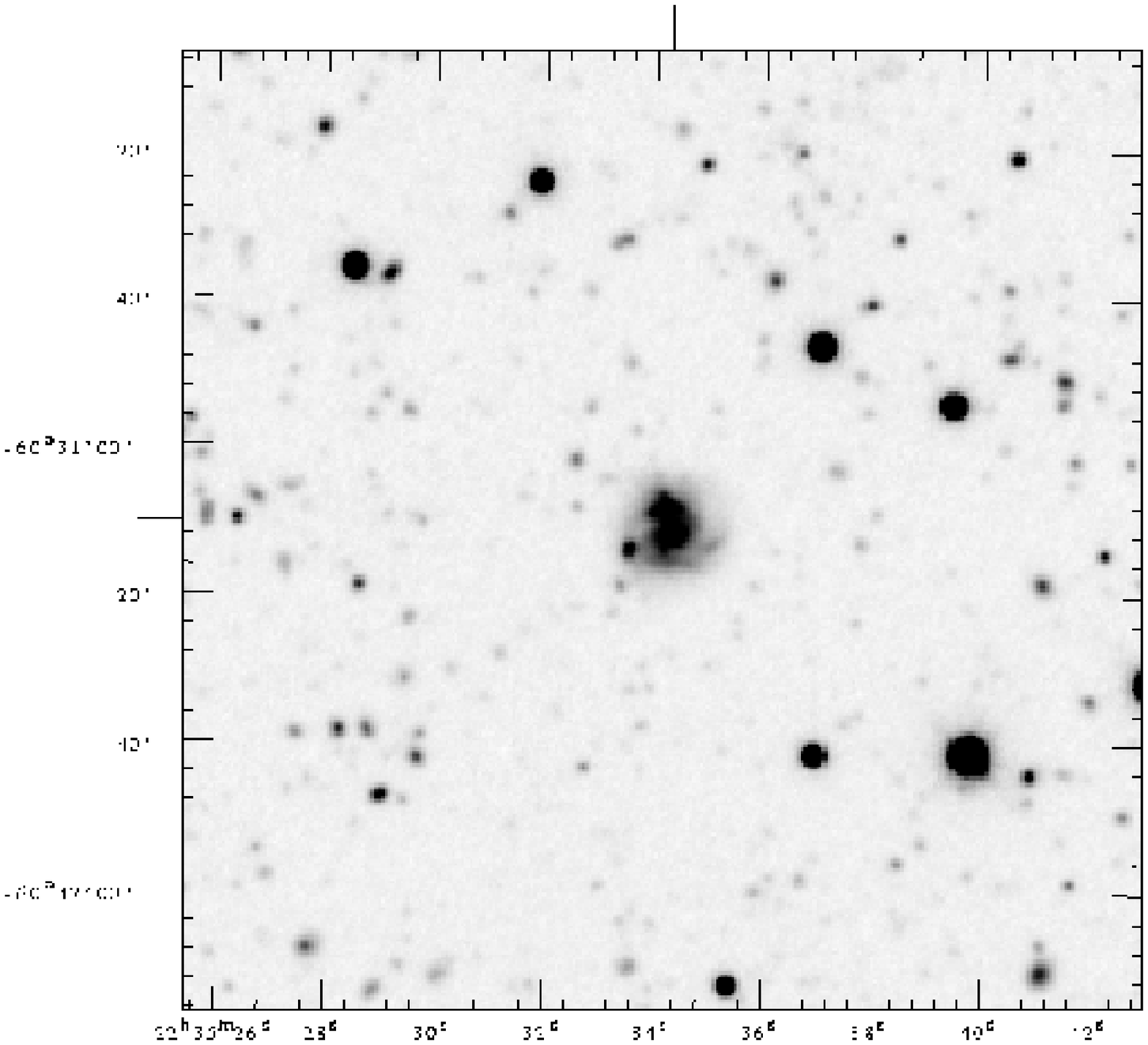}}
\resizebox{8cm}{!}{\includegraphics{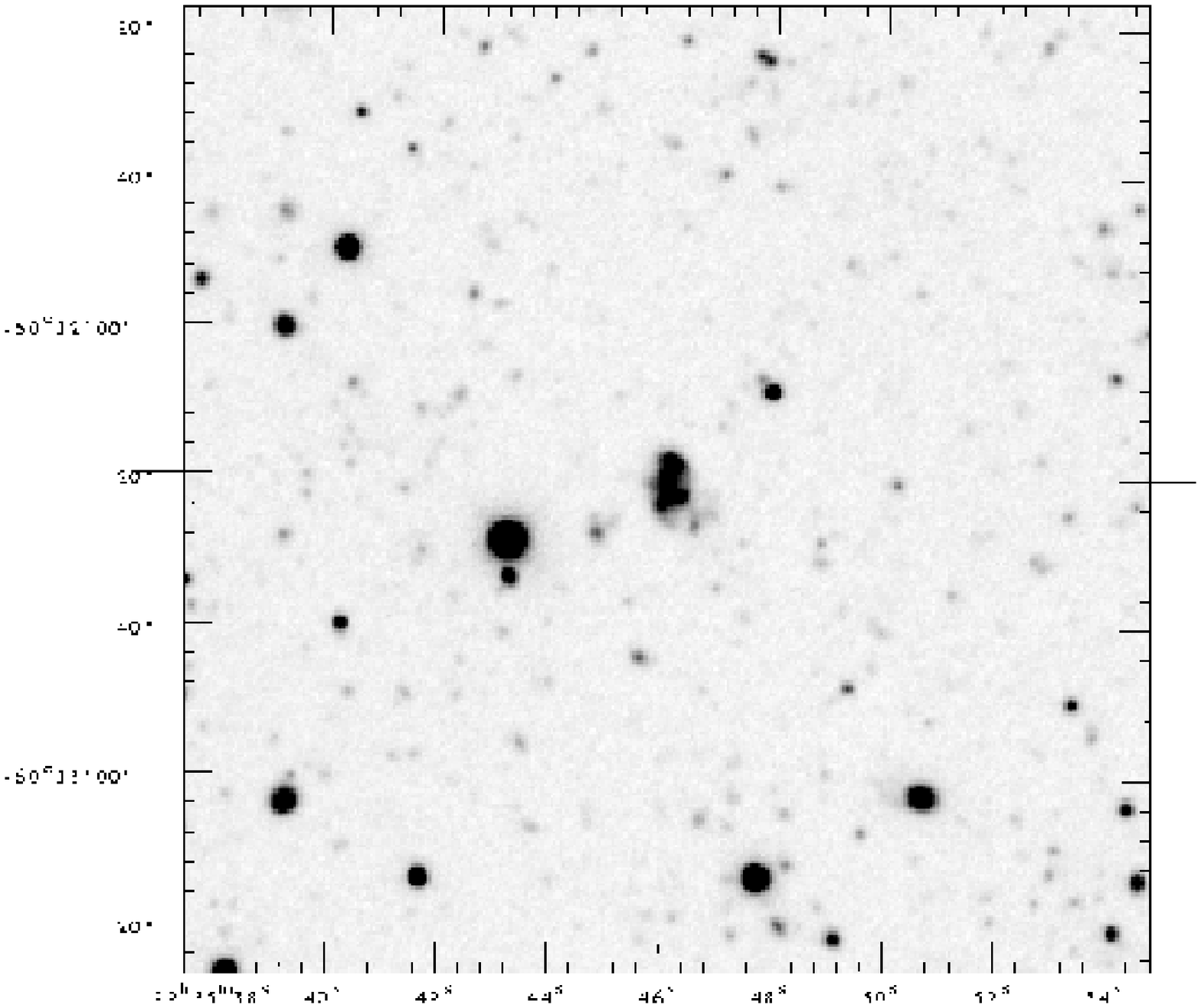}}
\resizebox{8cm}{!}{\includegraphics{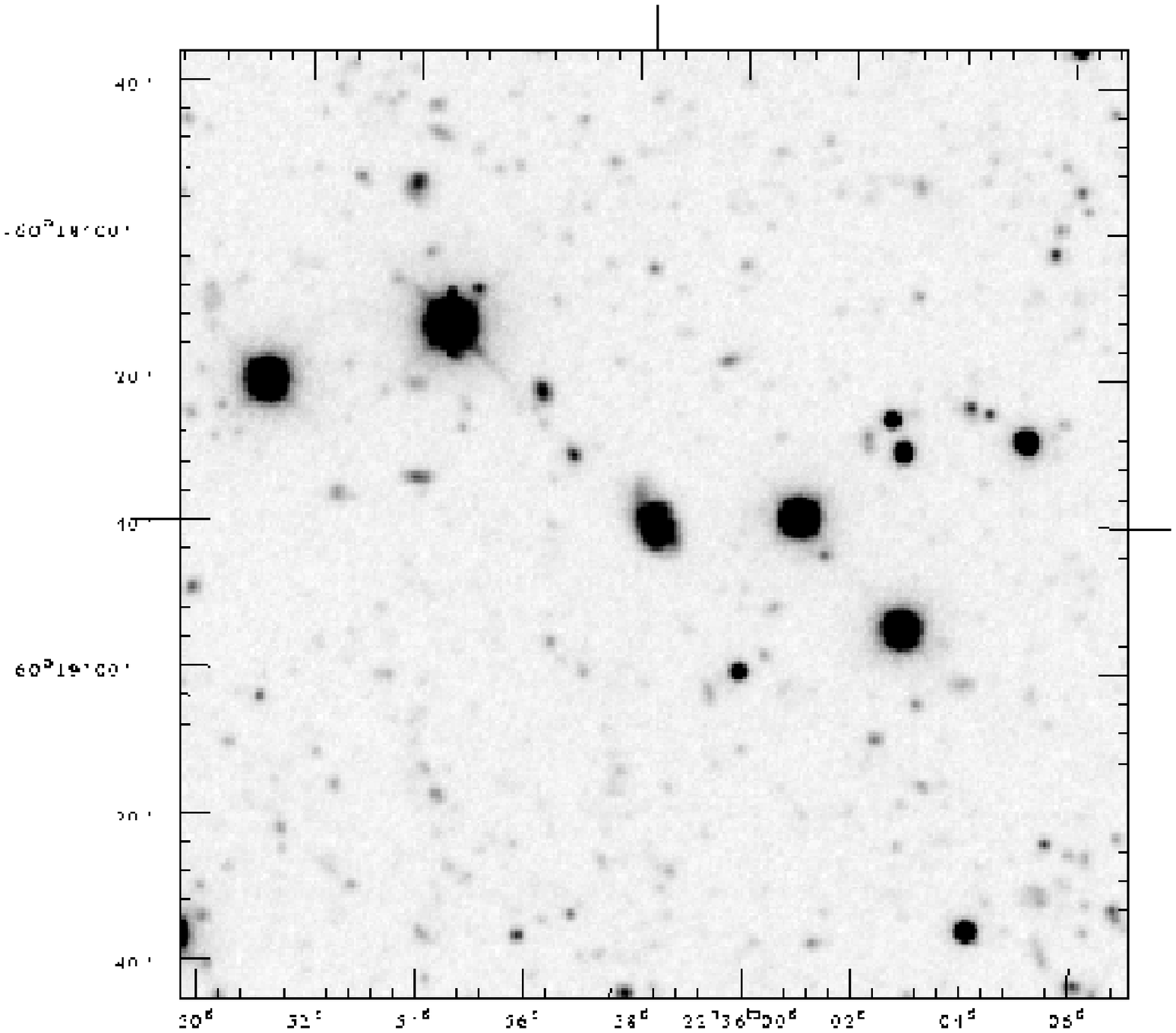}}
\resizebox{8cm}{!}{\includegraphics{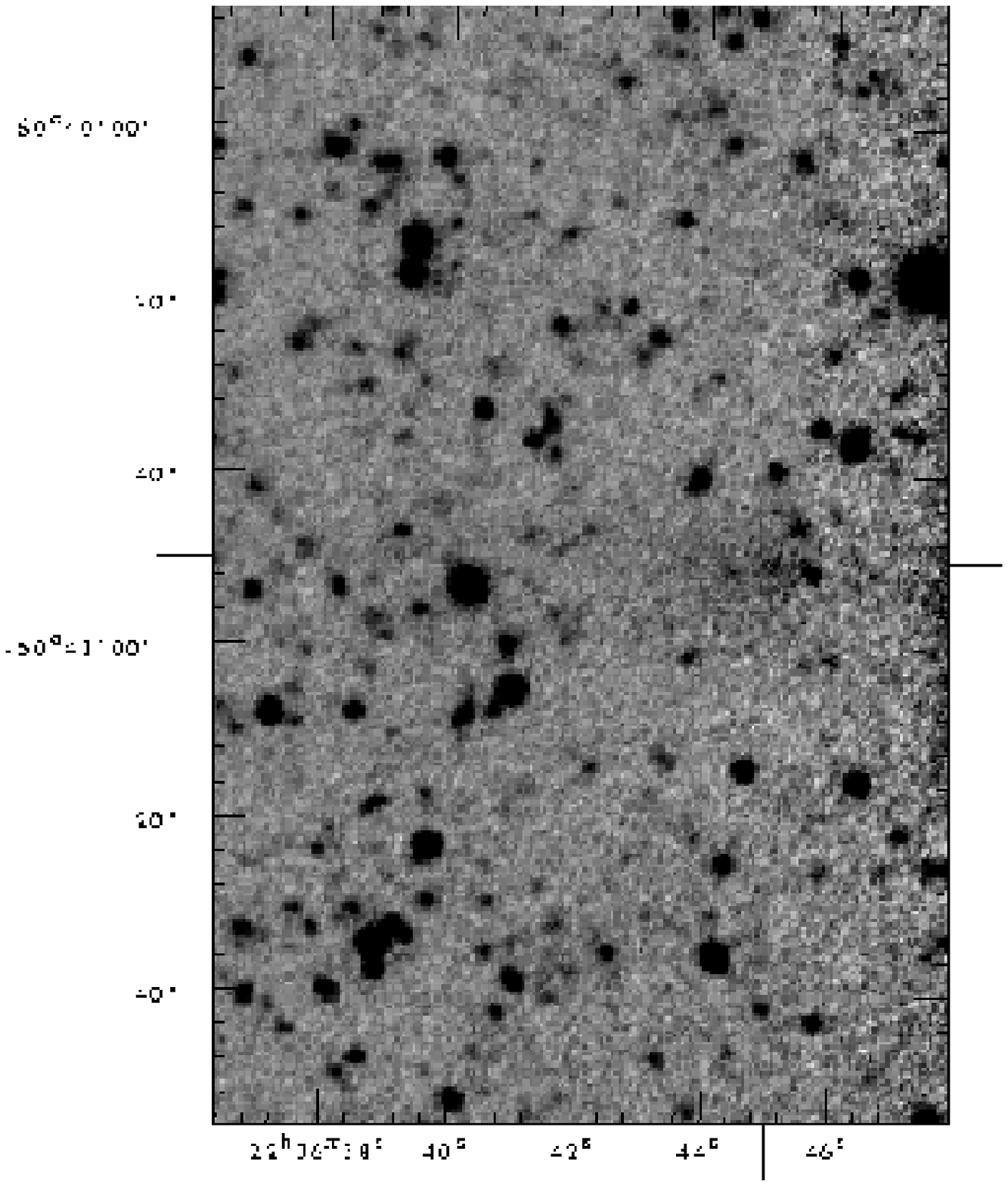}}
\parbox{18cm}{{\bf Fig. 2 continued:} From left to right and from top to bottom
  \object{LSB J22343-60222}, \object{LSB J22352-60420},
  \object{LSB J22353-60311}, \hspace{1cm} \object{LSB J22354-60122},
  \object{LSB J22355-60183}, and \object{LSB J22364-60405}. }
\end{figure*}

In the following paragraph we give a few notes on the 12 most probable
  LSB galaxies:\\

{\bf \object{LSB\,J22311-60160}:} Very faint object
($\mu_\mathrm{0,B_{W}}$\,=\,26.3\,mag\,arcsec$^{-2}$) found only in the deep
$B_\mathrm{W}$ image in a region with a lower signal to noise level and it
could not be fitted very well by an exponential profile. 

{\bf \object{LSB J22311-60503}:} Disk galaxy showing a bright central region
and a very faint disk with some spiral structure.

{\bf \object{LSB J22320-60381}:} Faintest LSB candidate in our
sample with central surface brightness of
$\mu_\mathrm{0,B_{W}}$\,=\,26.9\,mag\,arcsec$^{-2}$. Only found in the deep
$B_\mathrm{W}$ image. The light profile is very well fitted by an exponential
profile. 

{\bf \object{LSB J22324-60520}:} Disk galaxy with a very bright
core, a central surface brightness at our upper limit
($\mu_\mathrm{0,B_{W}}$\,=\,22.2\,mag\,arcsec$^{-2}$), and very faint and
diffuse structure in the outer region. Maybe highly inclined.   

{\bf \object{LSB J22325-60155}:} Looks like a star forming
irregular galaxy, showing diffuse disk-like structure in the outer region.  

{\bf \object{LSB J22330-60543}:} Disk galaxy with bright core and diffuse spiral
structure. 

{\bf \object{LSB J22343-60222}:} Highly inclined galaxy with
bright inner part and faint diffuse disk like structure. The radial profile
shows a steep decline in the outer regions.

{\bf \object{LSB J22352-60420}:} Highly inclined galaxy showing
diffuse disk like structure. The radial profile shows a steep drop in the outer
region. 

{\bf \object{LSB J22353-60311}:} Appears to be a star forming
galaxy with a disk like structure .

{\bf \object{LSB J22354-60122}:} Star forming irregular
galaxy. The radial profile shows a steep decline in the outer region.

{\bf \object{LSB J22355-60183}:} Galaxy showing a bright core and
some hints for spiral structures. The radial profile has a steep decline in the
outer region.  

{\bf \object{LSB J22364-60405}:} Very faint LSB galaxy candidate
($\mu_\mathrm{0,B_{W}}$\,=\,25.82\,mag\,arcsec$^{-2}$) found only in the deep
$B_\mathrm{W}$ image in a region with variable signal to noise level. 

\section{Results and Discussion}

\subsection{Selection effects and parameter space}
  One important aspect that still has to be discussed in more detail are
  the selection biases that affect our final LSB candidate sample. For
  our search we
  only used a diameter and surface brightness but no magnitude limit. 
  We selected all galaxies having diameters larger than 10.8\,arcsec and
  central surface brightness fainter than
  22.0\,$B_\mathrm{W}$\,mag\,arcsec$^{-2}$. Our
  object detection limit in surface brightness is about 27\,
  $B_\mathrm{W}$\,mag\,arcsec$^{-2}$ and the limiting surface brightness of
  the data is about 29\, $B_\mathrm{W}$\,mag\,arcsec$^{-2}$.  
  The object selection was done using the $B_\mathrm{W}$ data of 
  the NOAO Deep Wide-Field survey, being the most sensitive data at our
  disposal. We just selected galaxies having diameters larger than 
  10.8\,arcsec (measured by eye on the images), in order to reduce the
  contamination due to higher redshifted 
  and, therefore, cosmologically dimmed HSB galaxies. Estimating the 
  diameter by eye can lead to an incompletness of galaxies
  near the diameter limit. Therefore, it is possible that our sample is
  not representative for the real number of objects close to the diameter
  limit of our survey. With the surface brightness selection criteria 
  ($\mu_0$\,$\ge$\,22.0\,mag\,arcsec$^{-2}$) we just restrict our sample
  against the high surface brightness objects found in the data. Choosing an
  upper surface brightness limit of
  $\mu_\mathrm{0,B_{W}}$\,=\,22.0\,mag\,arcsec$^{-2}$ gives 
  us a sufficient overlap to the region of the higher surface brightness
  galaxies, so that we will not lose objects close to the LSB galaxy surface
  brightness limit. 
  
  \indent
  A second selection bias concerning the size selection results from the
  median filter method. For the filtering process we used a kernel with a size
  of 25 pixel which represents our diameter selection limit. However, this
  kernel size biases our sample against objects with 
  much larger or smaller sizes. We used the median filter method in order to
  search for extreme LSB candidates
  ($\mu_\mathrm{0,B_{W}}$\,$\ge$\,24.5\,mag\,arcsec$^{-2}$). Therefore, our
  sample is biased against large objects at the very low surface brightness
  end.
 
  \indent
  With regard to the described search criteria we found LSB galaxies having
  small diameters 
  between 12\arcsec\,$\le$\,d\,$\le$\,36\arcsec. The scale-lengths
  calculated from the profiles are distributed over the range
  0\farcs9\,$\le$\,$\alpha_\mathrm{B_{W}}$\,$\le$\,6\arcsec.
  \begin{figure}
  \begin{center}
  \includegraphics[width=8.5cm]{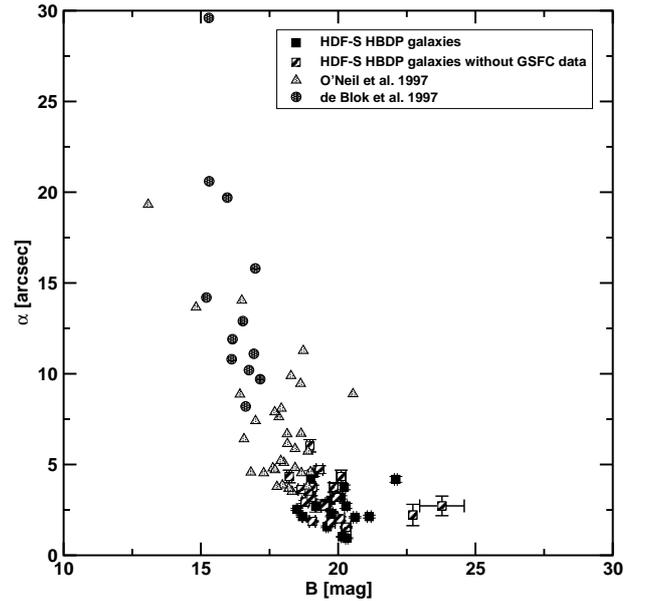}
  \caption{Disk scale-length against total magnitude of the
    galaxies detected in the present study, compared with the LSB
    galaxies from \citet{deblok.e...1997} and \citet{1997AJ....113.1212O}. For
    those LSB J galaxies, where no $B$ data is available we used the
    $B_\mathrm{W}$ scale-length instead (hatched squares).}
  \label{mubBtot}
  \end{center}
  \end{figure}  
  Our LSB candidate sample consists mainly of galaxies which could not be
  detected in other surveys, which are less sensitive for very low
  surface brightness outer regions, and they therefore fall below the
  diameter limits of these surveys.  
  In Fig.~\ref{mubBtot} we compare our sample with the LSB galaxy samples of
  \citet{1997AJ....113.1212O,1997AJ....114.2448O} and \citet{deblok.e...1997}.
  This graph shows that we predominantly find LSB galaxy candidates with
  small angular sizes and low total magnitudes. This would 
  indicate that our sample consists of more distant or rather dwarfish LSB
  galaxies if compared to the sample of
  \citet{1997AJ....113.1212O,1997AJ....114.2448O}. 

\subsection{Central surface brightness distribution}
\begin{figure*}[ht]
\begin{center}
\includegraphics[width=8cm]{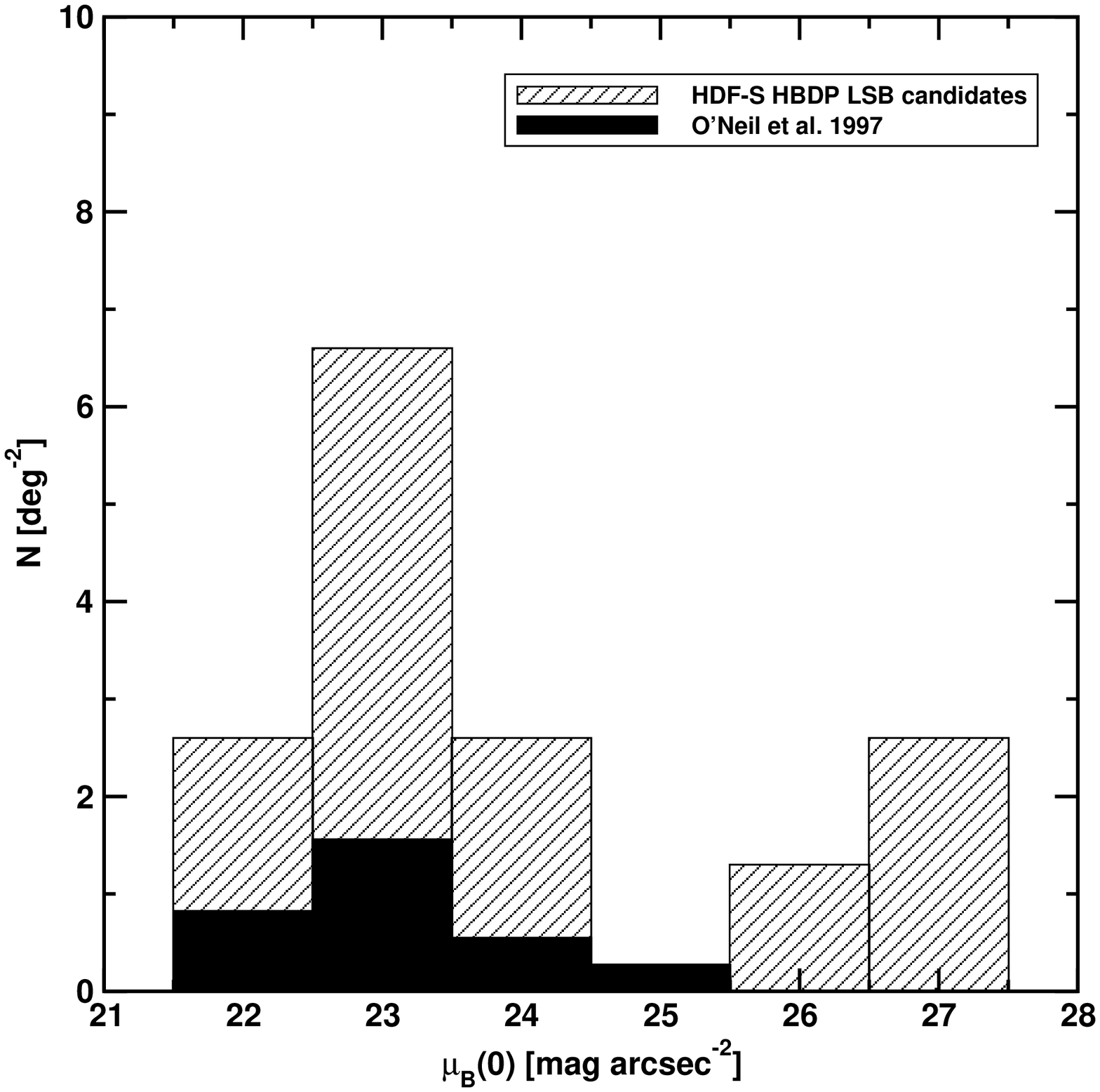}
\includegraphics[width=8cm]{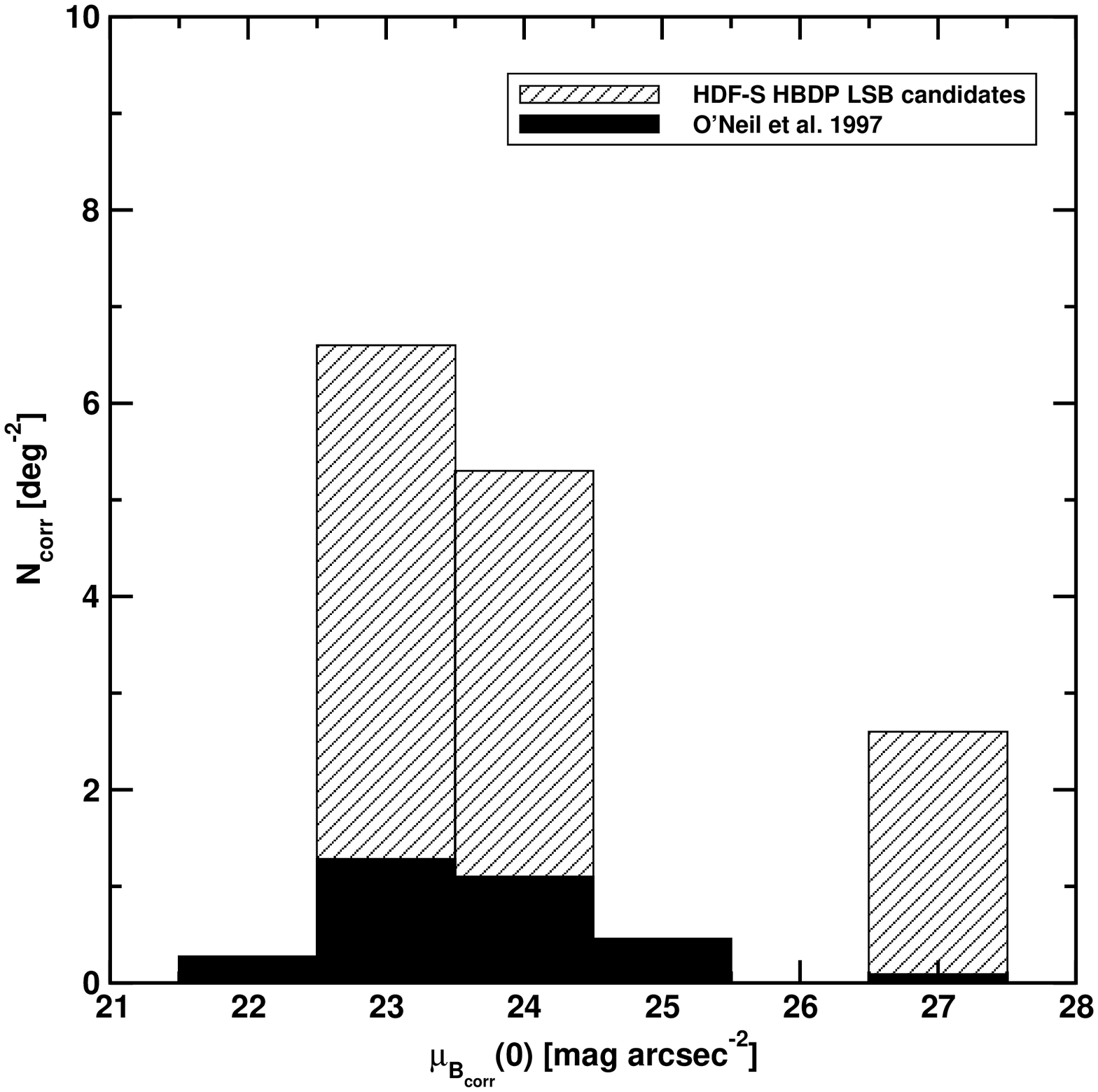}
\caption{In the left diagram we plot the normalized central surface brightness
distribution of the 9 LSB candidates in the Goddard Space Flight
Center/STIS-Field  and the 3 galaxies with extremely low
  central surface brightnesses
  ($mu_\mathrm{0,B_{w}}$\,$\ge$\,24.5\,mag\,arcsec$^{-2}$), which we only
  found in the NOAO field (hatched bars), together with a sample of LSB
galaxies from \citep{1997AJ....113.1212O,1997AJ....114.2448O} (filled bars). 
In the right diagram we show the inclination--corrected central surface
brightness distributions.}
\label{verte}
\end{center}
\end{figure*}
  In Fig.~\ref{verte} we show the $B$ band surface brightness
  distribution of the sample LSB galaxy candidates for the uncorrected (left
  panel) and the inclination corrected values (right panel). Both
  distributions show a maximum at a central surface brightness of
  $\mu_\mathrm{0,B}$\,=\,23.0\,mag\,arcsec$^{-2}$. We compare these
  distributions to those of
  \citet{1997AJ....113.1212O,1997AJ....114.2448O}. In spite of our 
  higher sensitivity in comparison to
  \citet{1997AJ....113.1212O,1997AJ....114.2448O} the maximum
  is not shifted to much fainter central surface brightness. However, we
  were able to detect galaxies down to much lower central surface
  brightnesses.

  \indent
  Furthermore, the distribution of our LSB candidate sample shows an
  gap around $\mu_\mathrm{0,B}$\,=\,24.5\,mag\,arcsec$^{-2}$),
  which could be an indication for a second population of very low surface
  brightness galaxies similar to the one
  proposed by \citet{2000AJ....120.1316K}. However, the gap could also
  be a result of selection effects by the use of two different methods to
  search for the LSB candidates (see Sect.~\ref{ser}). 
  The detection limit of our data
  ($\mu_{lim}$\,=\,27\,$B_\mathrm{W}$\,mag\,arcsec$^{-2}$) unfortunately does 
  not allow us to pursue this in more detail.
  We found three galaxies with very low surface brightness, and 
  especially 
  \object{LSB J22320-60381} has one of the lowest central surface brightness
  known today ($\mu_\mathrm{0,B_{W}}$\,=\,26.86\,mag\,arcsec$^{-2}$). From the
  fact that the volume over which galaxies can be detected is a strong
  function of the central surface brightness of the individual galaxies
  \citep{1997AJ....114.2178D} we assume that these three extreme LSB galaxies
  have a significant impact on the volume densities of LSB galaxies.   

\subsection{Radial Profiles}
\label{radprof}
  As described in Sect.~\ref{photpro} we derived azimuthally averaged  
  radial surface brightness profiles for all 37 LSB galaxy candidates
  found in the NOAO field (see Figs. in App.~\ref{appA}).
  The derived values for the two most extreme low surface
  brightness candidates (\object{LSB J22311-60160} 
  $\mu_\mathrm{0,B_{W}}$\,=\,26.52$\pm$0.29\,mag\,arcsec$^{-2}$ and
  \object{LSB J22364-60405}
  $\mu_\mathrm{0,B_{W}}$\,=\,25.41$\pm$0.28\,mag\,arcsec$^{-2}$) have
  higher uncertainties due to their location in regions
  with lower signal to noise levels.
  We find that 21 (57\%) of the 37 galaxies do not retain their
  exponential light distribution until fading into the noise. Their
  profiles are better fitted by a broken exponential.
  In all cases where GSFC $B$-band data is available we find the same
  structure in these B-band profiles, so we are not dealing with a
  problem of sky subtraction.
  For 17 of the 37 galaxies the inner exponential zone, with a break at
  $(1.3\pm0.4)$ times the inner scale-length, is followed by
  a downbending, steeper outer region.
  This structure could be linked to a truncation of the radial light
  distribution similar to the one observed for HSB disk 
  \citep{1979A&AS...38...15V,pohlenetal2002} and irregular galaxies
  \citep{2006ApJS..162...49H}.
  On the other side, the remaining 4 galaxies show an inner exponential
  zone, with a break at $\approx (2.8\pm0.6)$ times the inner scale-length,
  which is followed by an upbending, shallower outer region.
  These kind of profiles, called sometimes antitruncations, are also known for
  HSB and irregular galaxies 
  \citep[see][]{erwinetal2005,2006A&A...454..759P,2006ApJS..162...49H}. 
  For galaxies with truncated profiles the break appears to be too
  'early' compared to the mean value of 2.5\,$\pm$\,0.6 times the
  inner scale-length as observed by \citet{2006A&A...454..759P} for
  a large sample of nearby, late-type HSB galaxies.
  However, it is not yet clear what this ratio should be for LSB
  galaxies. 
  In a sample of irregular galaxies, \cite{2006ApJS..162...49H} 
  find typical breaks in the range of $1.5\!-\!1.7$ times the
  inner scale-length for their mostly low surface brightness 
  Im systems. 
  Nevertheless, we know that these truncations are also observed for 
  HSB galaxies at higher redshift \citep{perez2004,2005ApJ...630L..17T} 
  and show up there in the profile spatially slightly `earlier' 
  compared to local galaxies, at $1.8\,\pm\,0.5$ times the inner scale 
  length compared to $2.5\,\pm\,0.6$
  times locally \citep[see][]{perez2004,2006A&A...454..759P}.
  %
  %We know that these truncations are also observed for
  %galaxies at higher redshift \citep{perez2004,2005ApJ...630L..17T}
  %and show up there rather early in the profile \citep[see][]{perez2004}.
  %
  Just concentrating on the 19 galaxies for which GSFC multi-color data
  is available we find that 6 have radial profiles showing a
  truncation. Four of these are selected as highly probable LSB
  candidates and for three of the four LSB candidates the break appears at a
  mean surface brightness of
  $\mu_\mathrm{{\rm br},B}$\,=\,24.4\,$\pm$\,0.1\,mag\,arcsec$^{-2}$.
  This is consistent with the value of
  $\mu_\mathrm{{\rm br},B}$\,=\,24.1\,mag\,arcsec$^{-2}$ given
  by \citep{2006A&A...454..759P} for truncations probably
  related to star-formation thresholds.

\section{Summary and Conclusions}

  We presented the results of a search for LSB galaxies in a 0.76\,deg$^2$
  NOAO $B_\mathrm{W}$--band dataset including the Hubble Deep
  Field-South, which resulted in a sample of 37 galaxies with sizes
  between 12$\arcsec$\,$\le$\,d\,$\le$36$\arcsec$ and central surface
  brightness of
  22\,mag\,arcsec$^{-2}$\,$\le$\,$\mu_\mathrm{0,B_{W}}$\,$\le$\,27\,mag\,arcsec$^{-2}$.
  Using the smaller (0.59\,deg$^2$ field of view) and less sensitive
  GSFC multi-color $UBVRI$--band
  dataset we were able to obtain colors for 18 galaxies. 

  \indent
  The galaxy selection was done in the $B_\mathrm{W}$--band NOAO
  data, using two different search methods: a median filter method
  for the faintest objects
  ($\mu_\mathrm{0,B_{W}}$\,$\ge$\,24.5\,mag\,arcsec$^{-2}$) source extractor
  program SExtractor for the brighter ones.

  \indent
  In order to constrain the contamination of our sample by high
  redshift HSB galaxies in the background, we applied two selection
  criteria. On the one hand, by setting a diameter limit of 10.8\,arcsec.
  On the other hand, by a comparison of the colors of the selected galaxies for
  which GSFC multi-color information were available (subsample of 18
  galaxies), to those of five standard galaxy types. 
  We were thus
  able to eliminate the 9 most distant objects in our sample, which are not
  intrinsic LSB galaxies. It was not possible, however, to
  derive distances for galaxies in the local (z\,$<$\,0.15) Universe 
  due to the large uncertainties of this photometric redshift approach, and 
  our sample will still have a small contamination of moderately
  (z\,$\sim$\,0.2--0.5) redshifted, cosmologically
  dimmed HSB galaxies. 
  
  \indent
  Using these selection criteria we were able to
  derive a sample of 9 highly probable intrinsic LSB galaxy candidates
  for which spectroscopic follow-up observation should be carried
  out. Comparing the colors of our LSB candidate
  sample to those of the five standard HSB galaxy types we found
  that 7 occupy a different locus in color-color space, bluer in
  the $B-V$ and redder in the $U-B$ color for the $U-B$ vs. $B-V$ color-color
  diagram and bluer in the $B-R$ and redder in the $U-B$ color for the $U-B$
  vs. $B-R$ color-color diagram. This seems to be
  a first hint for a different stellar
  population mix and, therefore, also for a different star formation history
  for these galaxies, which also have higher fluxes in the
  $B$-band (compared to HSB galaxies). This is a possible 
  indication for a more prominent Balmer-bump linked to a younger 
  stellar population.

  \indent
  We also identified three galaxies with very low central surface
  brightness ($\mu_\mathrm{0,B_{W}}$\,$\ge$\,25.5\,mag\,arcsec$^{-2}$),
  for which we have no color information. Due to their 
  relatively large sizes we expect them to have distances of 
  z\,$\le$\,0.5, and they cannot be redshifted--dimmed HSBs. In total we
  ended up with a final sample of 12 possible LSB galaxy candidates in a 
  0.76\,deg$^2$ field. Scaling this result only for the field size of the GSFC
  we derived a number density of 16 LSB candidates per deg$^2$, which is 4
  times higher than the number densities derived for former surveys e.g., the
  Texas Survey which resulted in 4 LSB galaxies per
  deg$^2$ \citep[][]{1997AJ....113.1212O,1997AJ....114.2448O}.   
 
  \indent
  We did not find any giant LSB galaxiy with diameter larger than 
  1$\arcmin$. In
  comparison to other surveys we only found galaxies from the small end of the
  size distribution, with scale-lengths smaller than 6\,arcsec (see
  Fig.~\ref{mubBtot}). 
  This indicates that the majority of the derived sample could consists
  of more distant or dwarf like LSB galaxies compared to other surveys
  (e.g. Texas Survey). However, this assumption needs verification from 
  spectroscopic observations. 

  \indent
  The sample contains galaxies with moderate color indices but also one galaxy
  with a very red color index (\object{LSB J22324-60520}:
  $U$-$B$=0.46\,mag,  $B$-$V$=1.28\,mag, $B$-$R$=1.99\,mag).

%__________________________________________________________________

\begin{acknowledgements}
This research was supported by DFG Graduiertenkolleg "The Magellanic Systems,
Galaxy Interaction and the Evolution of Dwarf Galaxies" (Universities
Bonn/Bochum). We thank the NOAO Deep Survey team for making the pilot survey
data immediately public, and the STIS team at GSFC for the second data set. 
\end{acknowledgements}
\appendix
\section{Radial Profiles}
\label{appA}
Here we show the $B_\mathrm{W}$ azimuthally averaged radial surface
brightness profiles of the full sample of 37 LSB candidate galaxies 
found in the NOAO field. They were derived by fitting ellipses to the 
galaxies using the IRAF
task {\it ellipse} from the {\it stsdas} package. The ellipses were choosen 
with a starting positions and allowing for only slight recentering 
by the {\it ellipse} task. Therefore the centers
of the galaxies could change slightly between fits in the $B_\mathrm{W}$- and
B-filter. For the 18 galaxies where $B$ filter data is available 
from the GSFC dataset, we also show the surface brightness
profiles in this band. The central surface brightnesses in both filters
was estimated fitting a single exponential 
profile. For this fitting process we did not include data points at radii
smaller 0.9\, arcsec which are influenced by seeing effects (rounding of
profiles). For those galaxies showing a truncation or antitruncation (see
Sect.~\ref{radprof}) in the outer region of the profile we only fit the inner
part using a pure exponential approach.
\begin{figure*}[ht]
\centering
\resizebox{7cm}{!}{\includegraphics{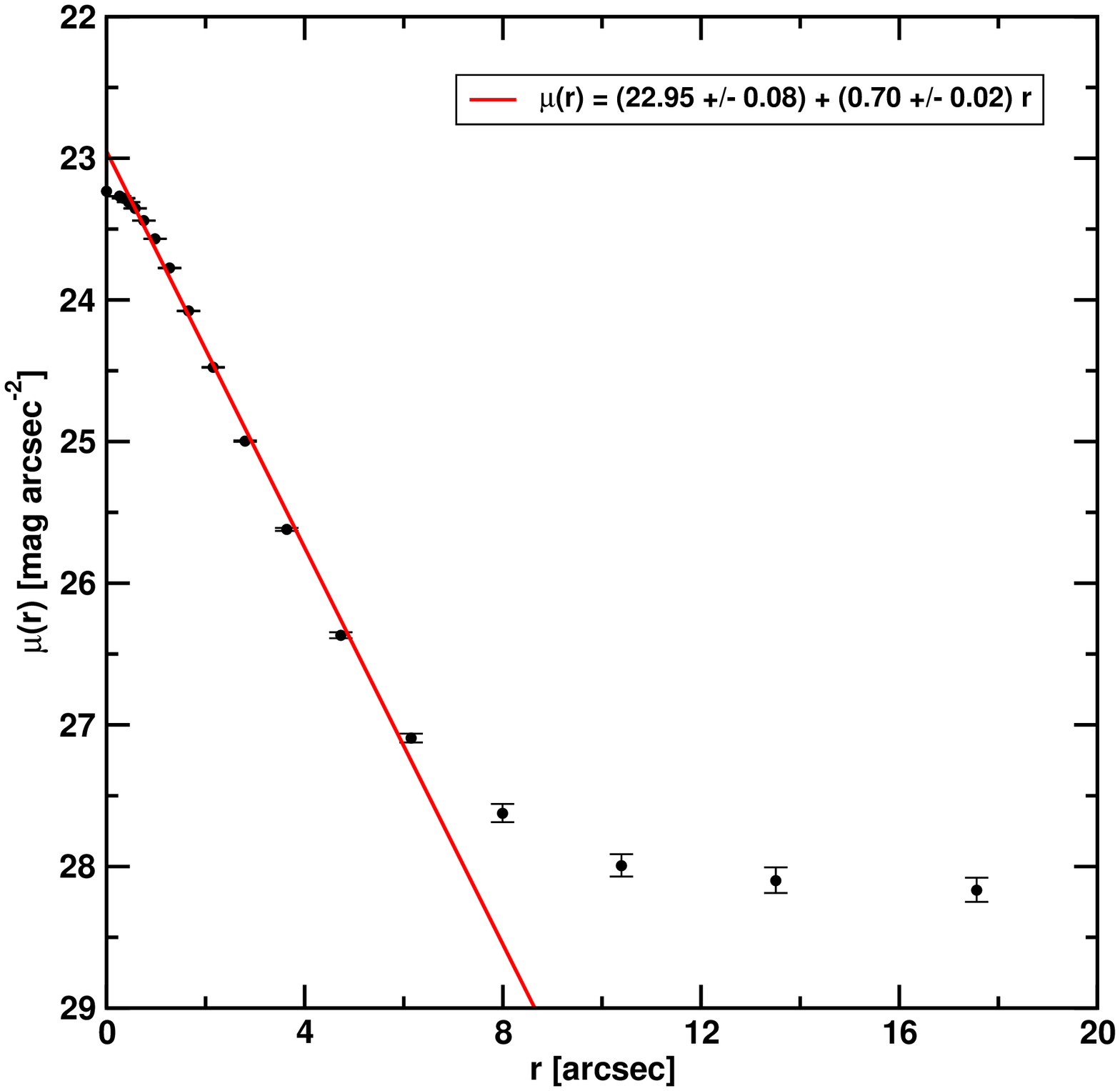}}
\resizebox{7cm}{!}{\includegraphics{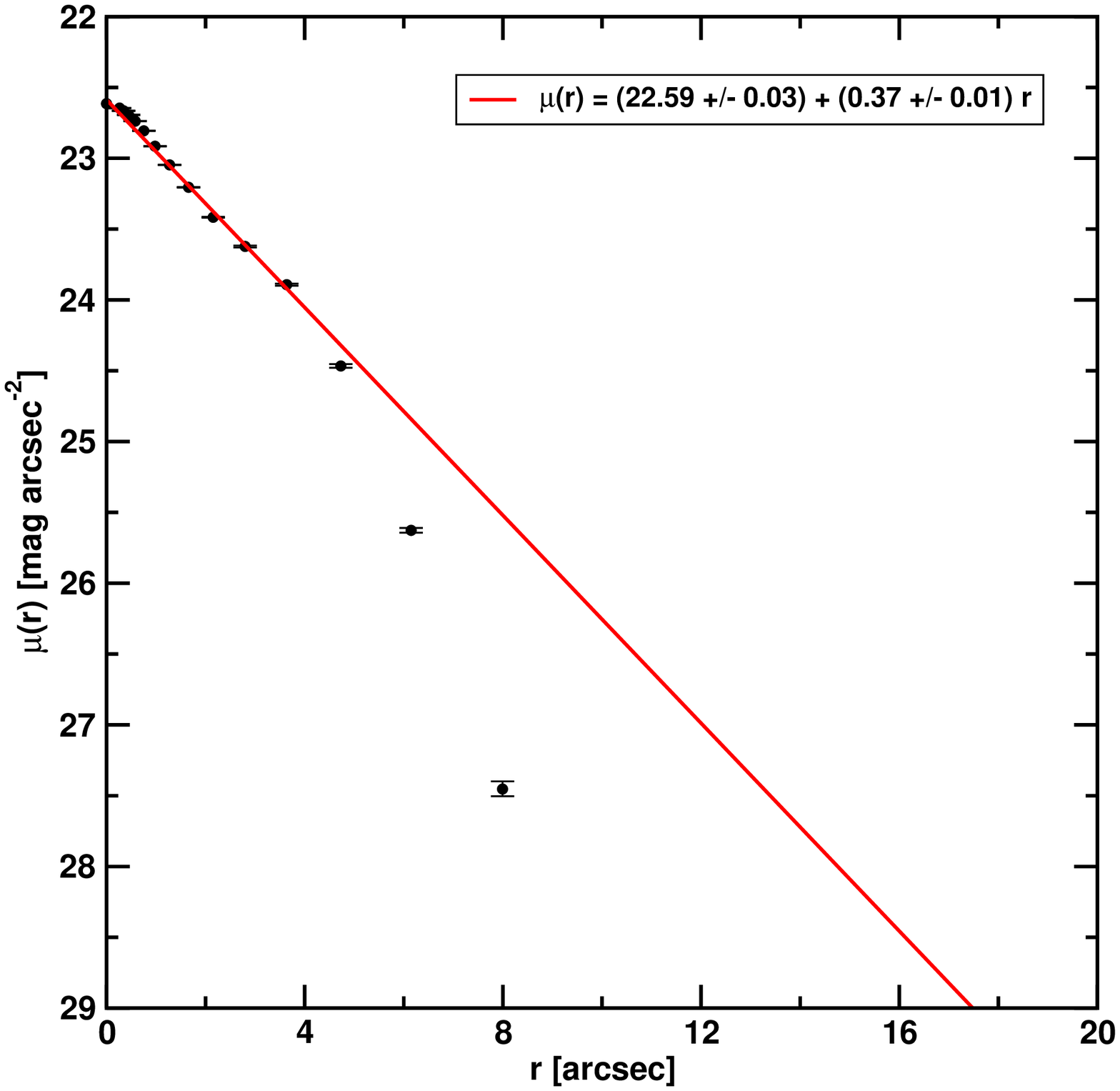}}
\caption{Surface brightness profiles of \object{LSB J22291-60303} (left panel)
  and \object{LSB J22291-60522} (right panel) are
  displayed. \object{LSB J22291-60303} show indications for a antitruncation
  around a radius of 5\,arcsec. For \object{LSB J22291-60522} a clear
  truncation of the profile in the outer region is visible.}
\resizebox{7cm}{!}{\includegraphics{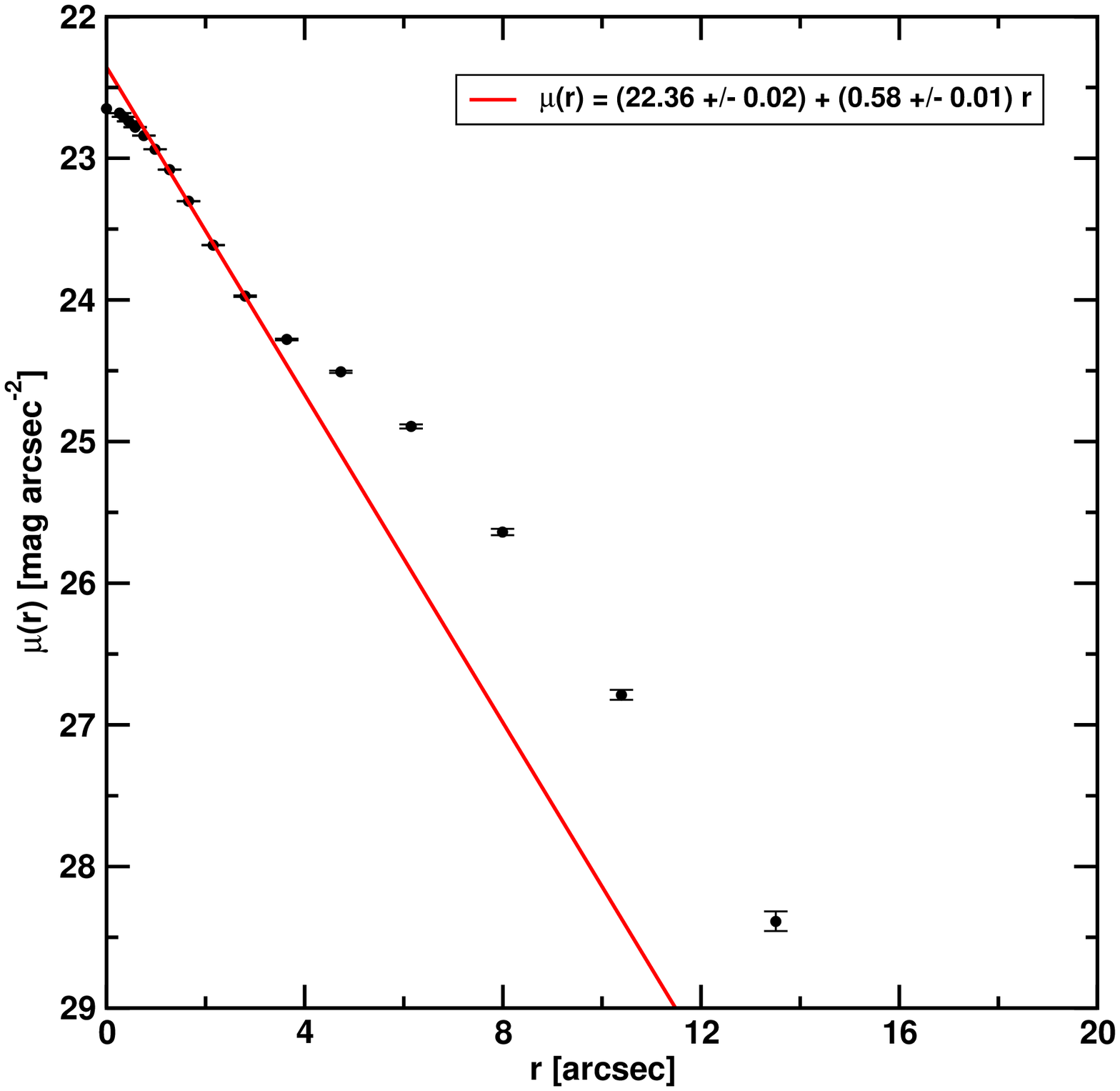}}
\resizebox{7cm}{!}{\includegraphics{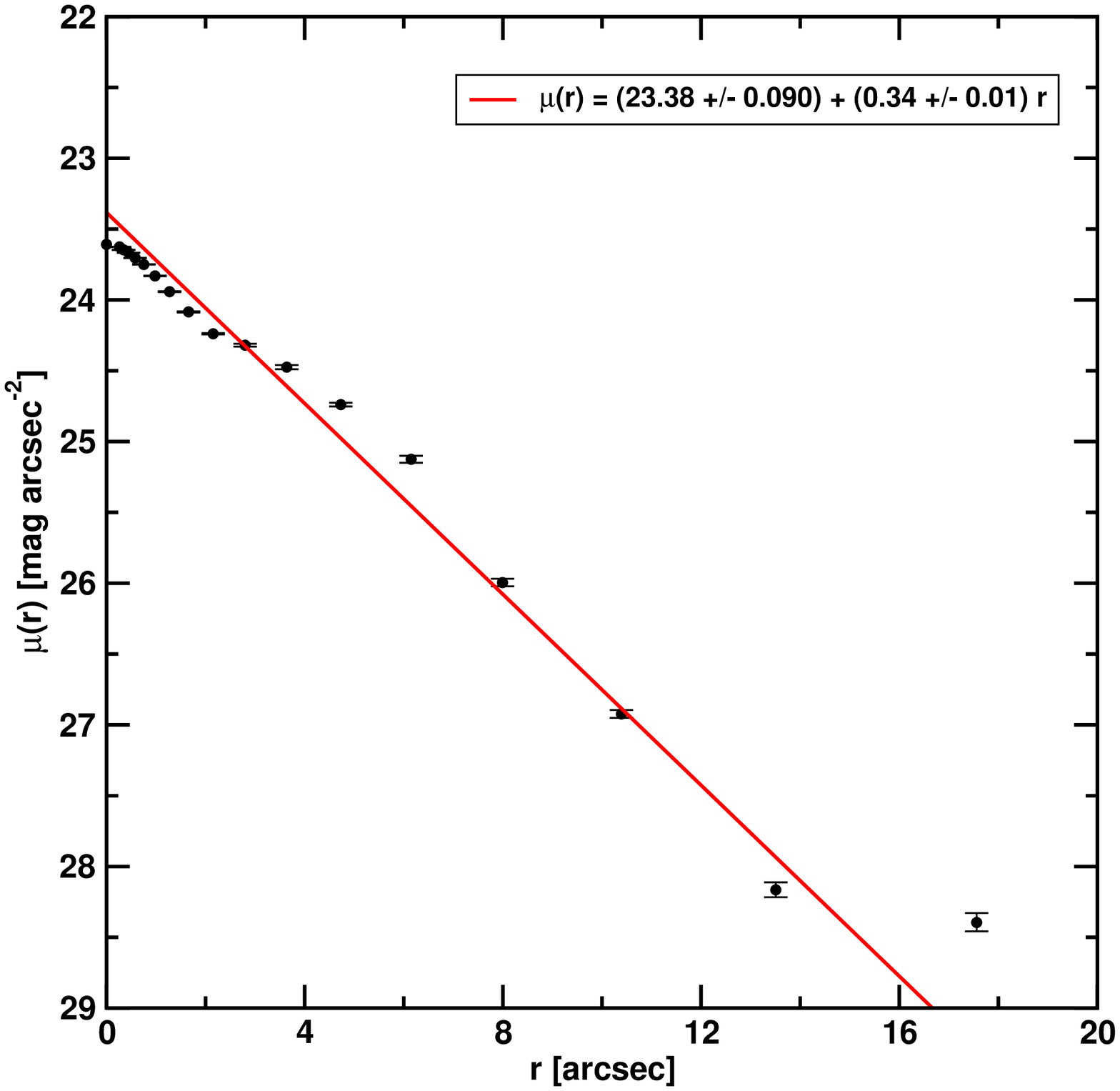}}
\caption{Surface brightness profiles of \object{LSB J22292-60540} (left panel)
  and \object{LSB J22293-60523} (right panel) are displayed. We excluded the
  outer part of the profile of \object{LSB J22292-60540} due to the innfluence
  of a spiral arm at 4\,arcsec.}
\resizebox{7cm}{!}{\includegraphics{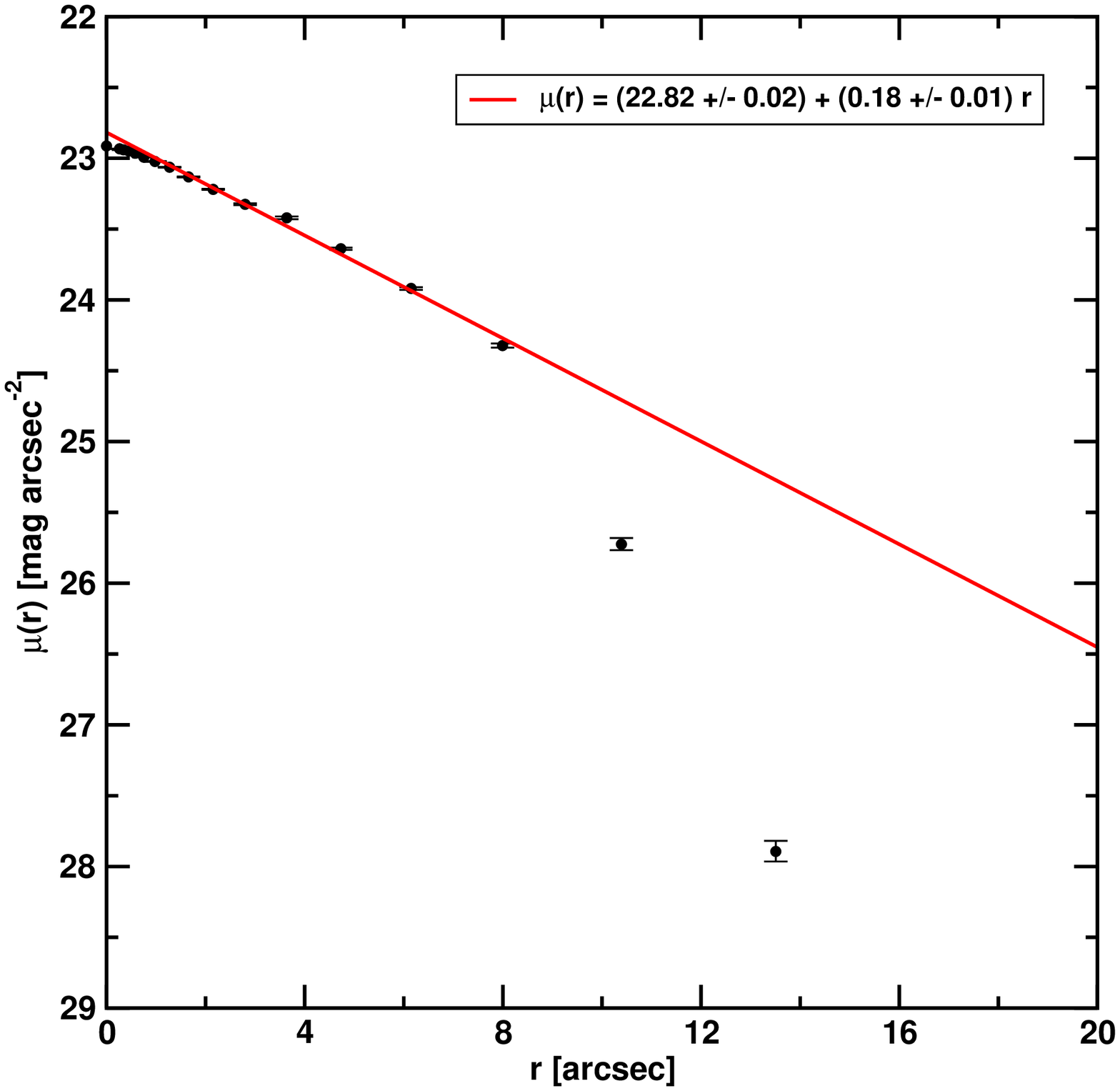}}
\resizebox{7cm}{!}{\includegraphics{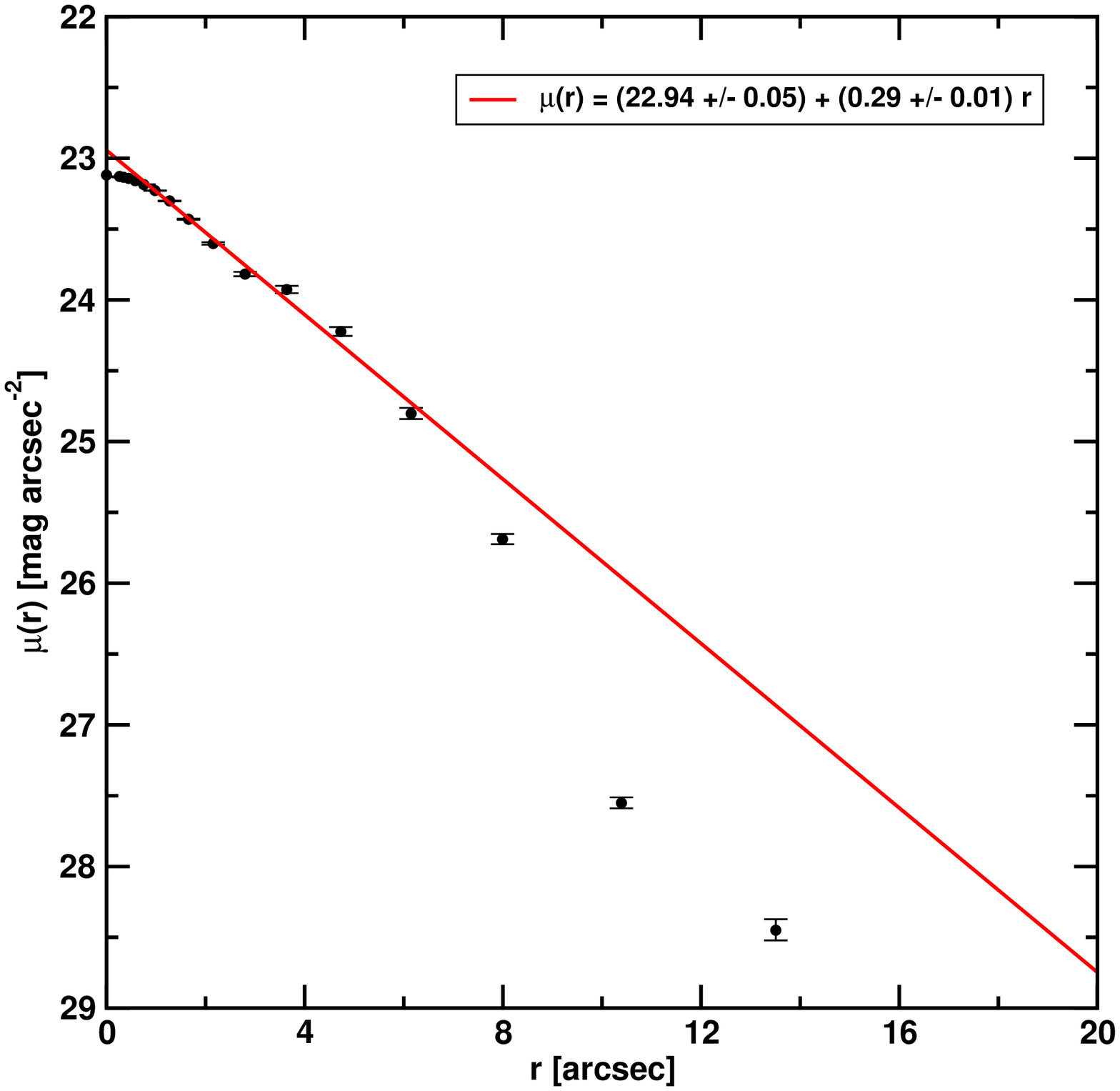}}
\caption{Surface brightness profiles of \object{LSB J22295-61001} (left panel)
  and \object{LSB J22300-60300} (right panel) are displayed. For both galaxies
  a clear truncation of the profiles in the outer region is visible.}
\label{radprof1}
\end{figure*}
\begin{figure*}
\centering
\resizebox{7cm}{!}{\includegraphics{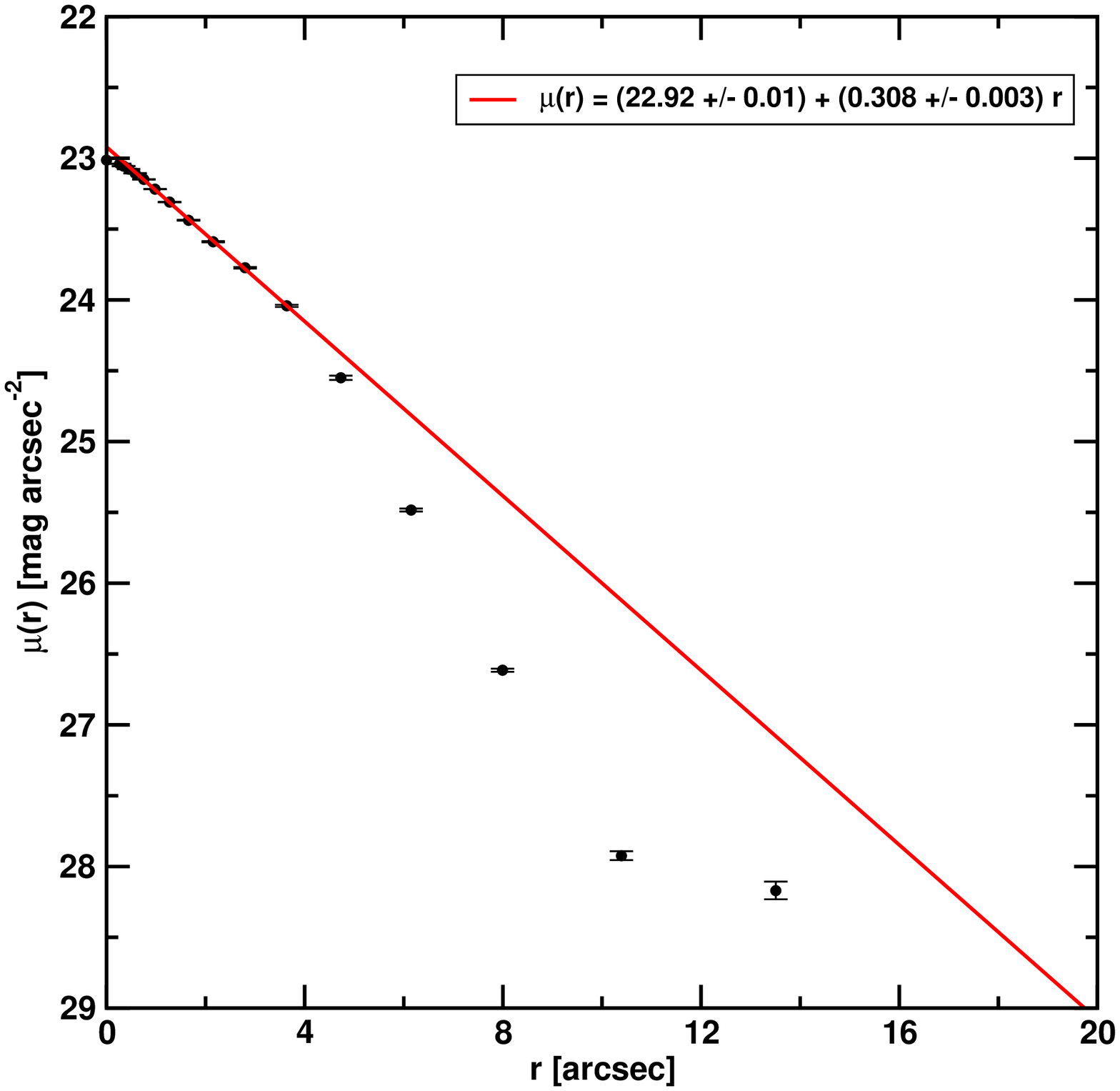}}
\resizebox{7cm}{!}{\includegraphics{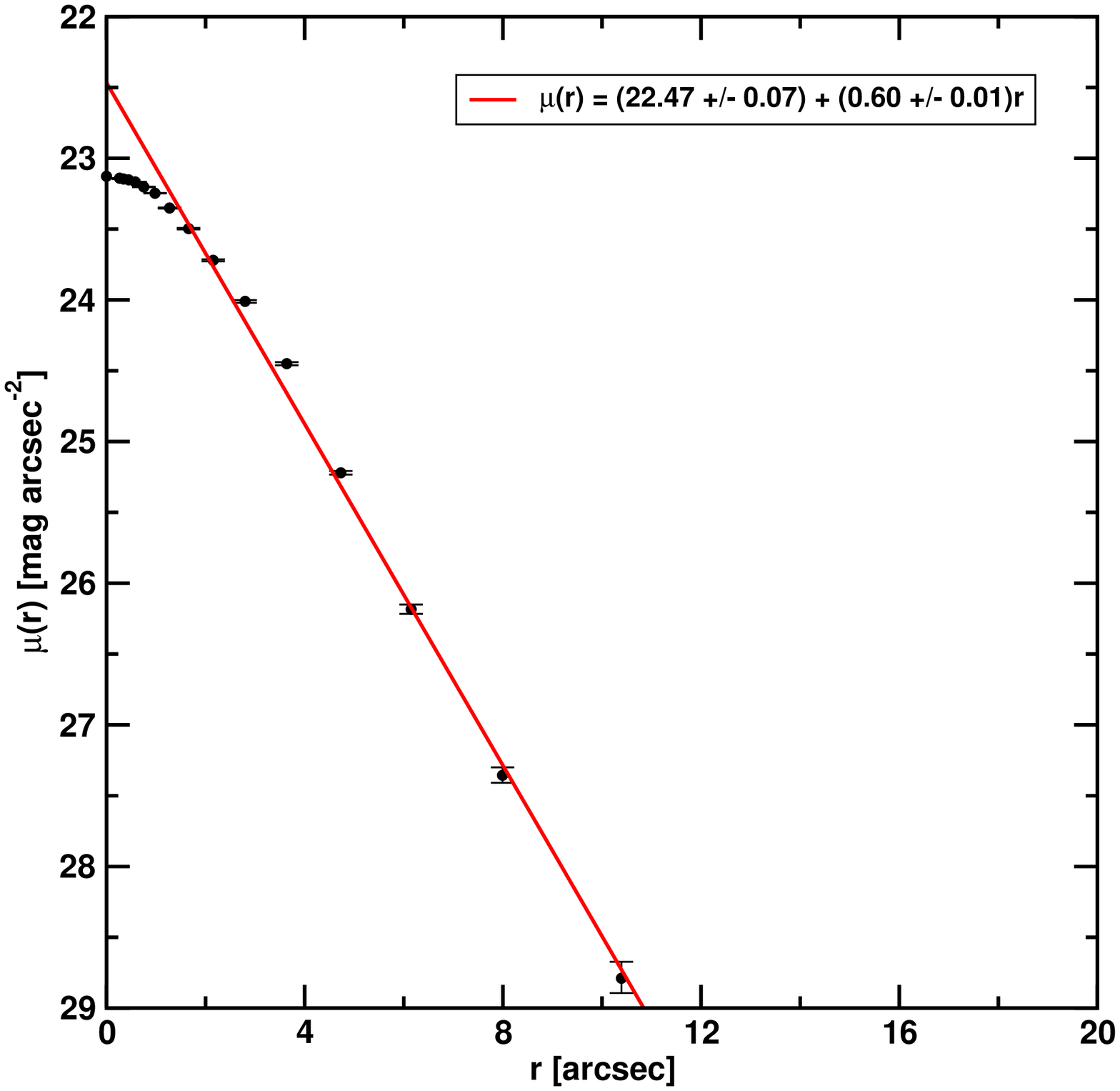}}
\caption{Surface brightness profiles of \object{LSB J22300-60380} (left panel)
  and \object{LSB J22301-60415} (right panel) are displayed. The profile of
  \object{LSB J22300-60380} show a clear truncation of the profile in the outer
  region.}
\resizebox{7cm}{!}{\includegraphics{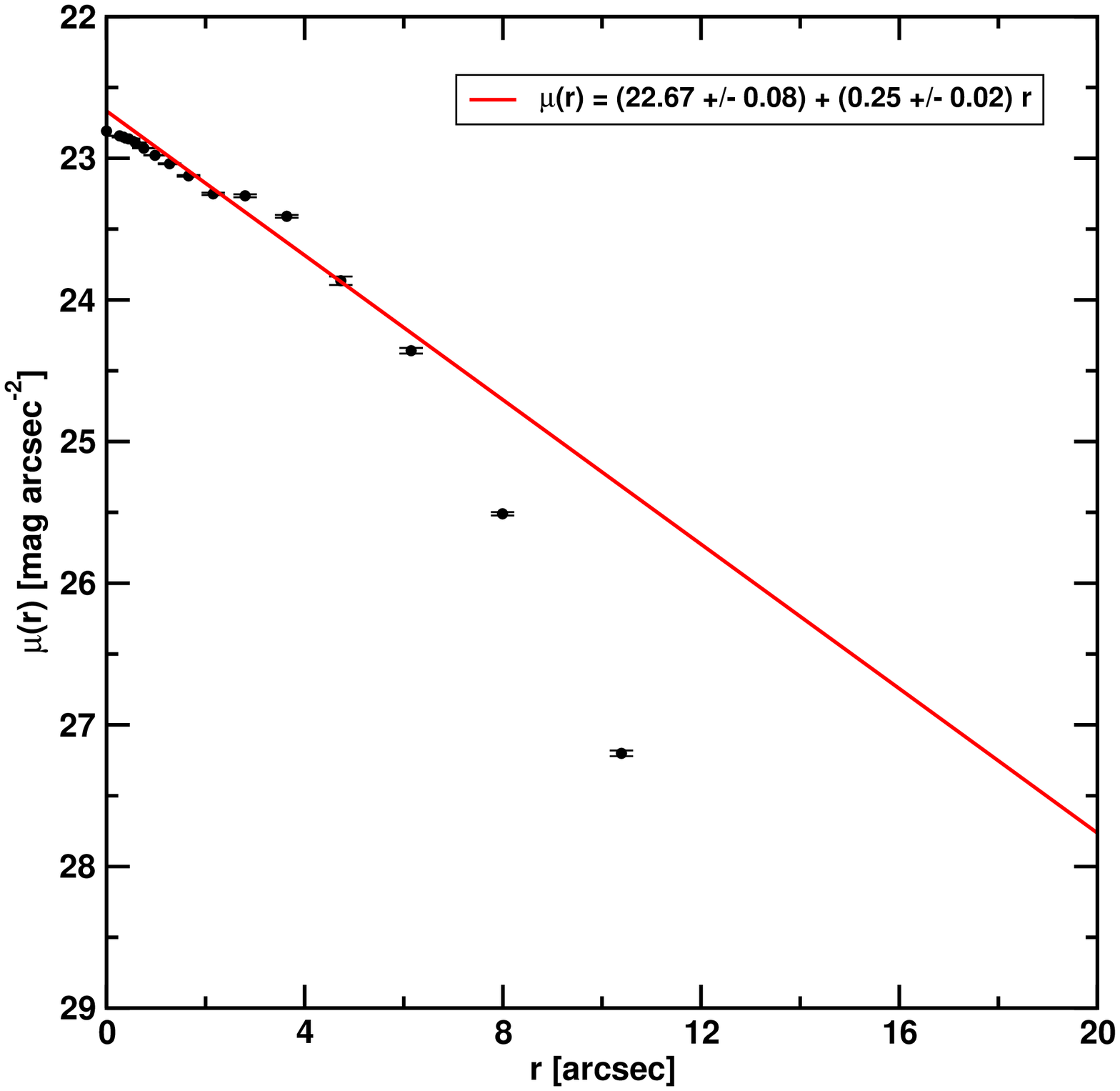}}
\resizebox{7cm}{!}{\includegraphics{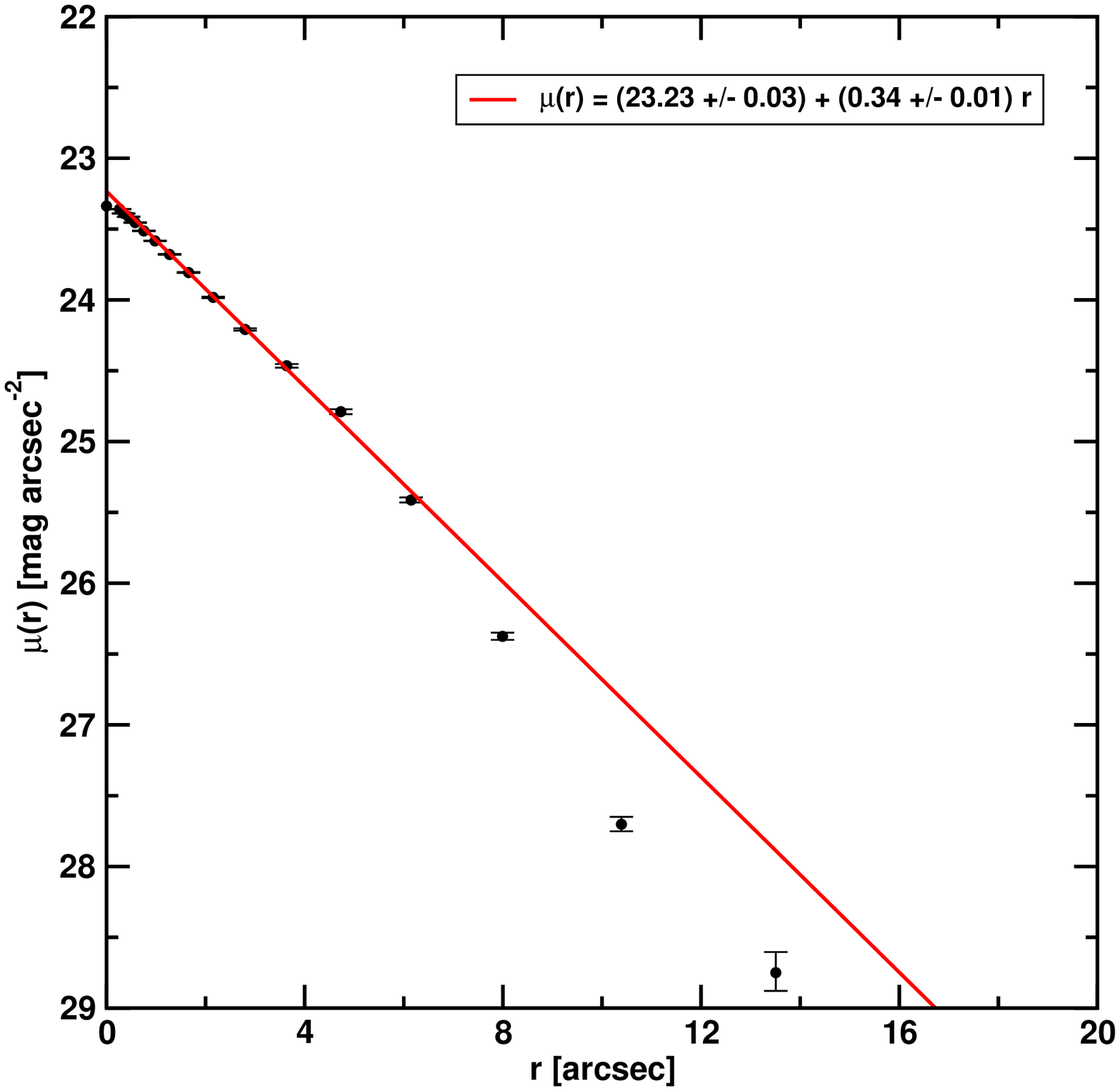}}
\caption{Surface brightness profiles of \object{LSB J22302-60352} (left panel)
  and \object{LSB J22302-60474} (right panel) are displayed. For both galaxies a
  clear truncation of the profile in the outer region is visible. }
\resizebox{7cm}{!}{\includegraphics{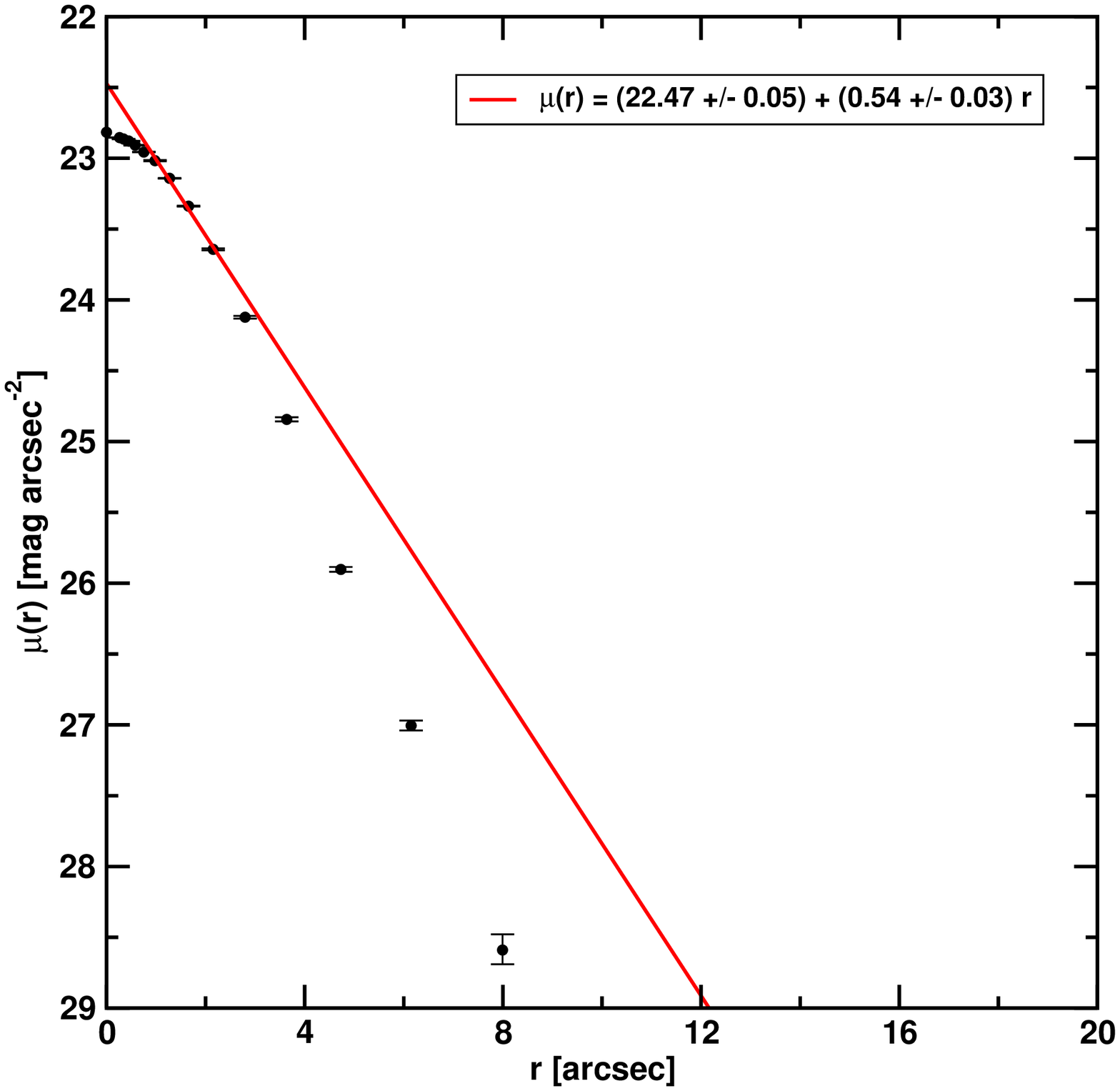}}
\resizebox{7cm}{!}{\includegraphics{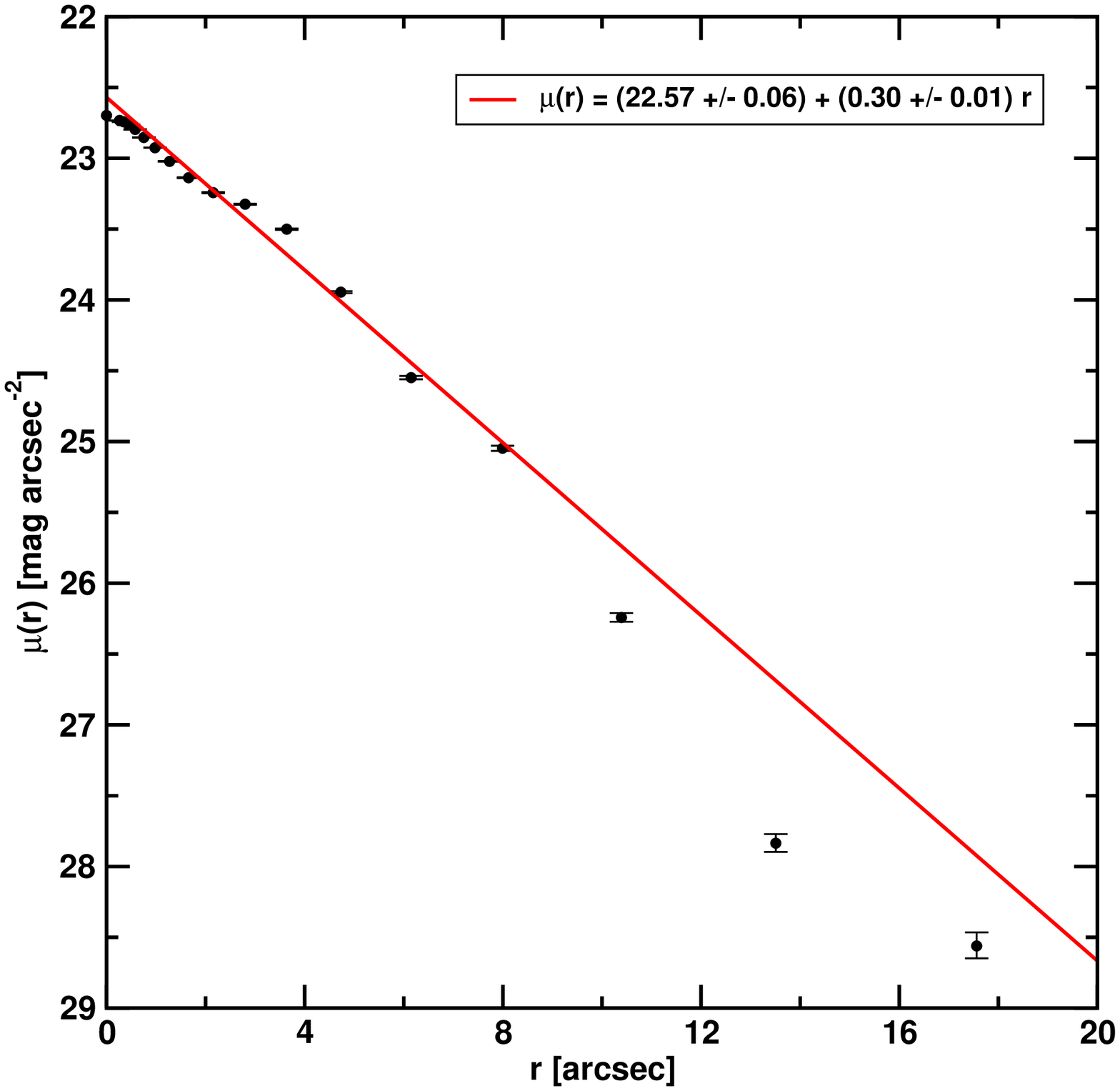}}
\caption{Surface brightness profiles of \object{LSB J22303-60514} (left panel)
  and \object{LSB J22304-61004} (right panel) are displayed. Both galaxies have
  a clear truncation of the profile in the outer region.}
\label{radprof2}
\end{figure*}
\begin{figure*}
\centering
\resizebox{7cm}{!}{\includegraphics{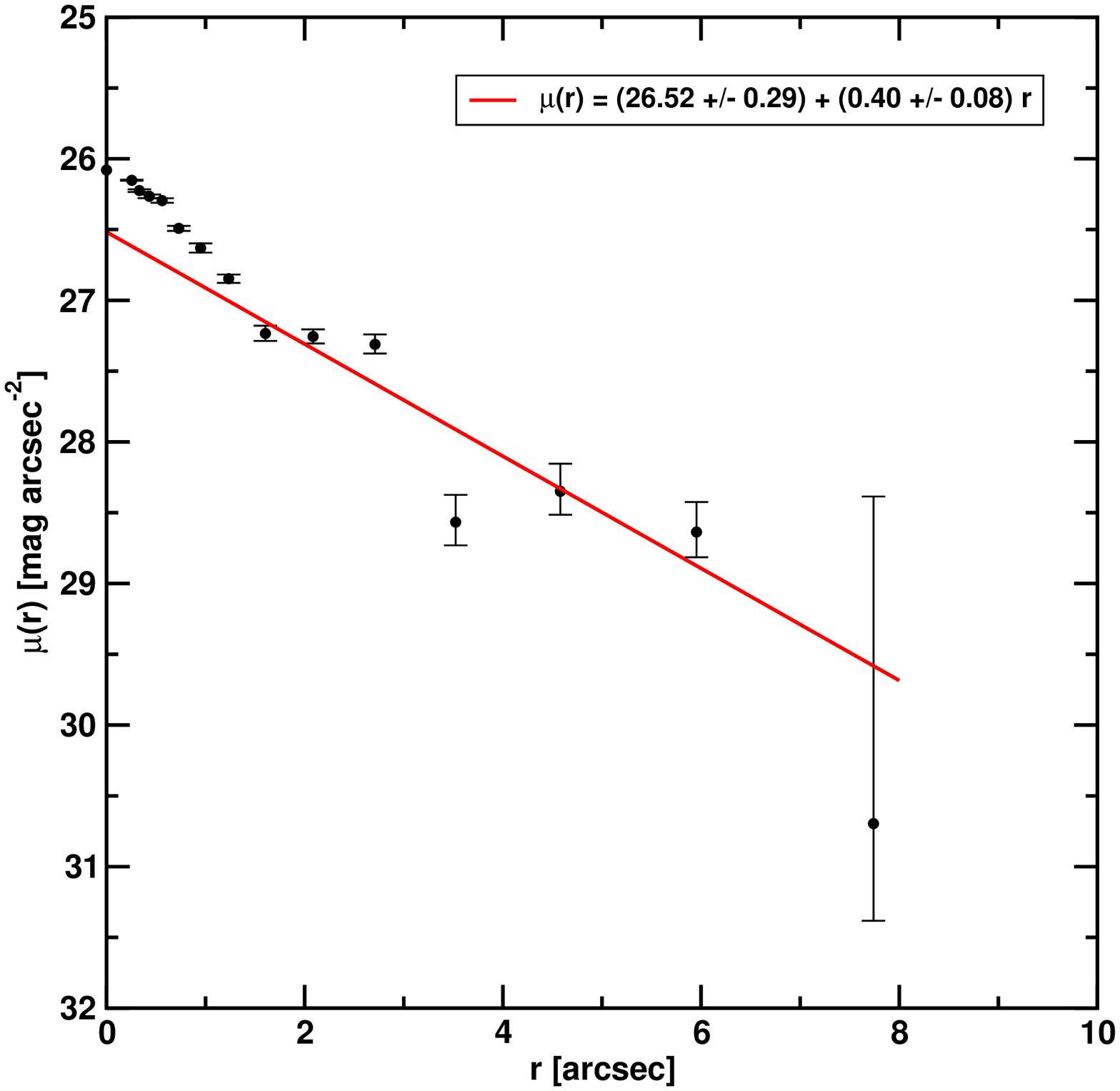}}
\resizebox{7cm}{!}{\includegraphics{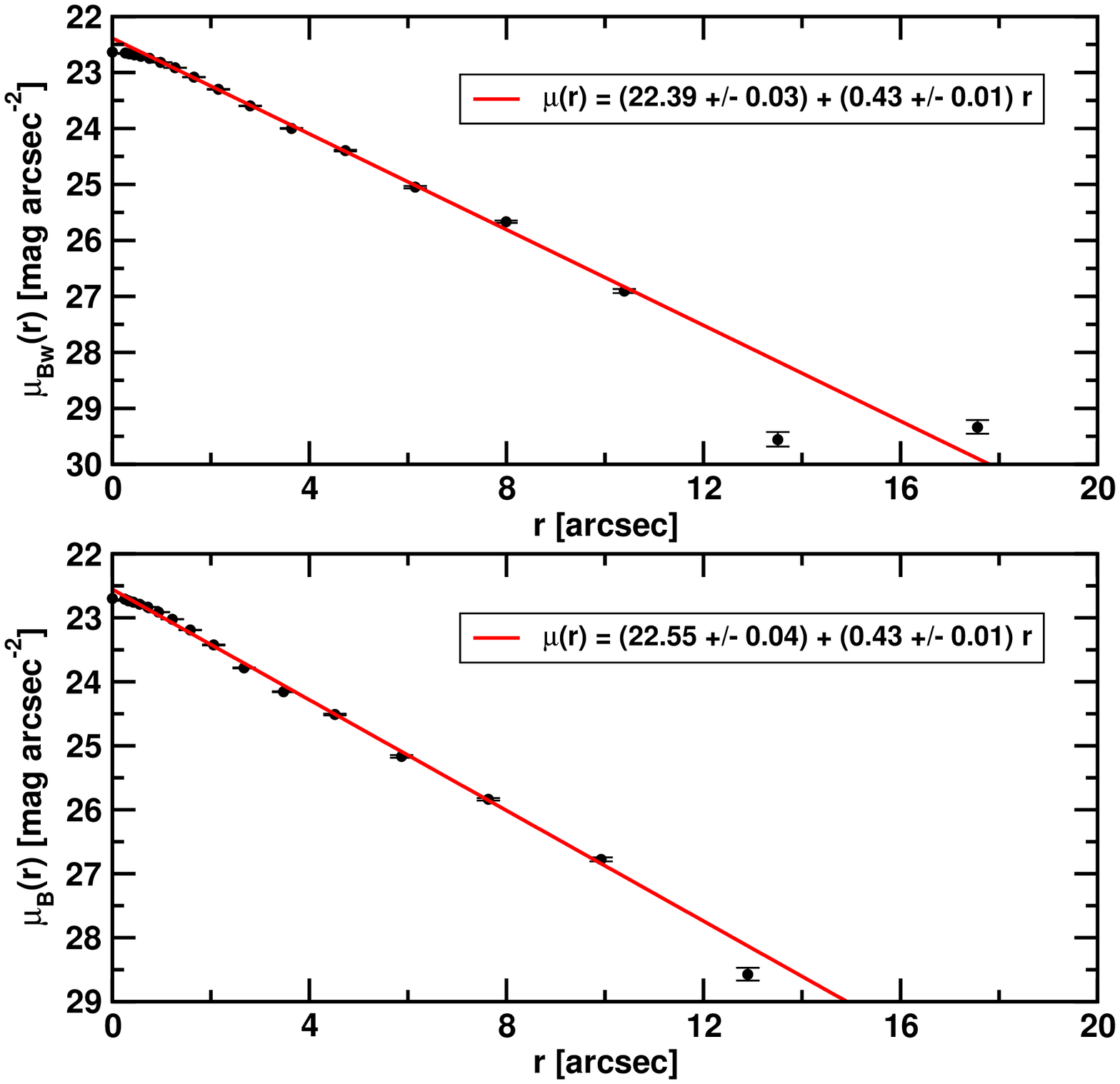}}
\caption{Surface brightness profiles of \object{LSB J22311-60160} (left panel)
  and \object{LSB J22311-60503} (right panels) are displayed. For
  \object{LSB J22311-60503} additionally the B-band surface brightness profile
  is shown. \object{LSB J22311-60160} belongs to the subsample of 3 extreme LSB
  galaxies.}
\resizebox{7cm}{!}{\includegraphics{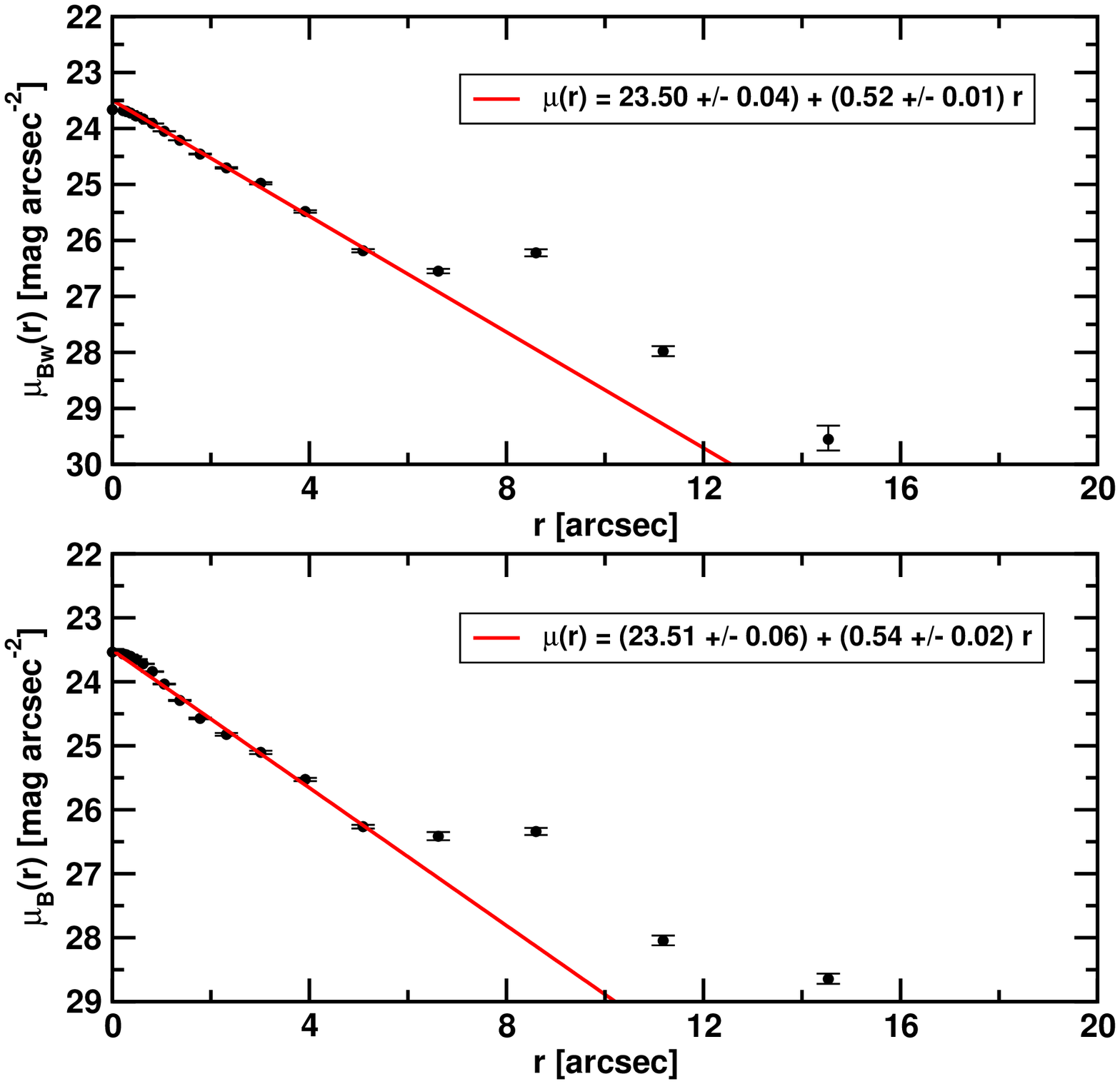}}
\resizebox{7cm}{!}{\includegraphics{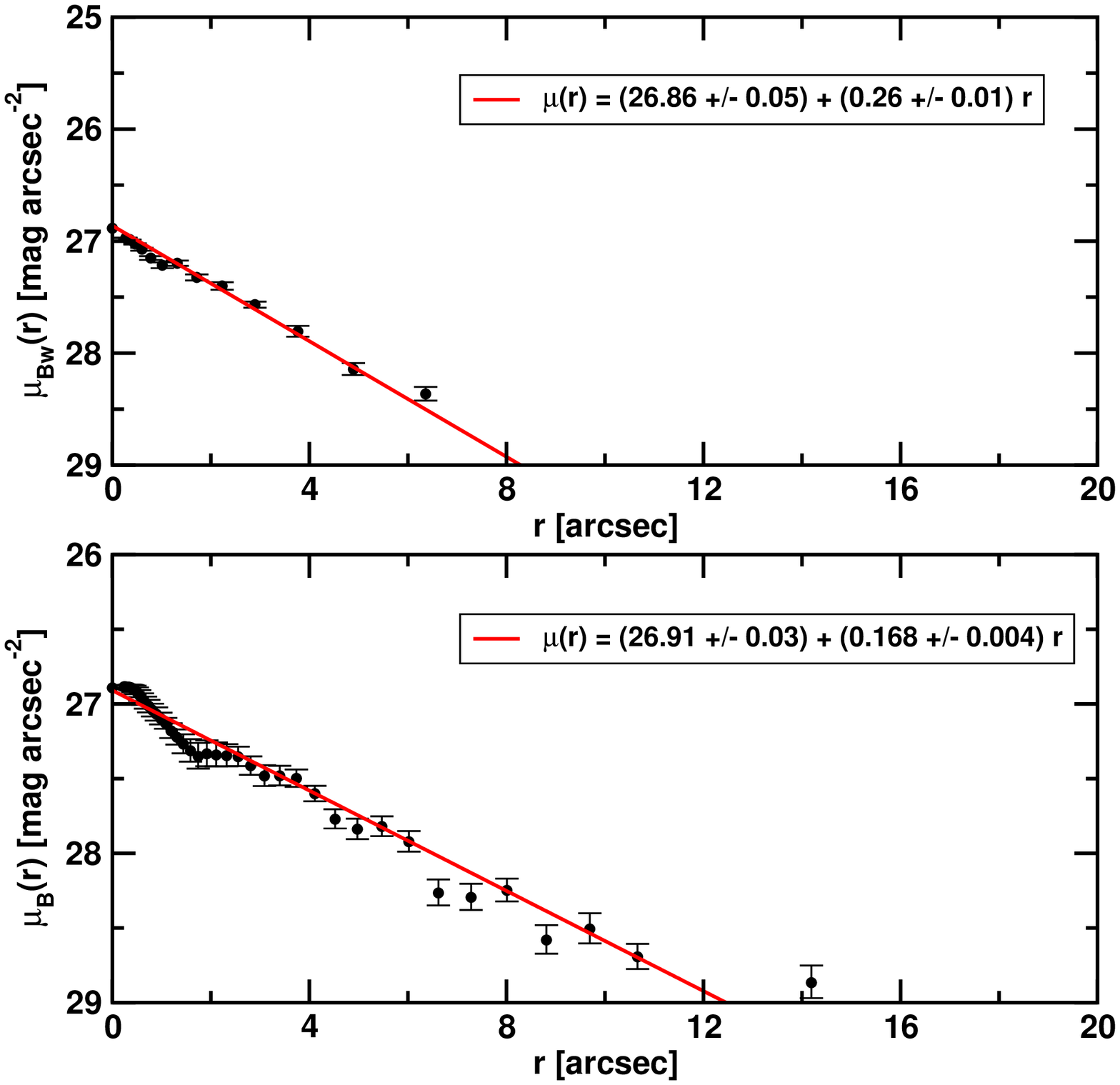}}
\caption{Surface brightness profiles in both filter bands of
  \object{LSB J22315-60481} (left panels) and \object{LSB J22320-60381} (right
  panels are displayed. The profile of \object{LSB J22315-60481} show
  indications for a antitruncation around a radius of
  4.5\,arcsec. \object{LSB J22320-60381} belongs to the subsample of 3 extreme
  LSB galaxies having one of the lowest measured central surface brightness
  known today ($\mu_\mathrm{0,B}$\,=\,26.91\,mag\,arcsec$^{-2}$).}
\resizebox{7cm}{!}{\includegraphics{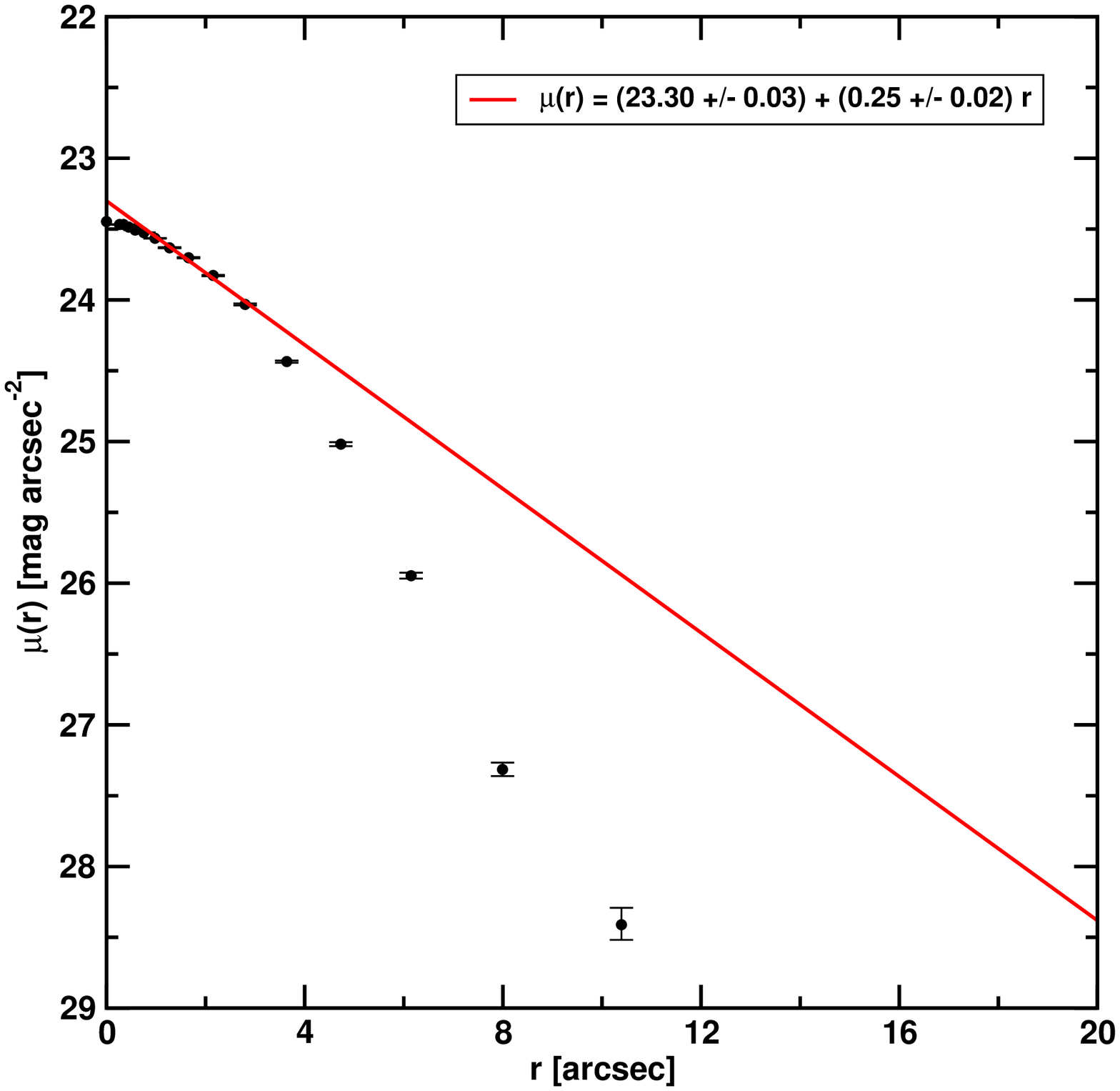}}
\resizebox{7cm}{!}{\includegraphics{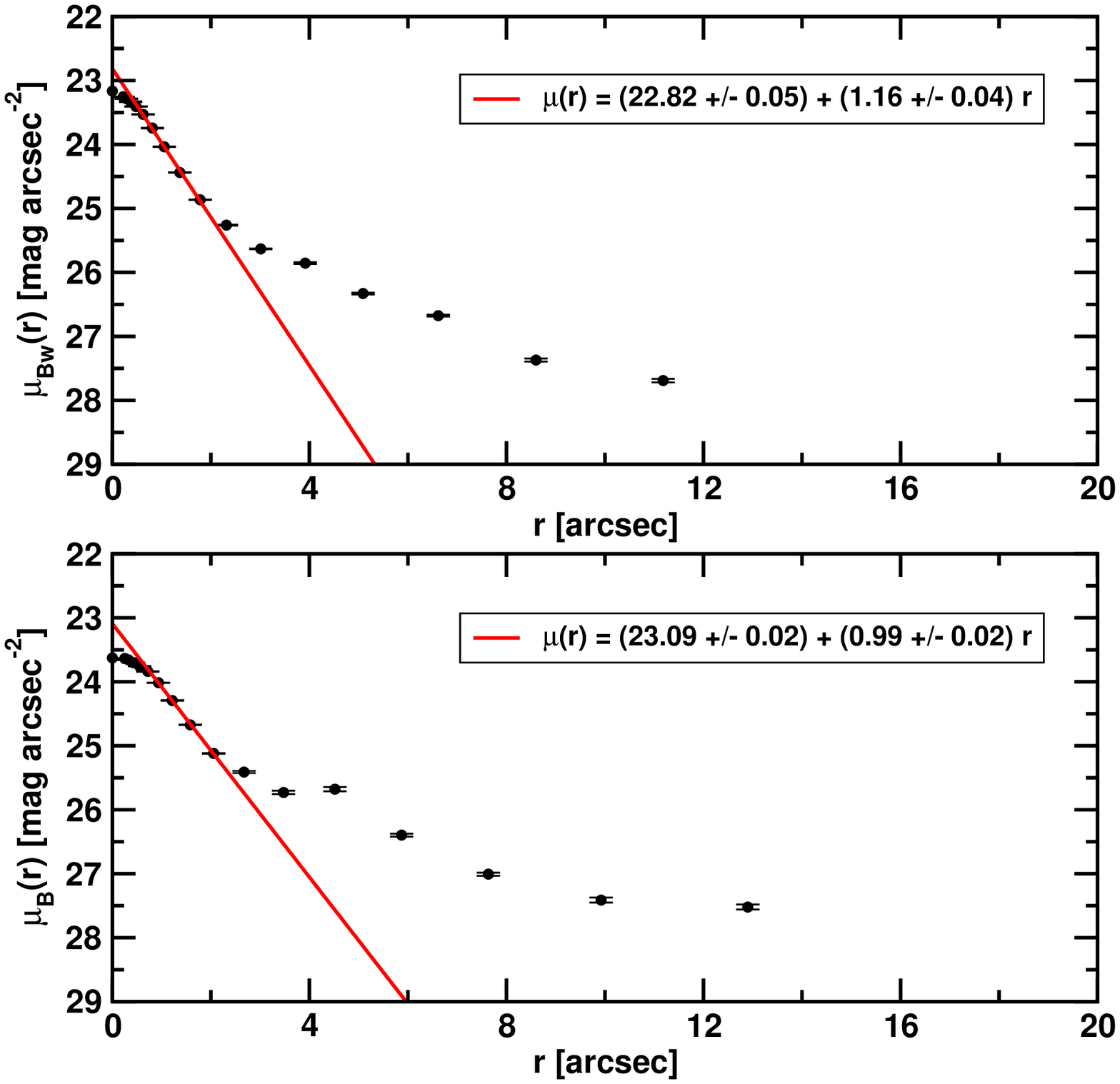}}
\caption{Surface brightness profiles of \object{LSB J22321-61015} (left panel)
  and \object{LSB J22322-60142} (right panels) are displayed. The radial
  profile of \object{LSB J22321-61015} shows a clear truncation in the outer
  region. For \object{LSB J22322-60142} additionally the B-band surface
  brightness profile is shown. The profiles in both filter bands show an
  indication for an antitruncation around a radius of 2\,arcsec.}
\label{radprof3}
\end{figure*}
\begin{figure*}
\centering
\resizebox{7cm}{!}{\includegraphics{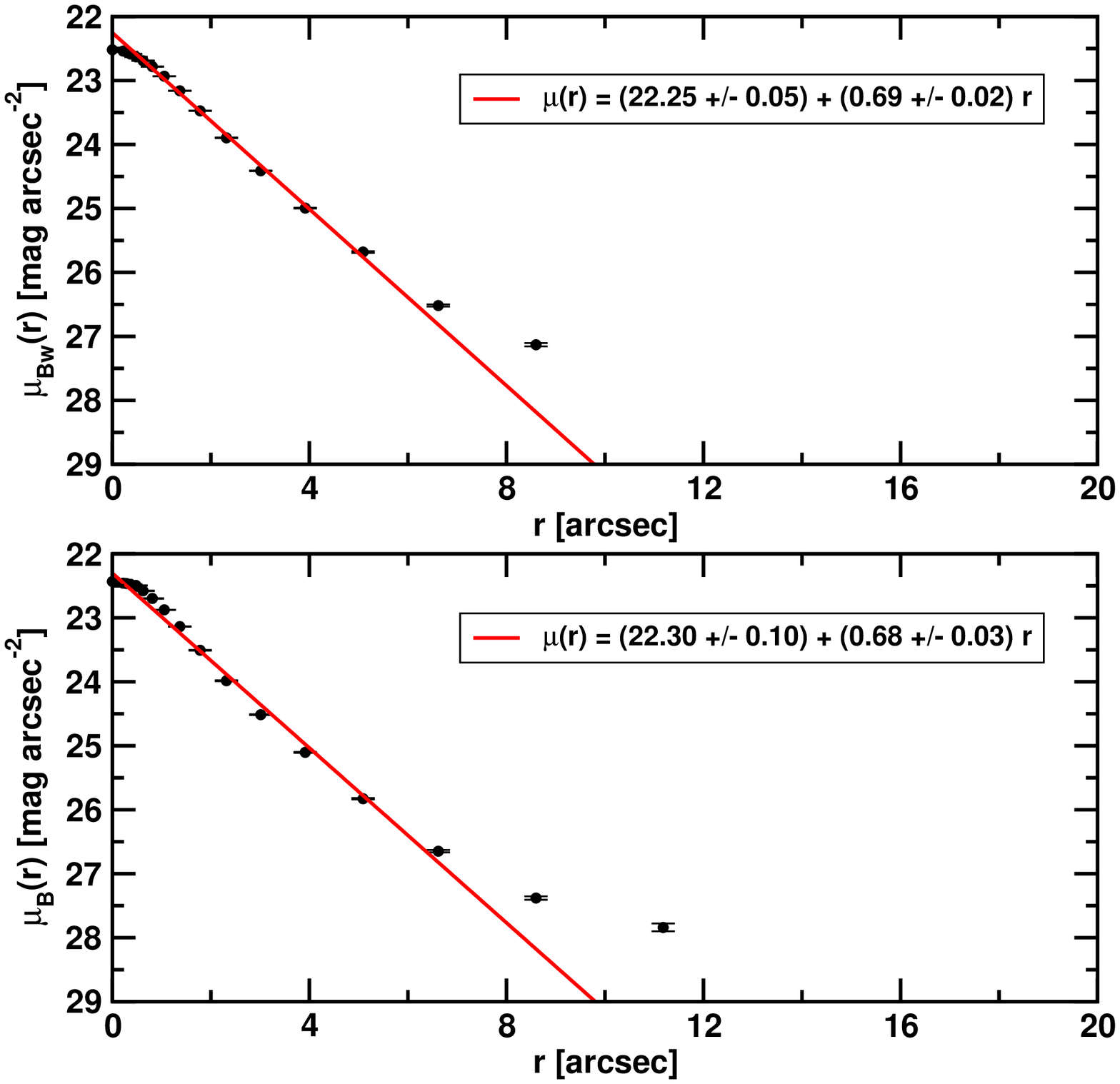}}
\resizebox{7cm}{!}{\includegraphics{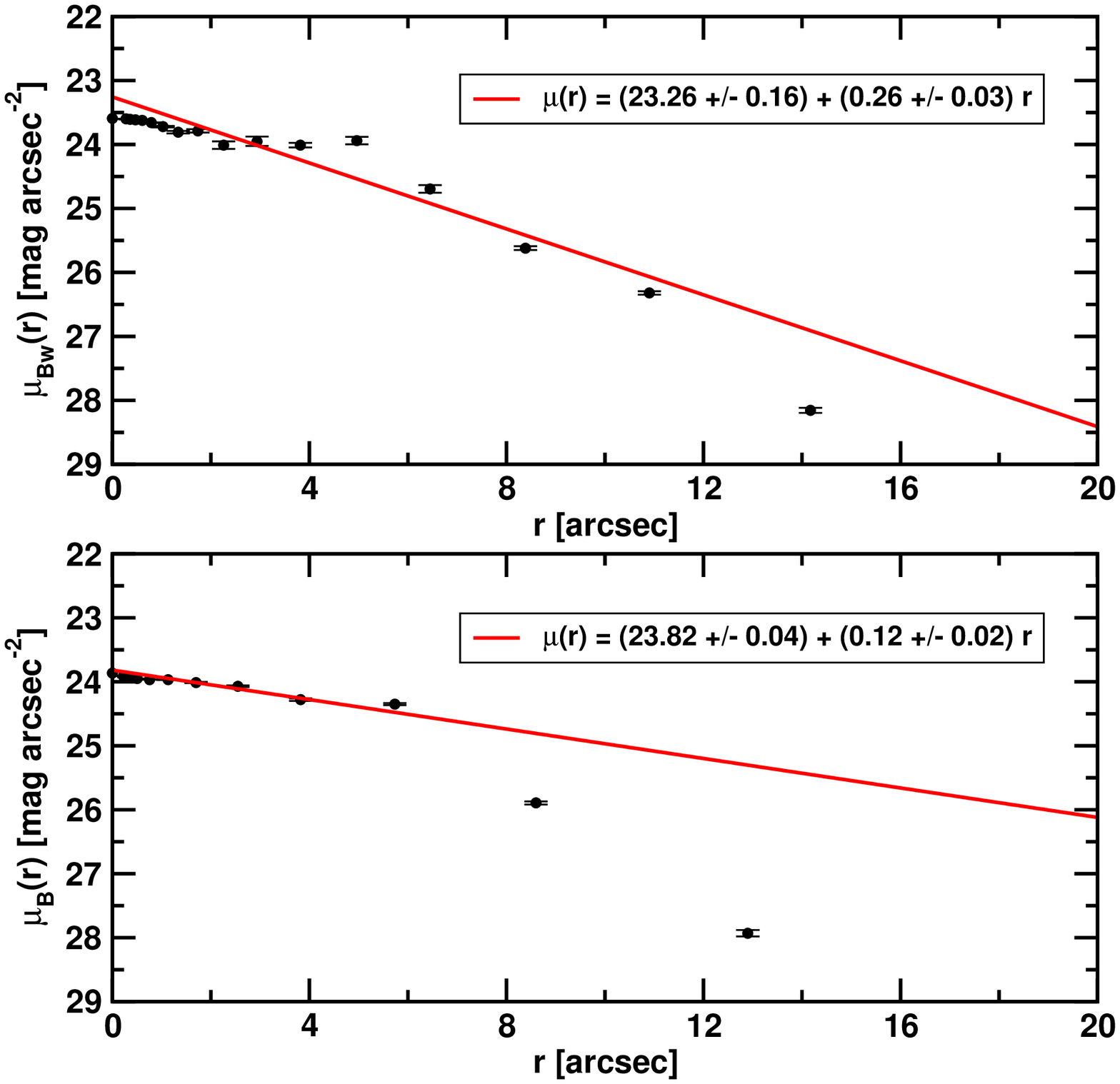}}
\caption{Surface brightness profiles in both filter bands of
  \object{LSB J22324-60520} (left panels) and \object{LSB J22325-60155} (right
  panels) are displayed. No truncation is visible for the B-band profile of
  \object{LSB J22325-60155}. The shape of the 
  profile is the result of the irregular structure of this galaxy, showing
  several bright spots.}
\resizebox{7cm}{!}{\includegraphics{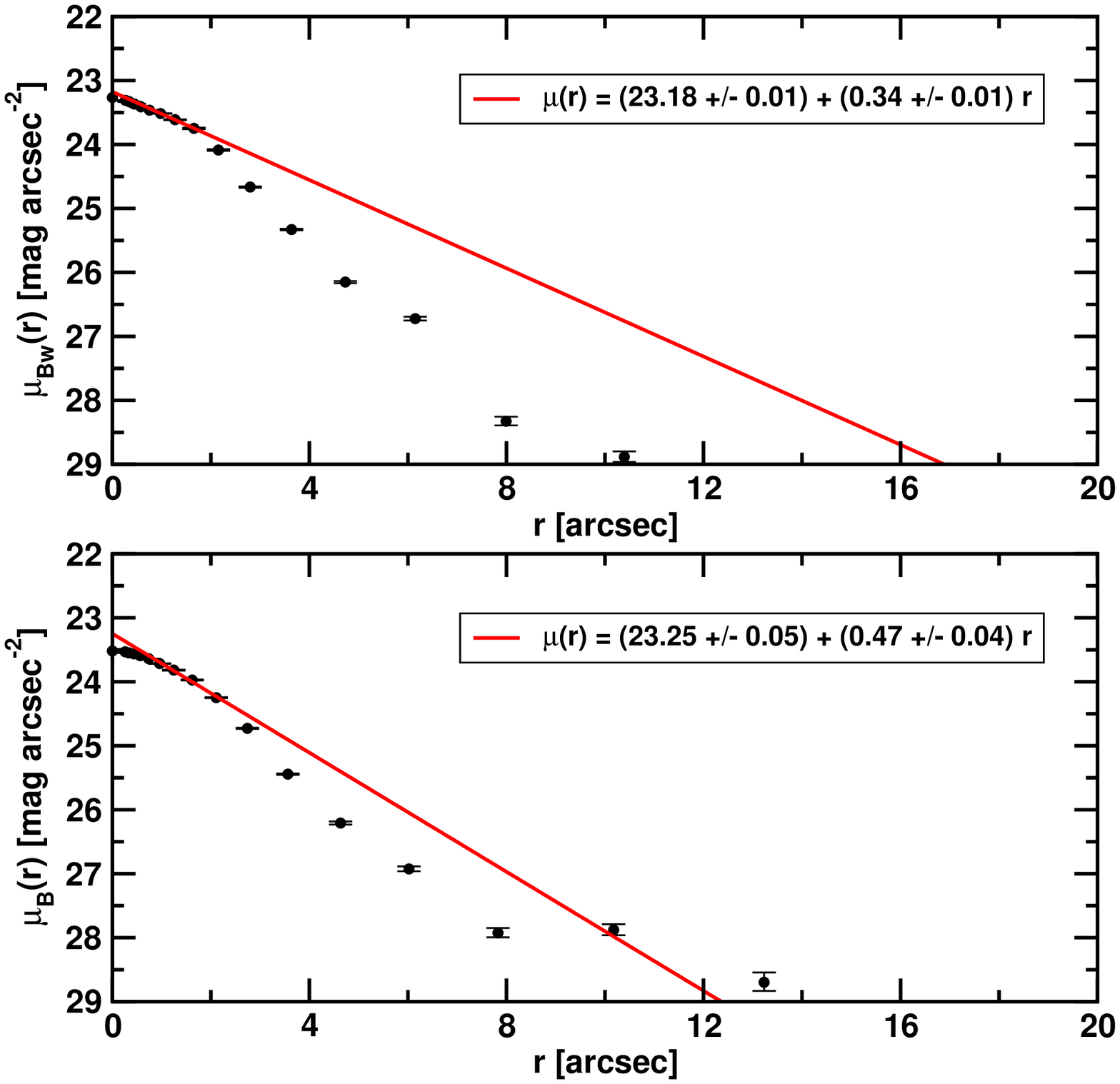}}
\resizebox{7cm}{!}{\includegraphics{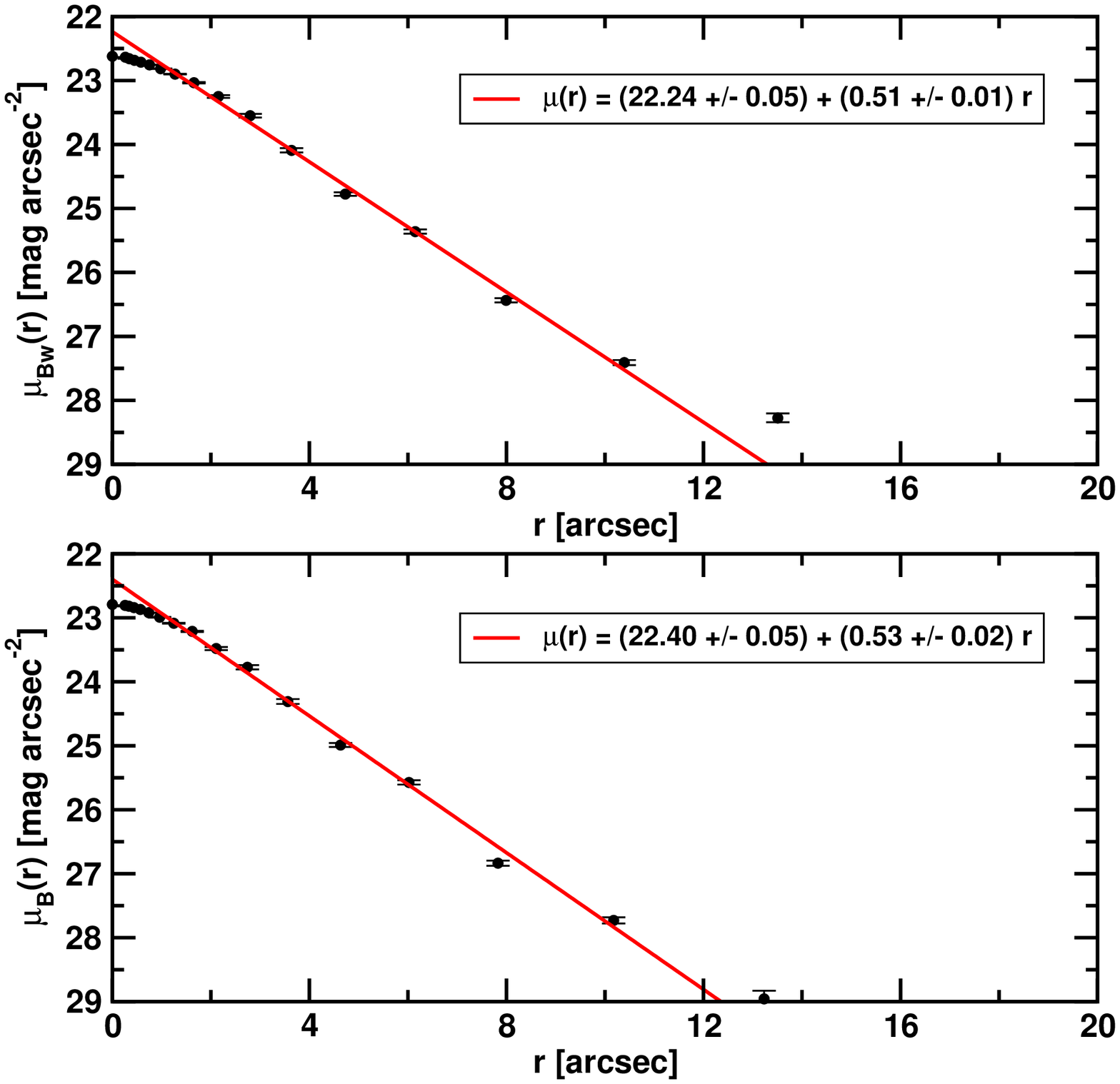}}
\caption{Surface brightness profiles in both filter bands of
  \object{LSB J22325-60211} 
  (left panels) and \object{LSB J22330-60543} (right panels) are displayed. For
  \object{LSB J22325-60211} a clear truncation of the profile is visible in
  both filter bands starting at the same position.}
\resizebox{7cm}{!}{\includegraphics{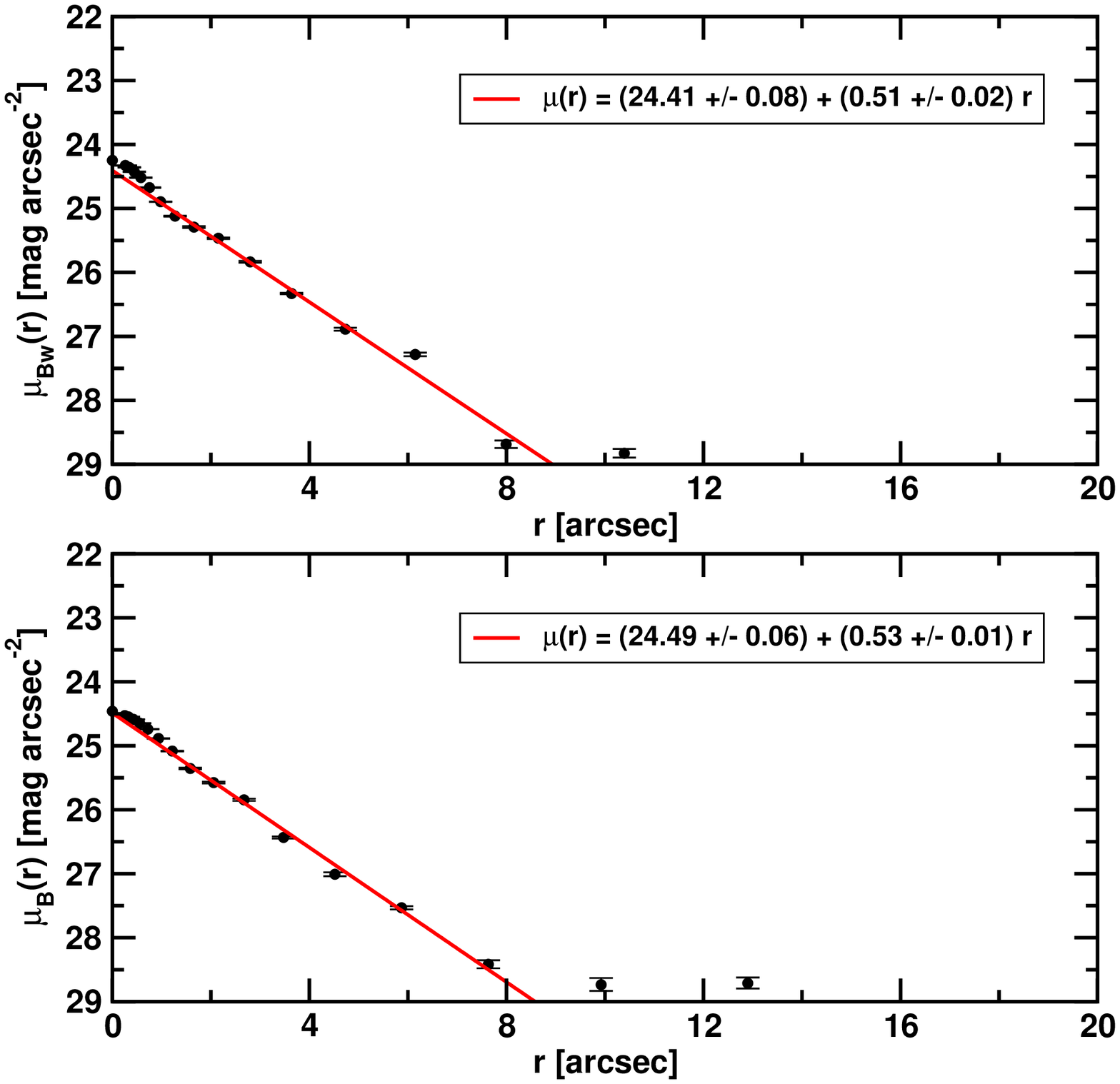}}
\resizebox{7cm}{!}{\includegraphics{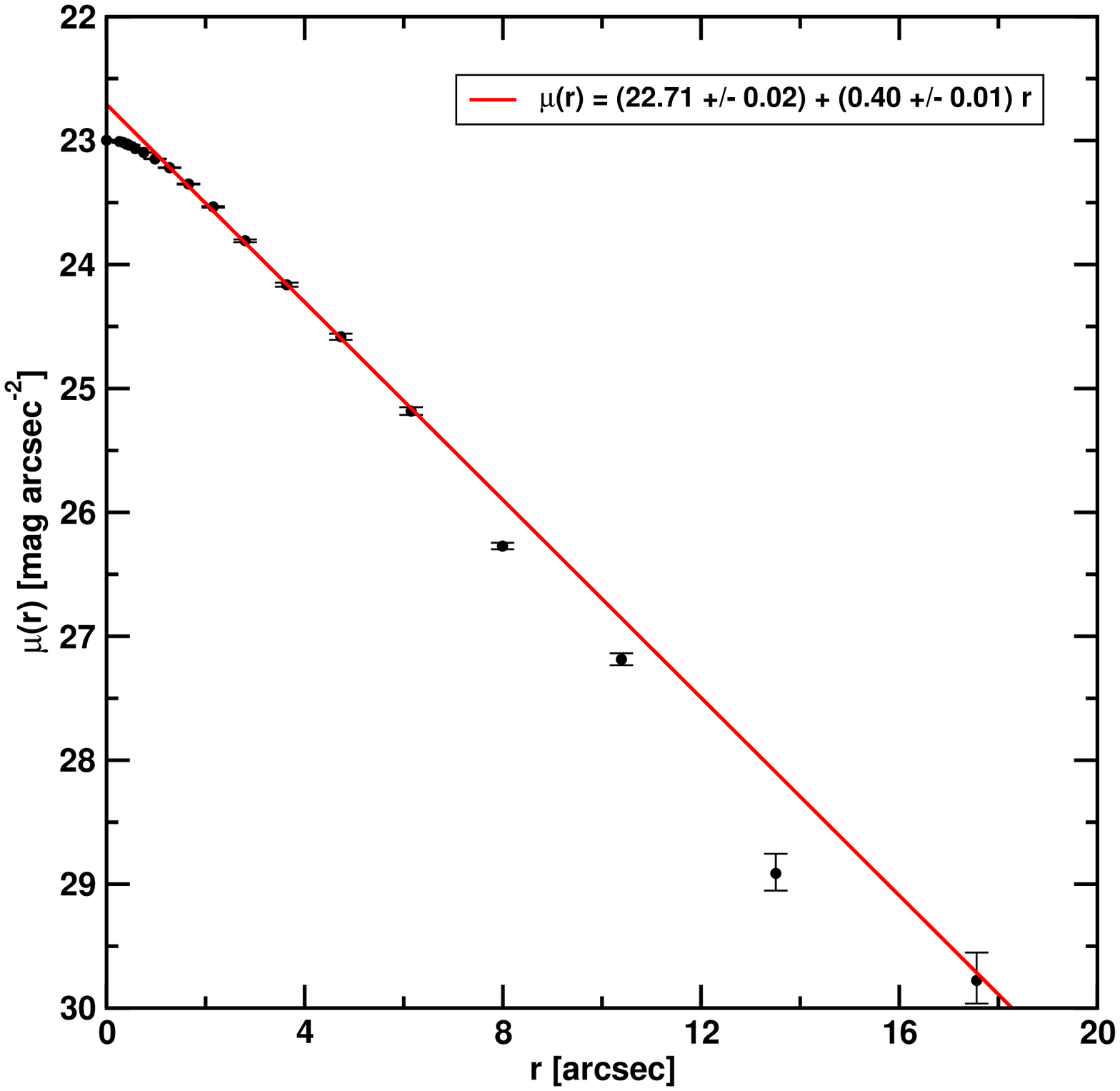}}
\caption{Surface brightness profiles of \object{LSB J22331-60340} (left panels)
  and \object{LSB J22332-60561} (right panel) are displayed.}
\label{radprof4}
\end{figure*}
\begin{figure*}
\centering
\resizebox{7cm}{!}{\includegraphics{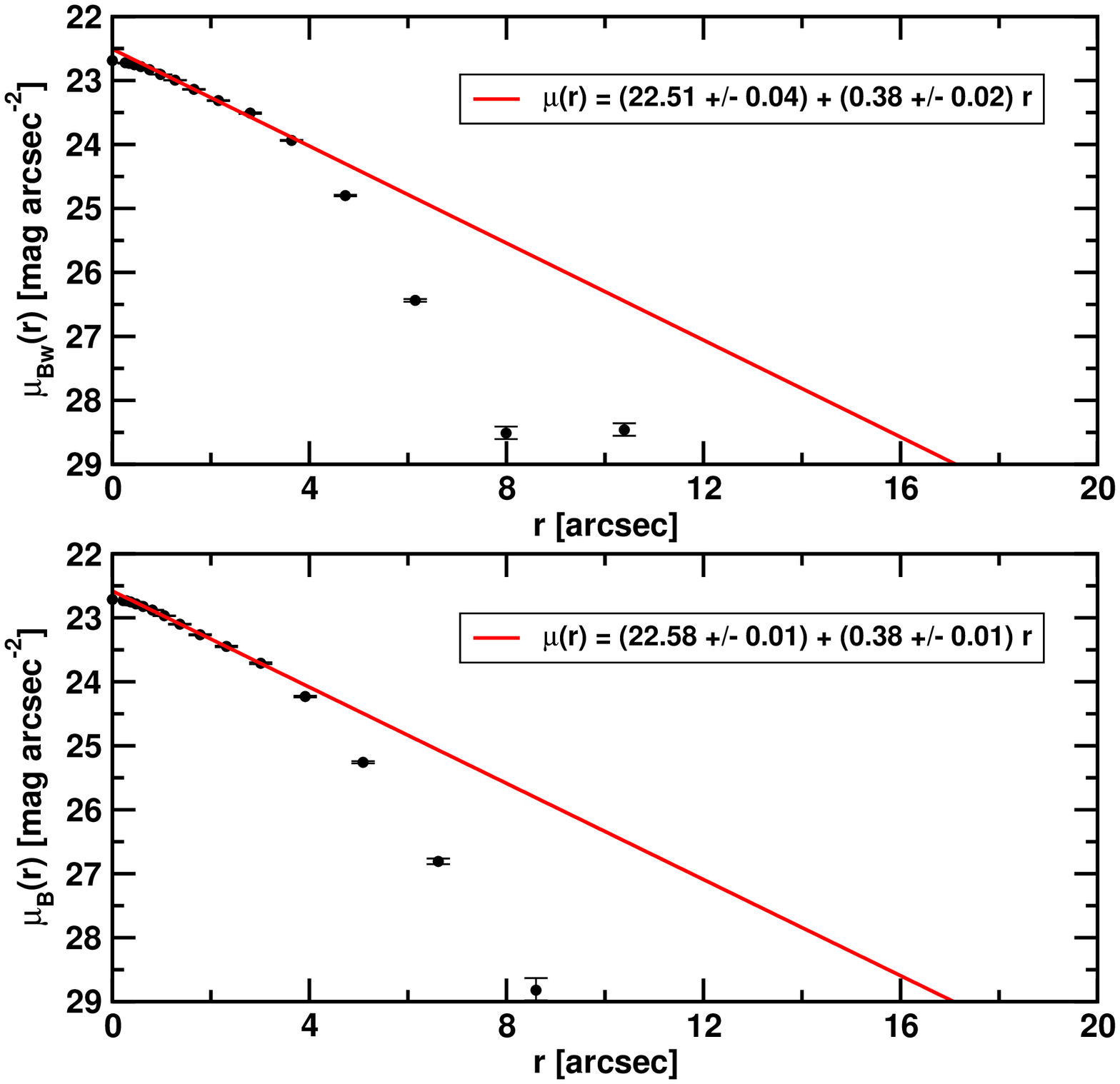}}
\resizebox{7cm}{!}{\includegraphics{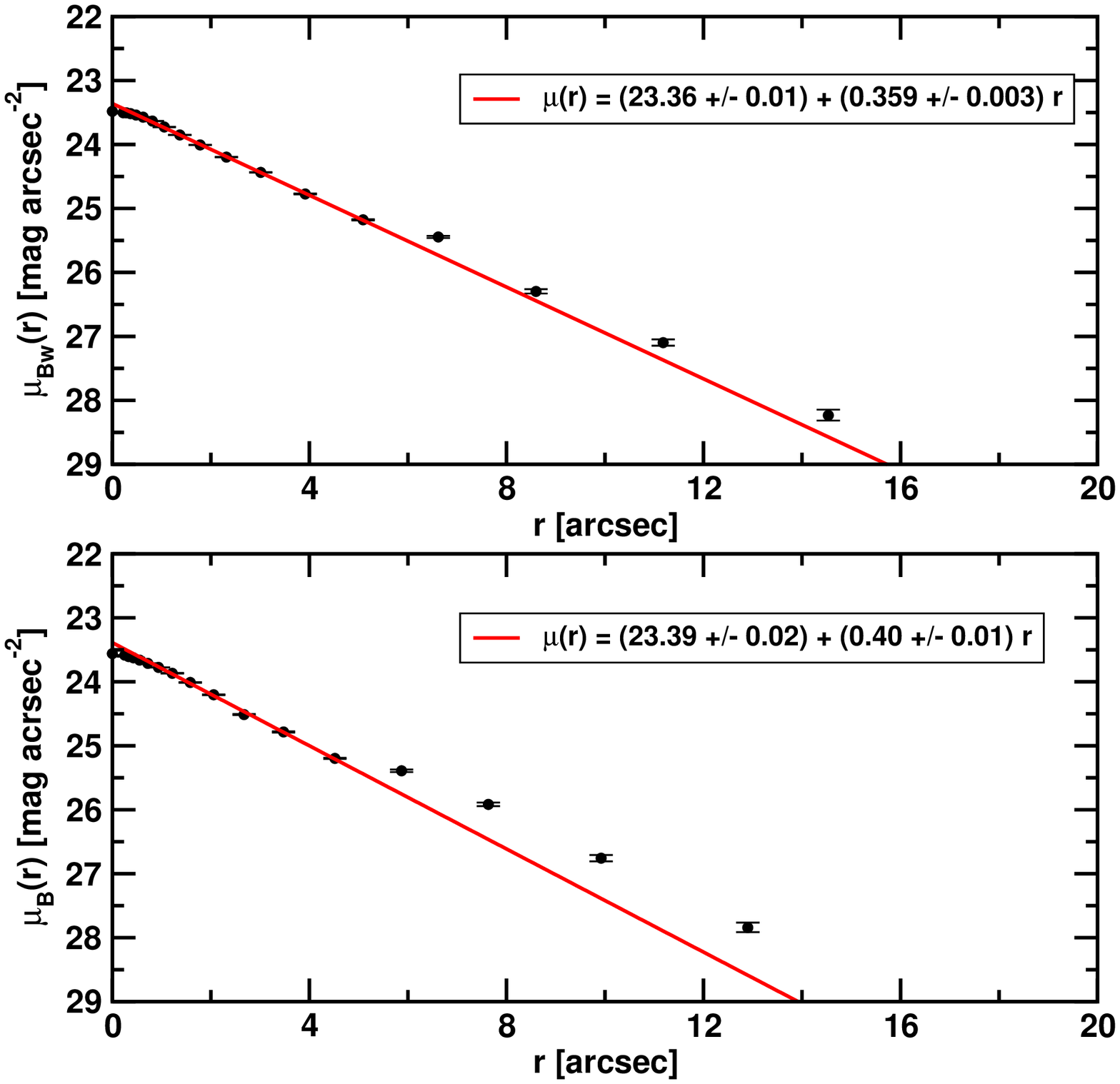}}
\caption{Surface brightness profiles in both filter bands of
  \object{LSB J22341-60475} (left panels) and \object{LSB J22342-60505} (right
  panels) are displayed. For \object{LSB J22341-60475} a clear truncation of
  the profile is visible in both filter bands starting at the same position.}
\resizebox{7cm}{!}{\includegraphics{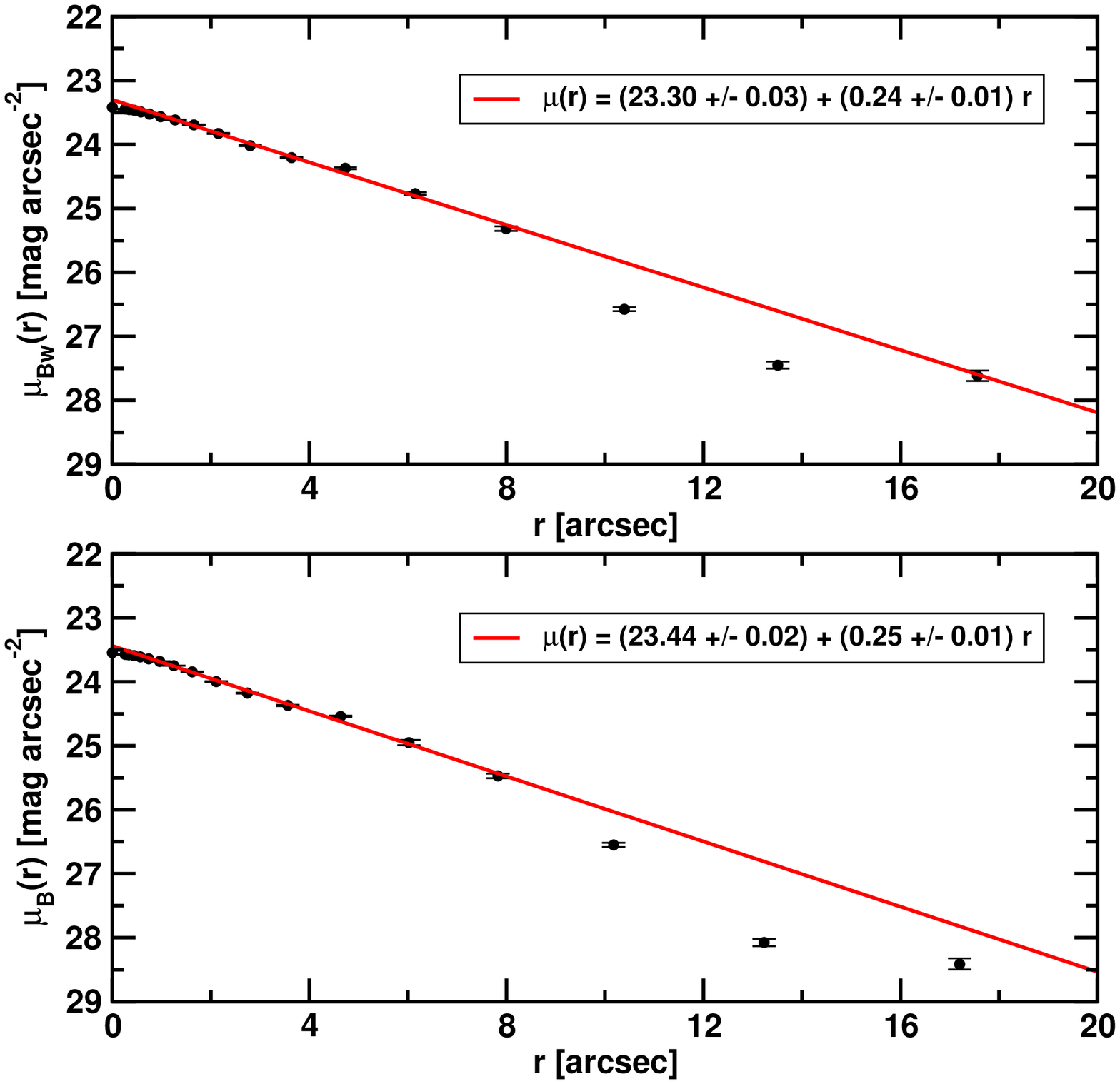}}
\resizebox{7cm}{!}{\includegraphics{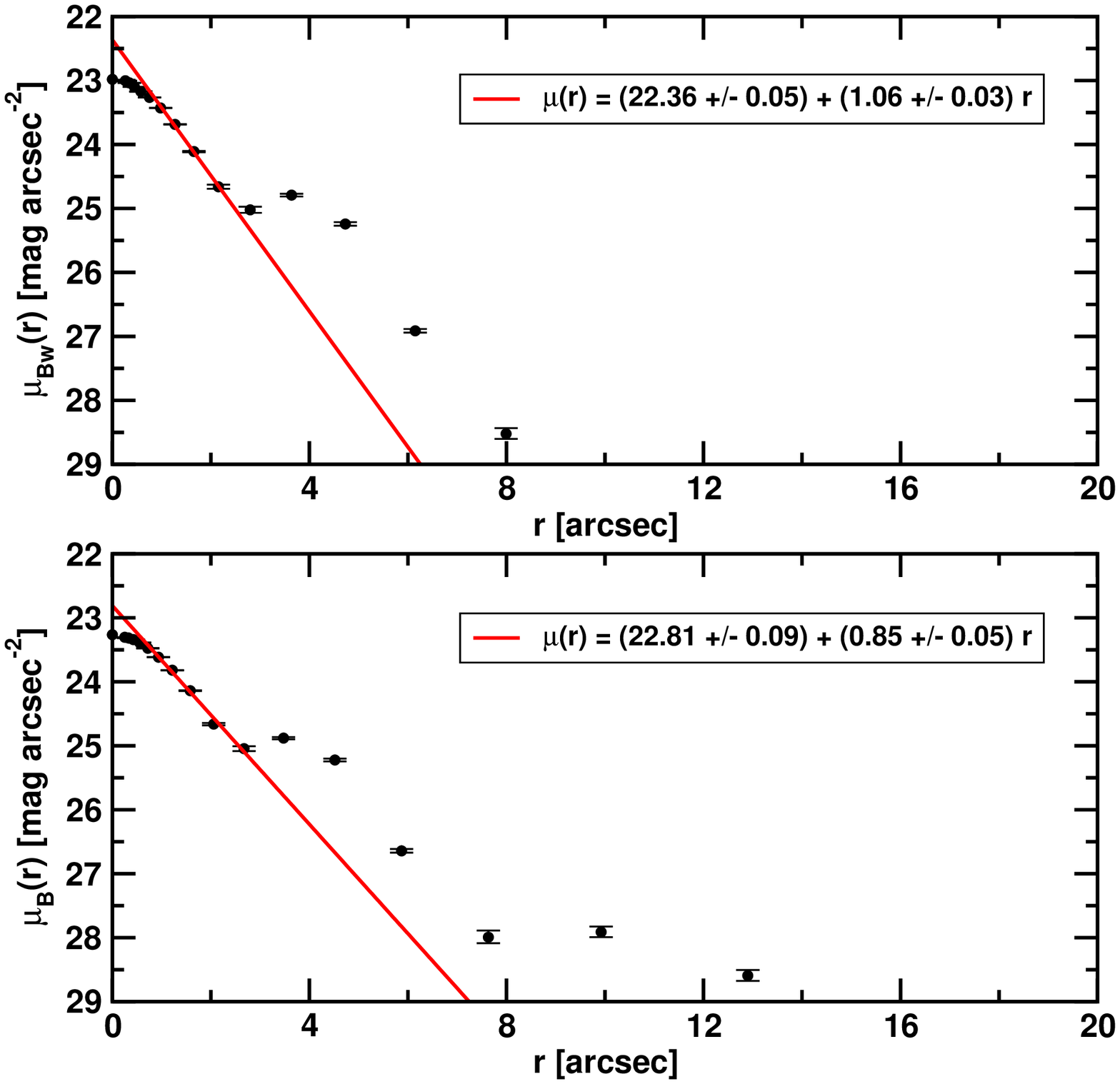}}
\caption{Surface brightness profiles in both filter bands of
  \object{LSB J22343-60222} (left panels) and \object{LSB J22345-60210} (right
  panels) are displayed. For \object{LSB J22343-60222} a truncation of the
  profile is visible in both filter bands
  starting at the same position.}
\resizebox{7cm}{!}{\includegraphics{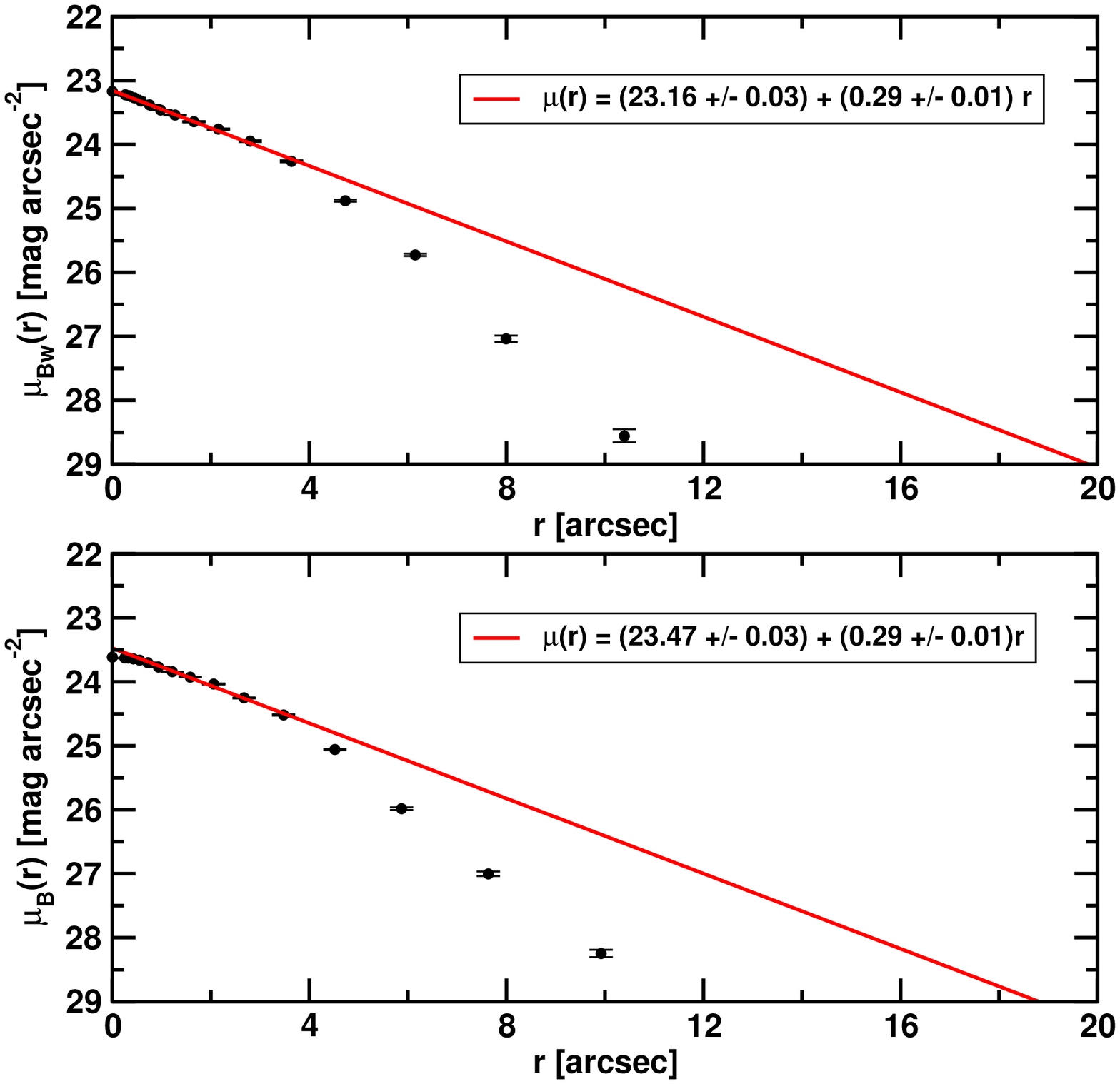}}
\resizebox{7cm}{!}{\includegraphics{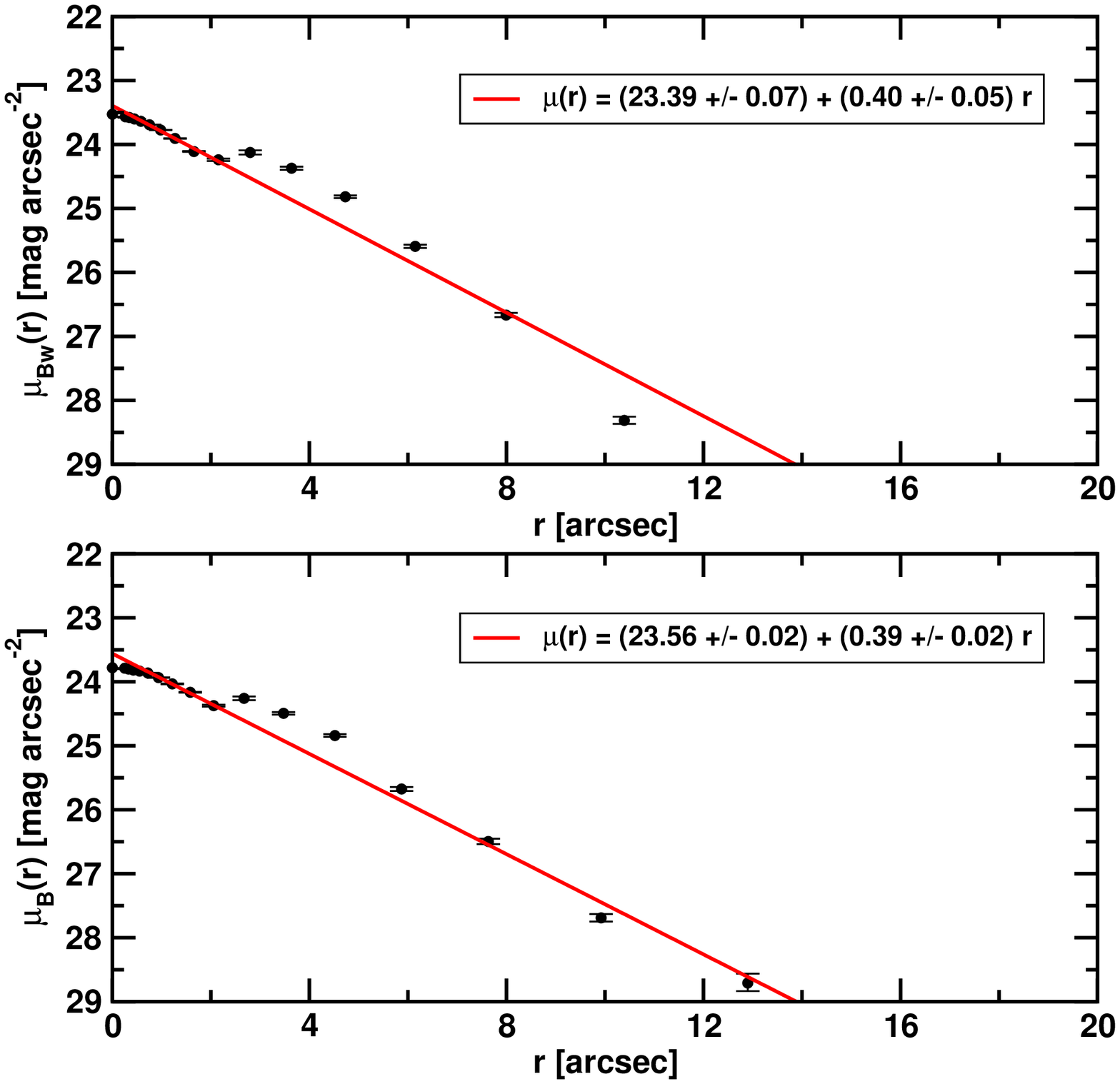}}
\caption{Surface brightness profiles in both filter bands of
  \object{LSB J22352-60420} (left panels) and \object{LSB J22353-60311} (right
  panels) are displayed. For \object{LSB J22352-60420} a clear truncation of
  the profile is visible in both filter bands starting at the same position.}
\label{radprof5}
\end{figure*}
\begin{figure*}
\centering
\resizebox{7cm}{!}{\includegraphics{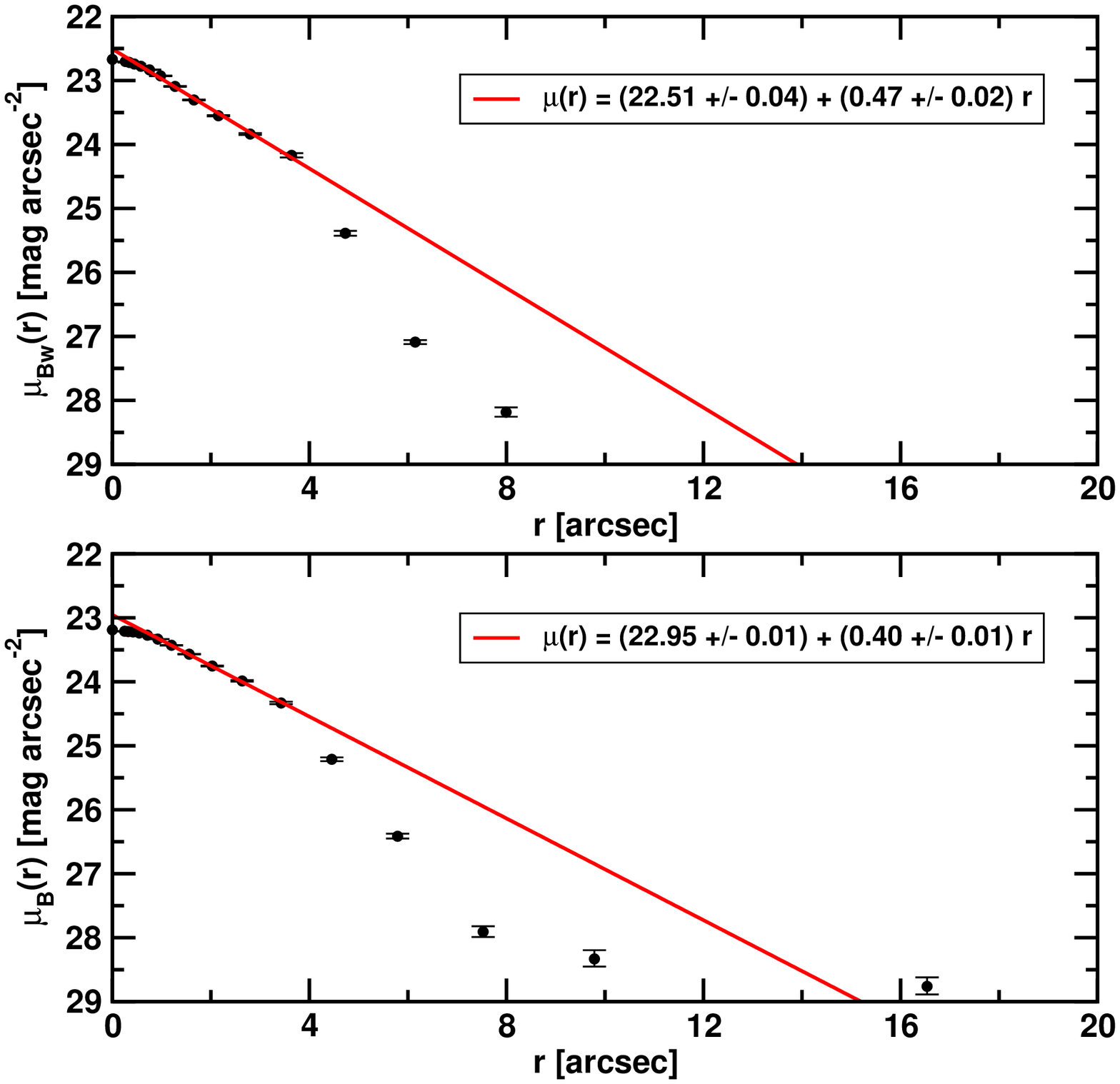}}
\resizebox{7cm}{!}{\includegraphics{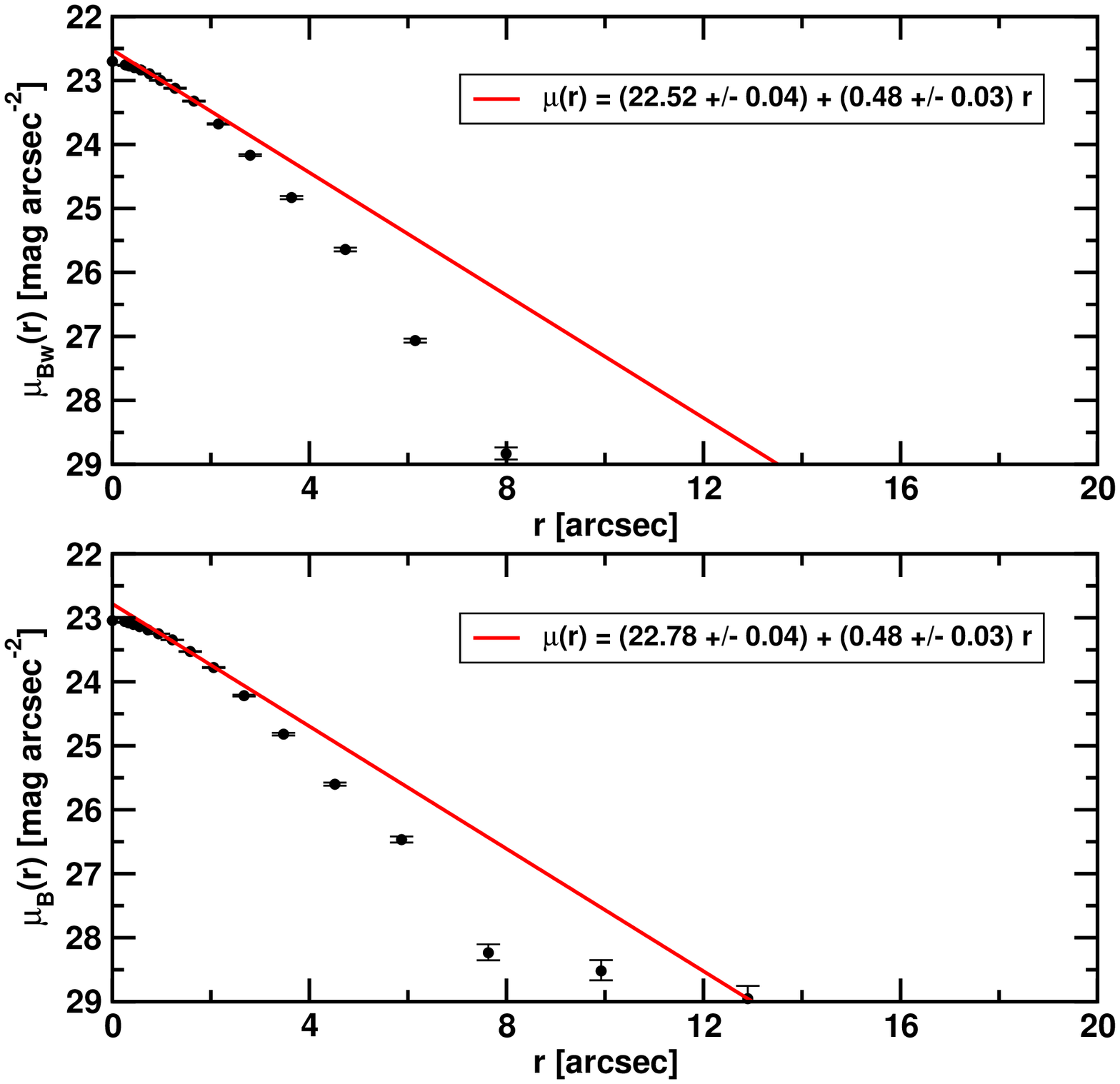}}
\caption{Surface brightness profiles in both filter bands of
  \object{LSB J22354-60122} (left panels) and \object{LSB J22355-60183} (right
  panels) are displayed. For both 
  galaxies a clear truncation of the profile is visible in both filter bands
  starting at the same position.}
\resizebox{7cm}{!}{\includegraphics{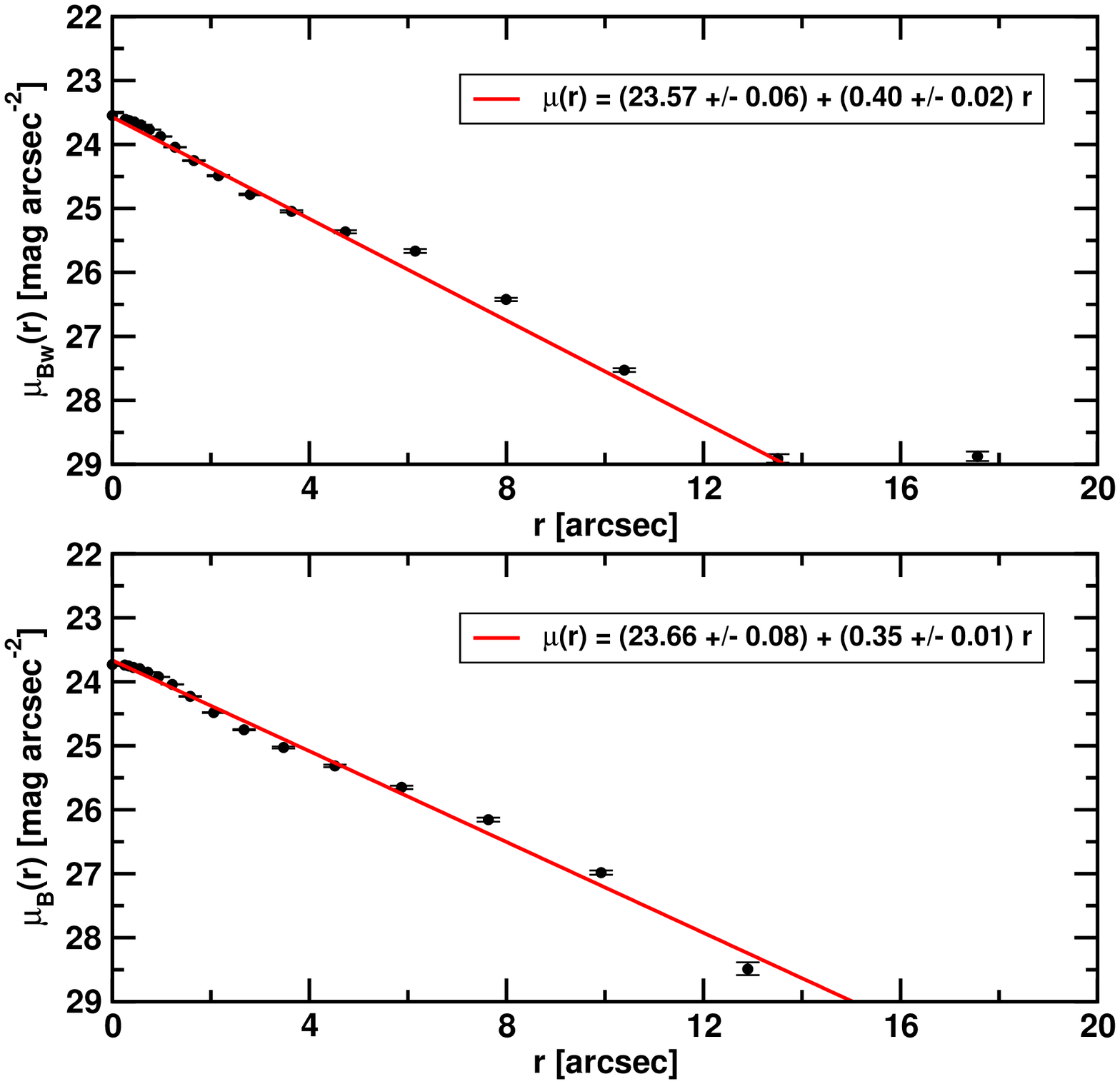}}
\resizebox{7cm}{!}{\includegraphics{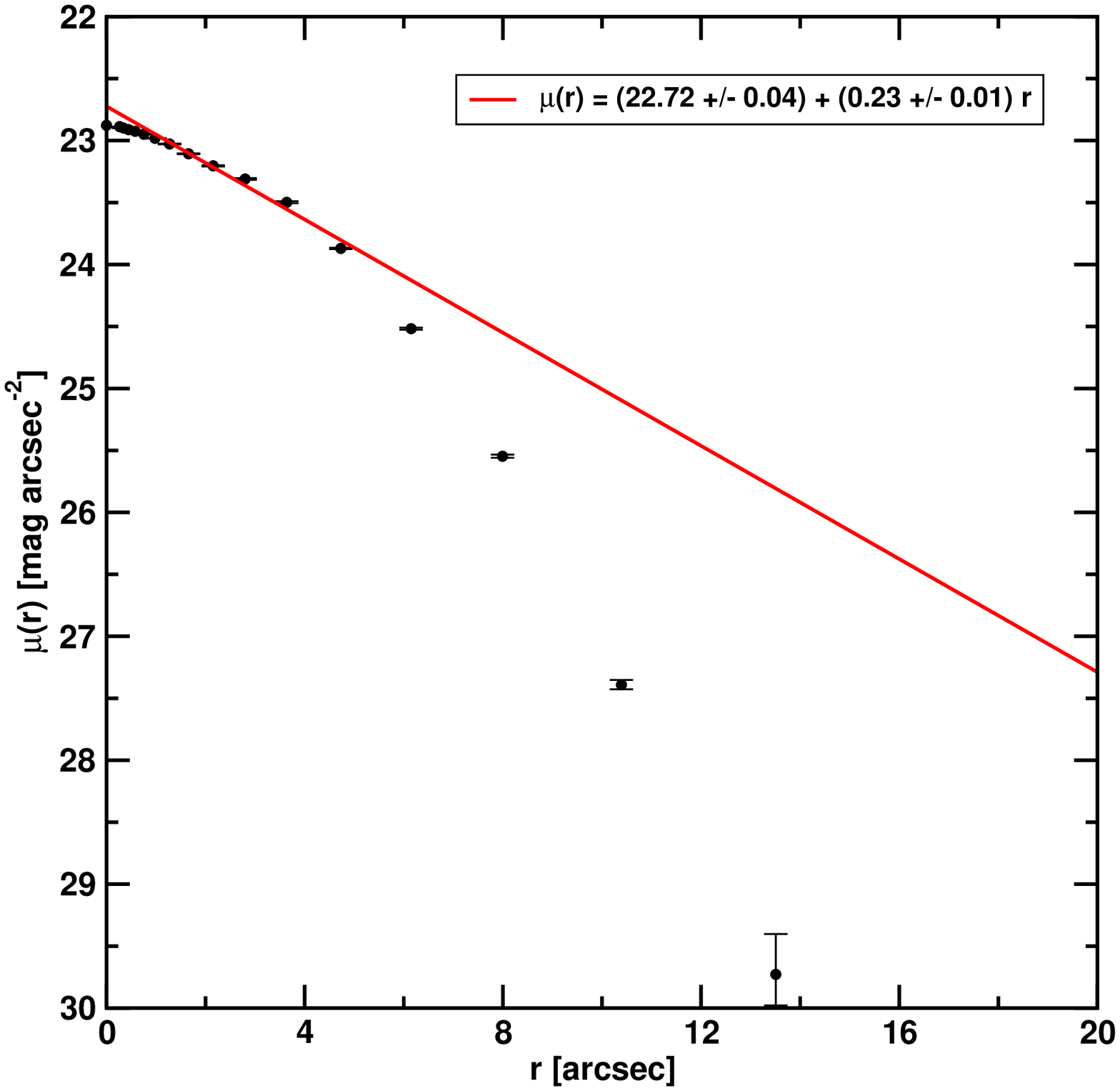}}
\caption{Surface brightness profiles of \object{LSB J22355-60390} (left panels)
  and \object{LSB J22360-60561} (right panel) are displayed. The profile of
  \object{LSB J22360-60561} is clearly truncated in the outer region.}
\resizebox{7cm}{!}{\includegraphics{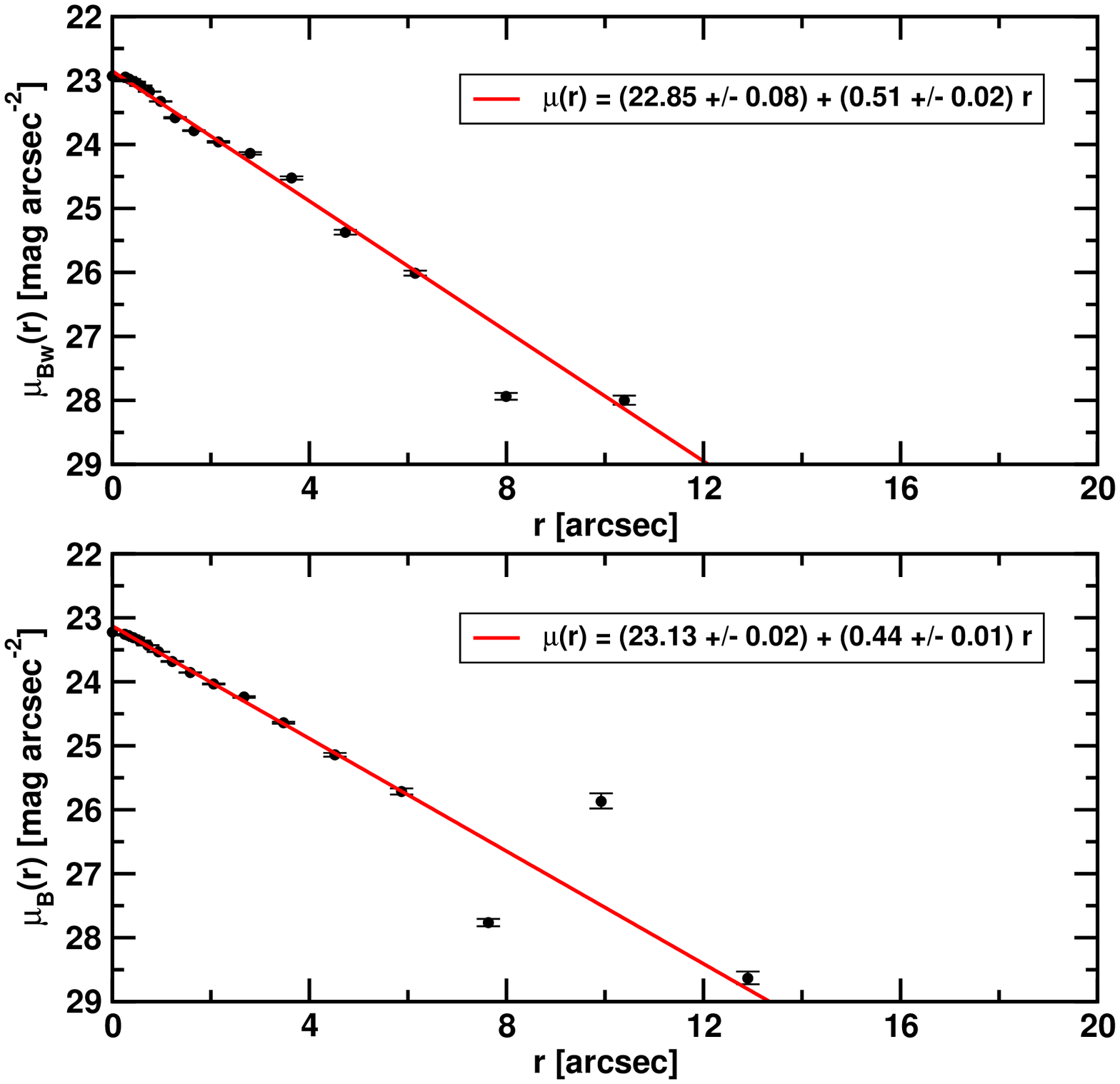}}
\resizebox{7cm}{!}{\includegraphics{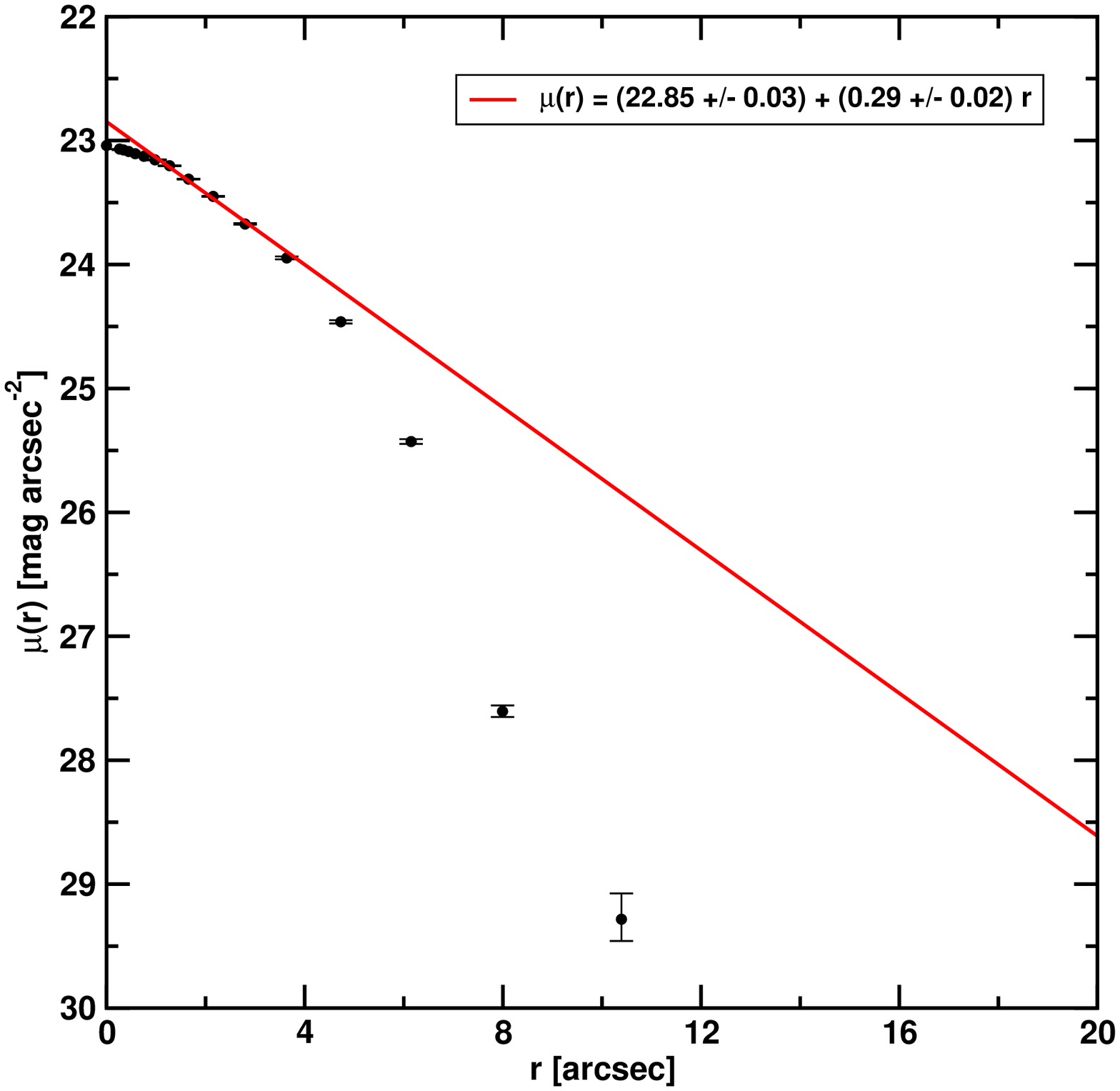}}
\caption{Surface brightness profiles of \object{LSB J22361-60223} (left panels)
  and \object{LSB J22361-60562} (right panel) are displayed. The profile of
  \object{LSB J22361-60562} is clearly truncated in the outer region.}
\label{radprof6}
\end{figure*}
\begin{figure*}
\centering
\resizebox{7cm}{!}{\includegraphics{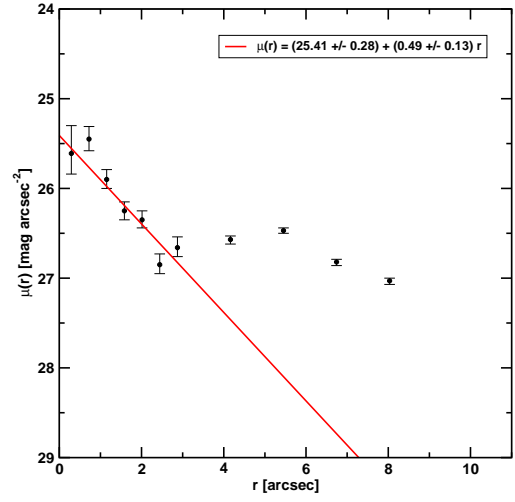}}
\caption{Surface brightness profile of \object{LSB J22364-60405} is
  displayed. This galaxy belongs to the subsample of 3 extreme LSB
  galaxies. This galaxy was found in region with higher noise level. The
  profile below a surface brightness level of 26.5\,mag\,arcsec$^{-2}$ is
  dominated by the noise.}
\label{radprof7}
\end{figure*}

\bibliographystyle{aa}
\bibliography{lsb_bib}
\listofobjects
\end{document}